# Deep learning for pancreas segmentation: a systematic review

**Andrea Moglia**
Department of Electronics,
Information, and Bioengineering
Polytechnic University of Milan
Milan, 20133, Italy
andrea.moglia@polimi.it

**Matteo Cavicchioli**
Department of Electronics,
Information, and Bioengineering
Polytechnic University of Milan
Milan, 20133, Italy
matteo.cavicchioli@polimi.it

**Luca Mainardi**
Department of Electronics,
Information, and Bioengineering
Polytechnic University of Milan
Milan, 20133, Italy
luca.mainardi@polimi.it

**Pietro Cerveri**
Department of Industrial,
and Information Engineering
University of Pavia
Pavia, 27100, Italy
pietro.cerveri@unipv.it
Department of Electronics,
Information, and Bioengineering
Polytechnic University of Milan
Milan, 20133, Italy
pietro.cerveri@polimi.it

July 24, 2024

## Abstract

Pancreas segmentation has been traditionally challenging due to its small size in computed tomography abdominal volumes, high variability of shape and positions among patients, and blurred boundaries due to low contrast between the pancreas and surrounding organs. Many deep learning models for pancreas segmentation have been proposed in the past few years. We present a thorough systematic review based on the Preferred Reporting Items for Systematic Reviews and Meta-analyses (PRISMA) statement. The literature search was conducted on PubMed, Web of Science, Scopus, and IEEE Xplore on original studies published in peer-reviewed journals from 2013 to 2023. Overall, 130 studies were retrieved. We initially provided an overview of the technical background of the most common network architectures and publicly available datasets. Then, the analysis of the studies combining visual presentation in tabular form and text description was reported. The tables grouped the studies specifying the application, dataset size, design (model architecture, learning strategy, and loss function), results, and main contributions. We first analyzed the studies focusing on parenchyma segmentation using coarse-to-fine approaches, multi-organ segmentation, semi-supervised learning, and unsupervised learning, followed by those studies on generalization to other datasets and those concerning the design of new loss functions. Then, we analyzed the studies on segmentation of tumors, cysts, and inflammation reporting multi-stage methods, semi-supervised learning, generalization to other datasets, and design of new loss functions. Finally, we provided a critical discussion on the subject based on the published evidence underlining current issues that need to be addressed before clinical translation.

# 1    Introduction

The pancreas is a small J-like-shaped glandular organ, located inside the deep part of the abdomen, and subdivided into three regions, namely head, body, and tail. A healthy pancreas generally occupies around 0.5% of a computer tomography (CT) abdominal volume [Zhou et al., 2023]. In some patients, the healthy tissue may be affected by disorders such as inflammation, e.g., pancreatitis, while in more severe cases it may be affected by cysts and tumors. The latter are particularly insidious because they generate few symptoms and are often diagnosed at an advanced stage. In addition, they are very aggressive and lethal. Pancreas tumors are the fourth leading cause of death among all cancer types in the United States for the male gender and the third one for the female gender Siegel et al. [2024]. A five-year survival rate of 13% was reported in the United States in the period 2013-2019, which is the lowest one among all cancer types Siegel et al. [2024]. The diagnosis of a pancreatic mass involves clinical assessment, laboratory testing, and advanced imaging techniques. Patient history and physical examination are initially performed to identify symptoms and risk factors. Laboratory tests on blood samples are subsequently conducted to measure CA 19-9 marker, before imaging tests. Ultrasound is usually the preliminary imaging assessment tool, followed by CT or magnetic resonance imaging (MRI) to delineate the tumor size and location in more detail. In particular CT scans are pivotal for staging cancer, evaluating its resectability, and planning surgical interventions. In fact pancreas surgery requires accurate recognition of anatomical variations and the spatial relationships of the tumor location with the surrounding vessels and organs in order to determine the optimal location of the pancreas resection Miyamoto et al. [2024]. MRI provides excellent soft tissue contrast, highlighting vascular and ductal details Şolea et al. [2024]. The recent guidelines of the European Society for Medical Oncology recommended CT as the primary modality for detailing tumor characteristics and spread Conroy et al. [2023].

Given the rising demand for enhanced early detection of pancreatic diseases, precise segmentation from medical images has become imperative. In this regard, its segmentation from medical images is a prerequisite for accurate computer-assisted diagnosis, surgical navigation, post-surgical follow-up, and radiotherapy.

## 1.1    Challenges in pancreas segmentation

Traditionally, medical image segmentation, including pancreas segmentation, has relied heavily on manual delineation by expert radiologists. This poses critical challenges including inter- and intra-observer variability, time-consuming labor, and subjective interpretation. Limited availability of experts, human error, and scalability issues further complicate the process. Extensive training requirements and reproducibility concerns hinder the widespread adoption of manual segmentation methods [Chen et al., 2022a]. Thus, there is an urgent need for efficient and reliable approaches to pancreas segmentation. The segmentation of the pancreas is very challenging, but it is even more difficult in the case of tumors and inflammations since the conditions are exacerbated. Firstly, whereas the pancreas is very small, typically representing a small fraction of the CT volume, pancreatic tumors are even smaller, with most of them accounting for less than 0.1% of the entire CT abdominal volume. Secondly, the contrast between the pancreas and its surrounding organs in CT scans is weak, which is caused by the similar range of voxel intensities. As a consequence, the boundaries of the pancreas and tumors are blurred, and the contrast with surrounding tissues is low, especially at the head of the pancreas. As a result, it is difficult to distinguish not only between the pancreas and the duodenum but also between the tissue (parenchyma) and tumors of the pancreas Zhou et al. [2023], Dai et al. [2023]. Likewise, the segmentation of an inflamed pancreas is more challenging than a normal one since it invades the surrounding organs causing blurry boundaries, and it has higher shape, size, and location variability than the normal pancreas Deng et al. [2023]. As such, boundary errors remain critical in preoperative planning of the pancreas, such as tumor resections and organ transplantation. Thirdly, the pancreas exhibits an irregular shape and susceptibility to deformation, complicating accurate segmentation. Anatomical variations in size, shape, and tumor positioning among patients, particularly the diverse locations of pancreatic tumors, pose challenges in distinguishing parenchyma from cancerous masses Zhou et al. [2023], Dai et al. [2023]. Lastly, differences in commercial CT scanners and CT phases can lead to significant variances in organ appearances Ma et al. [2022].

## 1.2    Work motivation

Progress in the past decade in deep learning (DL) has led to continuous improvements in medical imaging, including pancreas segmentation. An overview of pancreas segmentation based on DL is depicted in 1. Even though in the last years several reviews have delved into pancreas segmentation from CT scans using AI [Ghorpade et al., 2023, Kumar et al., 2019, Huang et al., 2022a, Yao et al., 2019, Aljabri and AlGhamdi, 2022, Rehman and Khan, 2020, Senkyire and Liu, 2021], our preliminary literature search has unearthed a significant number of studies overlooked by them. These considerations underscore the necessity for an updated systematic review to comprehensively cover the latest advancements in the field. Consequently, the goal of this review is to present systematically an in-depth analysis of DL for the segmentation of the parenchyma, tumors, cysts, and inflammation of the pancreas starting from CT scans.

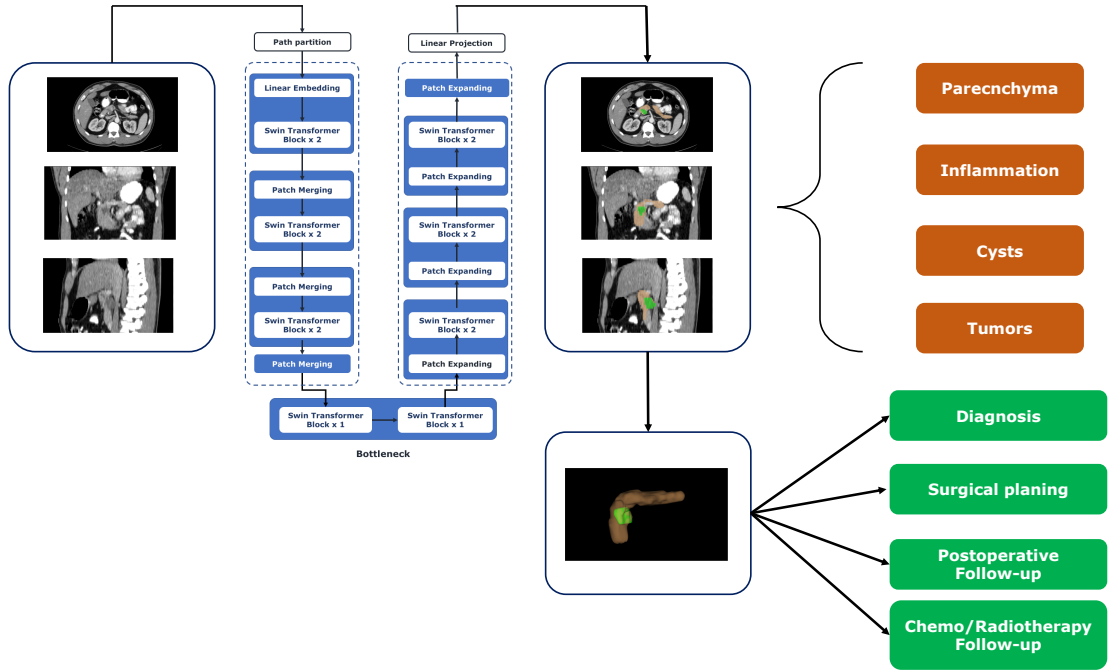

Figure 1: Overview of pancreas segmentation based on DL. Radiological images are processed by neural networks models outputting masks of the organ or lesions (e.g. cysts, and tumors). Applications include diagnosis, surgical planning, postoperative follow-up, and chemo/radiotherapy follow-up.

## 1.3 Structure and contribution of the work

The review is structured as follows. In Section 2 we describe the method to perform the literature search and extract the included studies. We also report the limitations of the published reviews in the field. In Section 3 we illustrate the main DL architectures, the available public datasets, metrics, and loss functions for pancreas segmentation. In Section 4 and Section 5 we present the results on DL for the segmentation of parenchyma, tumors, and other lesions of the pancreas. In Section 6 we discuss the findings of the review. Our major contributions are the following:

- description of the main DL architectures used for pancreas segmentation;

- systematic and extensive review on the technical advancements of DL for pancreas segmentation (parenchyma, tumors, cysts, and inflammation);

- visual presentation of all retrieved studies in tabular form in terms of application, dataset size, DL architecture, learning strategy, loss functions, results, and main contributions. The full list is available in Appendix;

- a thorough description of the proposed approaches in the studies;

- a comparison of the performances of the DL approaches for the various applications.

## 2 Methods

### 2.1 Literature search

In October 2023, a literature search was conducted on PubMed, Web of Science, and Scopus following the Preferred Reporting Items for Systematic Reviews and Meta-analyses (PRISMA) statement [Page et al., 2021]. The search was limited to articles in the English language with an abstract and published from January 1st, 2013 to October 31st, 2023. The following search terms were used: ("artificial intelligence" OR "deep learning" OR "convolutional neural network"

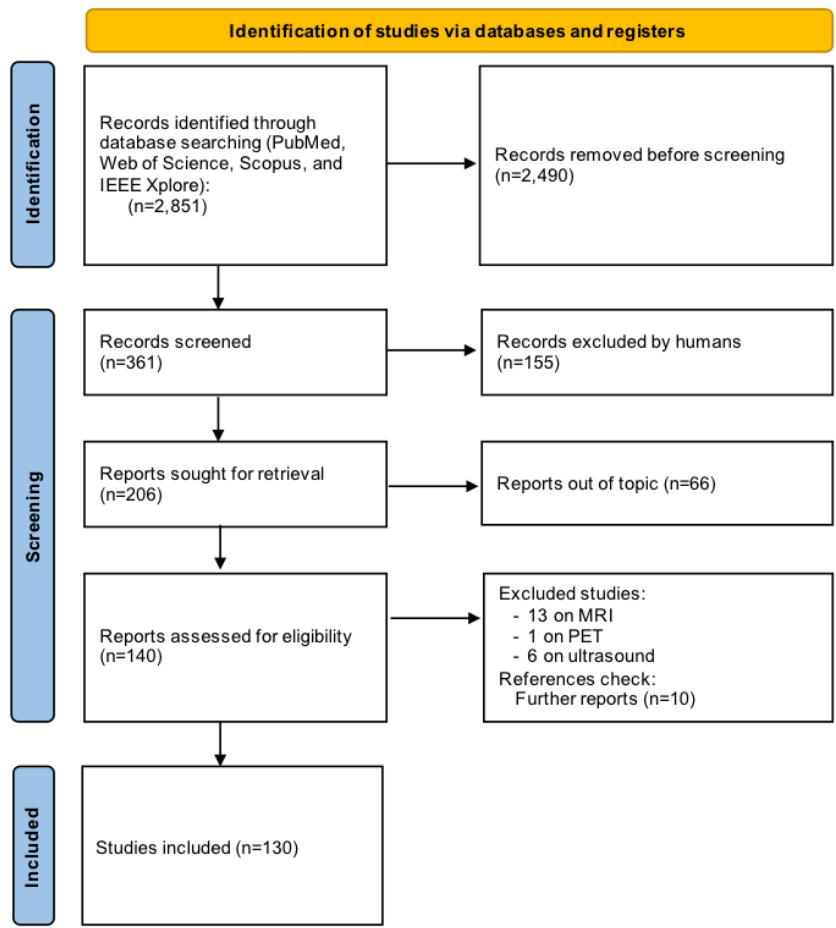

Figure 2: Flow chart of the study selection process according to the Preferred Reporting Items for Systematic Reviews and Meta-analyses (PRISMA) 2020 statement [Page et al., 2021]

OR "segmentation" OR "self-supervised learning" OR "supervised learning" OR "generative artificial intelligence" OR "encoder" OR "decoder") AND ("pancreas" OR "surgical planning pancreas" OR "preoperative planning pancreas"). Reviews, letters, non-peer-reviewed articles, conference abstracts, and proceedings were excluded from the analysis.

## 2.2 Data extraction

Identified articles were screened by title and abstract, followed by full-text review, data extraction, and review of references. Two reviewers (AM and MC) independently screened titles and abstracts for relevance. The sample, phenomenon of interest, design, evaluation, and research type (SPIDER) tool was used to organize relevant information for a subsequent visual presentation in the tabular form [Cooke et al., 2012]. In case of insufficient information, the corresponding authors of the articles concerned were contacted for further details. References were checked to retrieve further studies.

## 2.3 Data analysis

For each group, a table was prepared to visually present the data of the studies. A customized SPIDER tool was applied to the studies of each group, reporting: the dataset size (Sample), the application (Phenomenon of Interest), the model architecture, the learning strategy and loss function (Design), the results (Evaluation), and the main contributions of the study (Research).

## 2.4   Results of the literature search

The database search retrieved 2,851 results. After title and abstract screening, the full texts of 206 reported studies were analyzed, but only 140 were found eligible for inclusion. Twenty studies using imaging acquisition other than CT (magnetic resonance imaging (MRI), positron emission tomography (PET), and ultrasound) were excluded. The list of excluded articles and the reasons for exclusion are reported in Section 2.5. Ten additional studies were retrieved after a manual check of the references. A total of 130 studies were included for full-text analysis (Fig. 2). By considering the involved countries (Fig. 3, left panel), China led the ranking with a share of 54.4%, followed by the United States (17.1%), the United Kingdom (5.3%), Canada (3.5%), and Japan (3.5%). In the majority of studies, 3D neural networks (Fig. 3, central panel) were used (51.4%), followed by 2D models (42.7%), and 2.5D (5.8%). By considering the learning type, the vast majority concern studies on supervised learning (83.8%), followed by semi-supervised learning (9.5%), and unsupervised learning (4.4%). Other types of learning (reinforcement, weakly, and continual) are reported in 2.2% of the studies (Fig. 3, right panel). Overall, there is a positive trend in the number of published articles in peer-reviewed journals, included in the present review, even though the data for the year 2023 are available until October 31st (Fig. 4). Notably, there has been a surge in the number of studies on DL for the segmentation of pancreas tumors in 2023. The 130 reviewed studies were published in high-quality peer-reviewed journals with a mean 2023 impact factor of 5.39 (latest available data according to the Web of Science). As can be seen from Fig. 5 the studies were most frequently published in prominent journals in the medical imaging domain, like Medical Imaging Analysis and IEEE Transactions on Medical Imaging, with 11 and 12 publications, respectively. Of note, there are other studies published in leading journals like IEEE Transactions on Pattern Analysis and Machine Intelligence, IEEE Transactions on Image Processing, and Nature Methods.

## 2.5   Excluded studies on MRI, PET, and Ultrasound

The retrieval of the full-text articles included also 13 studies on MRI [Mazor et al., 2024, Yang et al., 2022a, Ding et al., 2022, Zhang et al., 2022, Kart et al., 2021, Chen et al., 2020a, Fu et al., 2018, Li et al., 2023a, Jiang et al., 2023, Liu et al., 2023, Li et al., 2023b, 2022a, Liang et al., 2020], one on PET [Zhang et al., 2023], and six on ultrasound [Yao et al., 2021, Fleurentin et al., 2023, Iwasa et al., 2021, Tang et al., 2023a,b, Seo et al., 2022]. After analysis, they were all excluded since they did not introduce technical advancements in terms of DL architectures, design of loss functions, semi-supervised, or unsupervised learning. In contrast, one study combining CT and MRI [Li et al., 2022b], and two combining CT and PET [Sundar et al., 2022, Wang et al., 2023] were included.

## 2.6   Limitations of published reviews

The published reviews are reported in Table 1. The most recent one was performed by [Ghorpade et al., 2023] and published in 2023. It is a narrative review of 44 studies (32 on parenchyma and 12 on tumors of the pancreas). The only systematic review on pancreas segmentation was performed by Kumar et al. [2019], which may be considered obsolete given the surge of published articles since 2020. It analyzed 19 studies (16 on CT and three on magnetic resonance). The review by Huang et al. [2022a] concerned artificial intelligence (AI) on pancreas cancer. Out of the included studies, only seven pertain to DL for pancreas segmentation. The review by Yao et al. [2019] discussed different approaches to pancreas segmentation, with 12 studies on AI. The other reviews reported the published literature on DL on medical images of several anatomical structures (organs, and bones) in addition to the pancreas [Aljabri and AlGhamdi, 2022, Rehman and Khan, 2020, Senkyire and Liu, 2021]. As can be seen from Table 1 the number of the included studies on the published reviews on pancreas segmentation is considerably lower than the results of our literature search.

## 2.7   Research questions

By using the SPIDER tool, the following research questions were elaborated to frame a thorough analysis of the published literature.

RQ1: Which datasets (publicly available and/or private) were used for pancreas segmentation based on DL?

RQ2: What are the approaches for pancreas segmentation based on DL?

RQ3: Which DL models were specifically designed?

RQ4: What are the performances of these models and how do they compare?

RQ5: What are the main contributions of the studies?

Table 1: Published reviews

| Reference | Type of review | Databases | Covered years | Reviewed studies on CT | Pancreas specific |
|---|---|---|---|---|---|
| Aljabri and AlGhamdi [2022] | Systematic | Google Scholar | 2014 - 2021 | 4 (parenchyma) | No |
| Ghorpade et al. [2023] | Narrative | PubMed and Web of Science | 2013 - 2023 | 32 (parenchyma) 12 (tumors and cysts) | Yes |
| Huang et al. [2022a] | Narrative | PubMed, Embase, and Web of Science | Until 2022 | 7 (tumors) | Yes |
| Kumar et al. [2019] | Systematic | MEDLINE, Espacenet, Google Patents, and the United States Patent and Trademark Office Patent | Until 2018 | 16 (parenchyma) | Yes |
| Rehman and Khan [2020] | Narrative | – | Until 2019 | 8 (parenchyma) | No |
| Senkyire and Liu [2021] | Narrative | PubMed, Scopus, and Web of Science | Until 2020 | 13 (parenchyma) | No |
| Yao et al. [2019] | Narrative | Web of Science | 2012 - 2018 | 12 (parenchyma) | Yes |

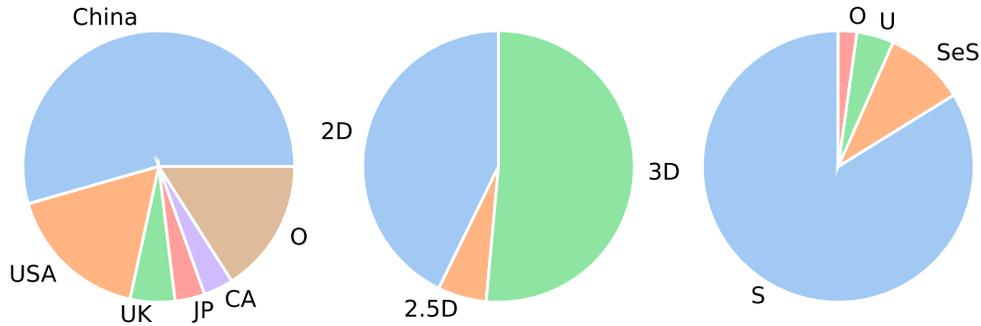

Figure 3: Share of reviewed studies by country of affiliated institutions of authors (left), class of network (middle), and type of learning (right). USA = United States of America, UK = United Kingdom, JP = Japan, CA = Canada, O = Other, S = supervised, SeS = semi-supervised, U = unsupervised

## 3 Technical background of deep learning techniques in pancreas segmentation

In this section, after an overview of abdominal organ segmentation methods, the DL architectures specifically used for pancreas segmentation are illustrated. They are foundational to the interpretation of the results of the reviewed studies.

### 3.1 Methods of abdominal organ segmentation

The methods of abdominal organ segmentation can be divided into model-based and learning-based ones [Ma et al., 2022]. The former generally reframe the image segmentation task as an energy functional minimization problem or explicitly match an atlas to a new image, such as variational models, statistical shape models, and atlas-based methods [Ma et al., 2022]. Statistical models involve the co-registration of images in a training dataset to derive anatomical correspondences, building a statistical model of the distribution of shapes and/or appearances of the corresponding anatomy in the training data, and fitting the resulting model to new images [Gibson et al., 2018]. The multi-atlas registration and label fusion method was proposed for automatic pancreas segmentation, to optimize organ labeling for each pixel by adopting a volumetric multiple atlas registration and robust label fusion [Li et al., 2020a]. Unfortunately, model-based approaches fail to segment the organs with weak boundaries and low contrasts like pancreas [Ma et al., 2022]. Learning-based methods extract meaningful features from annotated CT scans to distinguish target organs [Ma et al., 2022]. Learning-based methods can be categorized into supervised learning methods if the datasets are labeled; semi-supervised learning if a small amount of labeled is combined with a large amount of unlabelled data to extract knowledge from the unlabelled data, e.g. generating pseudo annotations for unlabeled examples, which are used jointly with labeled data to train the model (pseudo-labeling); unsupervised learning when the model learns the underlying

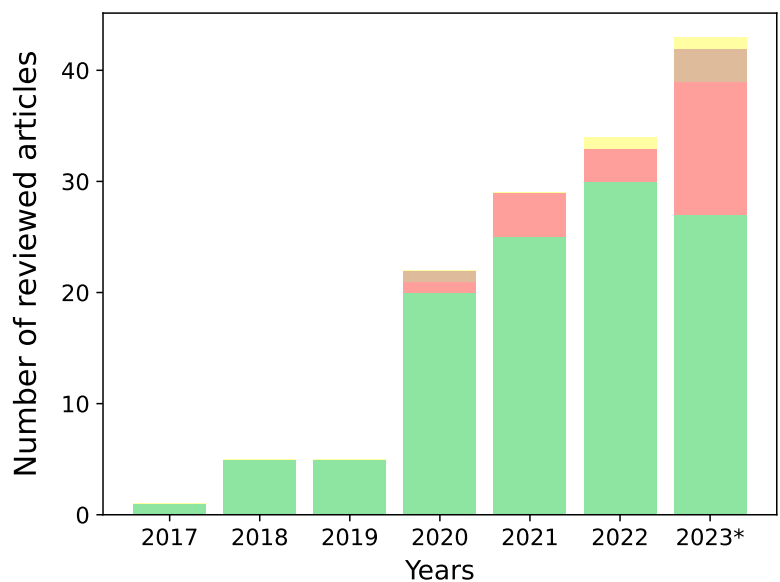

Figure 4: Annual distribution of the 130 reviewed articles. Studies on parenchyma (in green), tumors (in pink), cysts (in brown), and inflammation (in yellow). Note: some studies concerned more than one application, e.g. parenchyma, and tumor. *Data for the year 2023 are available until October 31st

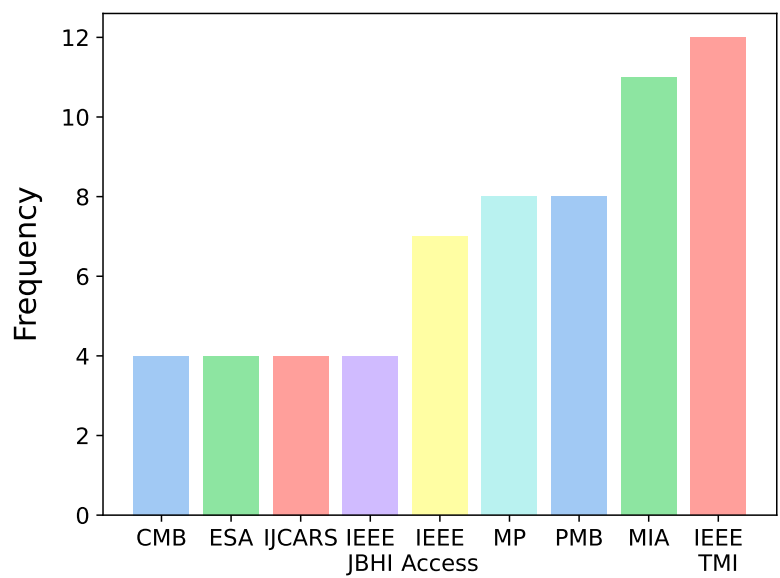

Figure 5: Most frequent journals publishing the reviewed studies. CMB = Computers in Biology and Medicine, ESA = Expert Systems with Applications, IJCARS = International Journal of Computer Assisted Radiology and Surgery, IEEE JBHI = IEEE Journal of Biomedical and Health Informatics, MP = Medical Physics, PMB = Physics in Medicine & Biology, MIA = Medical Image Analysis, IEEE TMI = IEEE Transactions on Medical Imaging

patterns or hidden data structures without labels; weakly supervised learning using weak annotations like scribbles; and continual learning to learn new tasks without forgetting the learned ones [Ma et al., 2022, Chen et al., 2022a].

With the recent advancements in DL, convolutional neural networks (CNNs) were proposed as a learning-based method and applied to different tasks of medical imaging, e.g. classification, detection, and segmentation [Chen et al., 2022a]. The CNNs for medical imaging segmentation can be categorized into 2D, 2.5D, and 3D models. In 2D networks, the data are sliced along one of the three image planes (axial, sagittal, and coronal). Then, the 2D slices are sent as the input to the DL model [Zhang et al., 2021a]. They are computationally efficient but lack the spatial context to extract the interslice information embedded in volumetric CT data [Wang et al., 2021a]. In contrast, 3D models use the entire CT volume as the input of the network, which can capture 3D spatial information of the CT volume. However, they are computationally expensive [Yan and Zhang, 2021]. In 2.5D models, three 2D models segment separately the input image in three image planes. Then, the segmentation is obtained by fusing the results of the three 2D models, for instance through voting [Zhang et al., 2021a]. 2.5D models represent a compromise between 2D and 3D ones, by making up for the lack of spatial context information of 2D models, but at the same time reducing the computational cost of 3D models [Dai et al., 2023].

## 3.2 UNet and its variants

UNet is a U-shape fully connected network (FCN) with an encoder and decoder. The encoder extracts features through convolutions, while the encoder restores the initial resolution of the input image through deconvolutions. The key innovation of UNet is represented by skip connections between opposing convolutional and deconvolutional layers [Ronneberger et al., 2015]. Skip connections successfully concatenate features learned at different levels to improve the segmentation performance, especially at the level of localization [Chen et al., 2022a]. 3D UNet is the counterpart of UNet, where the 2D operations were replaced by the corresponding 3D implementation [Çiçek et al., 2016]. In V-Net the forward convolutions were replaced by residual convolution units [Milletari et al., 2016]. DenseVNet introduced a cascade of dense feature stacks. In dense blocks, the feature maps are concatenated enabling a streamlined gradient backpropagation [Gibson et al., 2018]. A convolution is inserted into each skip connection to reduce the number of features. The maps generated in the decoding path are then concatenated and convolved. The result is added to a spatial prior, a low-resolution 3D map of trainable parameters bilinearly upsampled to the segmentation resolution, to generate the final result [Gibson et al., 2018]. DRINet was developed by merging dense blocks, residual inception blocks, and unpooling blocks [Chen et al., 2018].

However, the optimal depth of an encoder-decoder in the traditional UNet architecture can vary from one application to another, depending on the task complexity. A solution would be to train models of different depths separately and then aggregate the resulting models at inference time. However, this approach is inefficient since the separate networks do not share a common encoder. Moreover, the design of skip connections requires the fusion of the same-scale encoder and decoder feature maps. UNet++ was designed to overcome these limitations. It is based on an ensemble of several UNet networks with different depths partially sharing the same encoder but retaining their specific decoder. Densely connected skip connections enable dense feature propagation along horizontal and vertical skip connections and more flexible feature fusion at the decoders [Zhou et al., 2020]. The nnUNet framework is a cutting-edge DL framework for automating configuration across the segmentation pipeline, encompassing pre-processing, network architecture, training, and post-processing, adapting seamlessly to new datasets [Isensee et al., 2020]. nnUNet provides implementations of several UNet-based architectures, including 2D, 3D, and cascaded UNet designs. With open-source accessibility, nnUNet stands as a pivotal tool, delivering state-of-the-art performance and driving advancements in automated medical image analysis.

## 3.3 Attention and its variants

The concept of attention drew inspiration from human biological systems. For instance, the visual system focuses on some parts of an image rather than others [Chaudhari et al., 2021]. Basically, attention in DL can be explained as a mechanism incorporating the concept of relevance to pay attention to only certain parts of an input [Chaudhari et al., 2021]. The first use of attention in DL was presented by Bahdanau et al. [2014] for the encoder-decoder architecture for sequence-to-sequence tasks, like language translation. These models were based on recurrent neural networks for encoder and decoder, with the encoder compressing the input sequence into a single vector of fixed length at the last step of the encoding process, called hidden state. Unfortunately, in the case of long sequences, the compression may lead to loss of information [Chaudhari et al., 2021]. To overcome this limitation the key idea of attention was to introduce a structure called context vector equivalent to a weighted sum of the hidden states of the decoder (one for each encoding step) and the corresponding attention weights. This enables the decoder to access the entire sequence of the encoder and focus on the relevant positions in the input sequence thanks to the attention weights [Chaudhari et al., 2021]. Several types of attention were proposed for computer vision [Guo et al., 2022a]. Attention gate was developed to learn to suppress irrelevant regions in an input image while highlighting salient features useful for a specific task [Oktay et al., 2018]. Spatial attention can be performed by spatial transformers that are able to transform feature maps [Jaderberg et al., 2015]. The spatial transformers include three submodules: a localization network with feature maps as input and the predicted transformation parameters by regression as output; a grid generator to use the regressed transformation parameters to create a sampling grid, consisting of a set of points where the input feature map should be sampled to produce the transformed output; and a sampler using the input feature map and sampling grid to create the output map [Jaderberg et al., 2015]. Channel attention can be realized by squeeze and excitation (SE) block [Hu et al., 2017]. SE was designed to perform feature recalibration. Essentially, SE adds a parameter to each channel of a CNN block to adjust the relevance of each feature map. In the first part, the squeeze operation performs global average pooling to reduce each feature map along width and height to a numeric value, obtaining a channel descriptor. In the excitation part, the numeric values are then fed to two fully connected layers with ReLU and sigmoid activation functions to obtain new numeric values which are used to weigh the original feature maps and assign each channel a specific relevance [Hu et al., 2017]. The residual attention network is composed of a stack of several attention modules that generate attention-aware features [Wang et al., 2017]. Each attention module is divided into a trunk branch and a mask branch. Each trunk branch has its mask branch to learn attention specialized for its features. The trunk branch performs feature extraction and can be integrated into any network. The mask branch weighs output features from the truck branch [Wang et al., 2017]. The attention mask serves as a feature selector during forward inference and as a gradient update filter during backpropagation. Moreover, the mask branches prevent wrong gradients from updating trunk parameters. Inside each attention module, both spatial and cross-channel dependencies are modeled [Wang et al., 2017]. The convolutional block attention module (CBAM) was designed to emphasize meaningful features along channels and spatial axes in CNNs [Woo et al., 2018]. The idea behind CBAM is that the channel attention module solves the problem of learning "what" since each channel of a feature map can be considered a feature detector, while the spatial attention module solves the problem of learning "where" since it is based on the inter-spatial relationship of features [Woo et al., 2018]. Instead of computing the 3D attention map directly as in residual attention, CBAM decomposes the process of learning channel attention and spatial attention separately [Woo et al., 2018]. In addition to global max-polling as in SE, CBAM uses also max-pooling [Woo et al., 2018]. These two pooling methods are applied to an intermediate feature map. The results of both are forwarded to a shared network to produce a channel attention map. During the spatial attention process, average pooling and max-pooling are applied along the channel axis, and the results are concatenated. A convolution layer is then used to generate a spatial attention map. Channel and spatial attention can be arranged sequentially or parallelly, although the former provided better results [Woo et al., 2018].

### 3.4   Transformer and its variants

Since CNNs are not able to learn global and long-range semantic information due to the locality of convolution operations, transformers were introduced to overcome this limitation Azad et al. [2024]. Transformers were developed initially for natural language processing tasks. The original transformer consisted of an encoder and a decoder. The encoder converted an input sequence of tokens into a sequence of embedding vectors, called hidden state or context. The decoder used the encoder's hidden state to iteratively generate an output sequence of tokens, one token at a time [Vaswani et al., 2017]. The encoder was a stack of modules each of which included multi-head self-attention (MSA), layer normalization, feedforward layers, and a second layer normalization. MSA refers to the fact that these weights are computed for all hidden states in the same sequence, e.g., all the hidden states of the encoder. Positional embedding is added to retain positional information [Vaswani et al., 2017]. The decoder has several modules consisting of mask MSA and encoder-decoder attention blocks. The former ensures that the generated tokens are based on the past outputs and the current token being predicted, while the latter learns how to relate tokens from two different sequences, e.g. two different languages [Vaswani et al., 2017].

Inspired by the design of transformers for natural language processing, vision transformers (ViT) were proposed for imaging tasks [Dosovitskiy et al., 2020]. In this architecture, the image is split into a sequence of flattened 2D patches which are projected to obtain the patch embeddings. Positional embeddings are added to the patch embeddings to retain positional information. The resulting sequence of embeddings is fed as input to the encoder consisting of a series of standard transformer blocks with normalization, MSA, and a second normalization. A multi-layer perceptron is then added for the classification task [Dosovitskiy et al., 2020]. Since transformers lack translation equivariance and locality, they do not generalize well when trained on insufficient amounts of data. For this reason, ViT was pre-trained on ImageNet-21k to obtain satisfying results [Dosovitskiy et al., 2020]. In order to solve this issue data efficient image transformers (DeiT) were developed [Touvron et al., 2020]. Another limitation of ViT is its unsuitability when the image resolution is high due to the quadratic computation complexity of MSA w.r.t image resolution [Liu et al., 2021]. In fact, in standard transformers, MSA is obtained by computing globally the relationship between a token and the other tokens [Liu et al., 2021]. To solve this issue Shifted Window (Swin) Transformer was proposed [Liu et al., 2021]. This architecture builds hierarchical feature maps by starting from small-sized patches and gradually merging neighboring patches in deeper layers. The linear computational complexity is ensured by computing self-attention locally within non-overlapping windows that partition an image [Liu et al., 2021]. Additionally, the window in a layer is shifted w.r.t.

the previous layer, causing the self-attention computation in the new window to cross the boundaries of the previous window, thus providing connections among them [Liu et al., 2021].

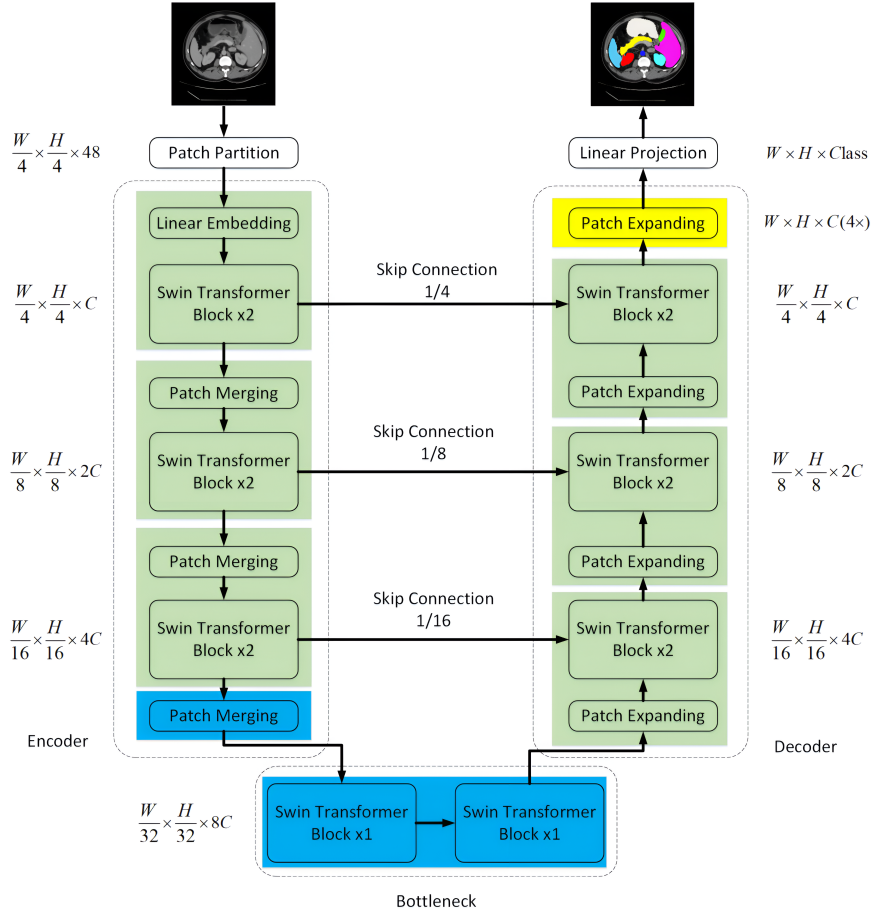

Figure 6: Architecture of Swin-UNet from Cao et al. [2023a]

In computer vision, transformers can be divided into pure and hybrid ones. In pure transformers, the MSA modules are used in both the encoder and decoder. Hybrid transformer architectures fuse the ViTs with convolution modules in the encoder, bottleneck, decoder, or skip connections to combine information about the global context and local details [Azad et al., 2024]. Swin-UNet is a pure transformer with a UNet-like architecture (Fig. 6) employing the Swin transformer block in the encoder, bottleneck, and decoder [Cao et al., 2023a]. CTUNet is a hybrid network (Fig. 7) for segmentation of the pancreas parenchyma with 3D channel transformer blocks inserted into the skip connection of a 3D UNet [Chen and Wan, 2022]. A pancreas attention module with a project and excite block was designed and added to each encoder to enhance the ability to extract context information, while cross attention was inserted between the output of each transformer and decoder to eliminate semantic inconsistency [Chen and Wan, 2022].

Residual transformer UNet (RTUNet) is a UNet-like network for pancreas parenchyma segmentation with convolutional blocks consisting of residual blocks, residual transformers, and dual convolution down-sampling. The residual transformer block adds progressive up-sampling to the basic transformer [Qiu et al., 2023]. UMRFormer-Net is a U-shaped encoder-decoder architecture (Fig. 8) with a hybrid CNN and transformer for segmentation of the pancreatic parenchyma and tumors [Fang et al., 2023]. It has five 3D CNN layers and a double transformer module inserted into the bottleneck and skip connection of the fourth layer to encode the long-range dependencies semantic information in a global space [Fang et al., 2023].

Convolutional pyramid vision is a hybrid network of CNN and hierarchical transformers for tumor segmentation. It generates multi-scale features by incorporating multi-kernel convolutional patch embedding and local spatial reduction to reduce computational cost. In this way, the model is able to capture the local information of multi-scale tumors [Viriyasaranon et al., 2023].

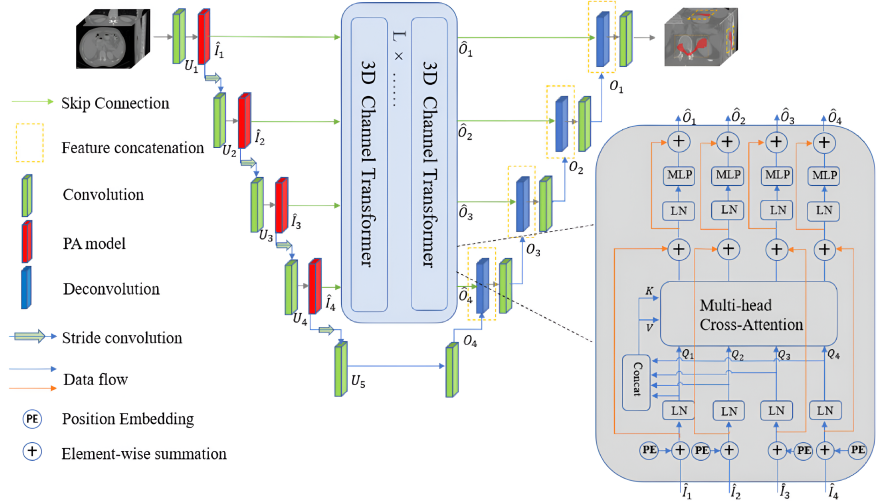

Figure 7: Architecture of CTUNet from Chen and Wan [2022]

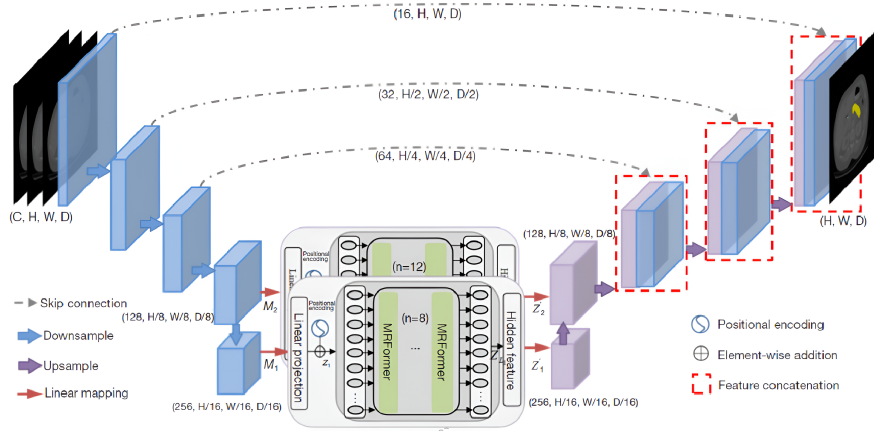

Figure 8: Architecture of UMRFormer-Net from Fang et al. [2023]

## 3.5 Generative Adversarial Network and its variants

Generative Adversarial Networks (GANs) are generative models with a generator and a discriminator network which are trained to compete and overcome each other. In GANs there is a minimax two-player game, where the generator network tries to fool a discriminator which has to distinguish between real images (coming from the training dataset) and false ones ( generated by the discriminator starting from a random noise distribution [Goodfellow et al., 2014]). CycleGAN networks were proposed for the image-to-image translation task, converting an image from one domain to another one [Zhu et al., 2017]. In contrast with previous approaches for image translation in computer vision with pair data between the two domains, in CycleGANs the the images are not paired [Zhu et al., 2017].

## 3.6 Dilated convolutions

The max-pooling and strides (downsampling) on CNNs layers result in feature maps with considerably reduced spatial resolution [Chen et al., 2016]. Inspired by the efficient computation of the undecimated wavelet transform, known as "algorithm a trous", Chen et al. [2016] proposed atrous convolution, replacing the downsampling in the last max pooling layers of CNNs with upsampling the filters by inserting holes ("trous" in French) between nonzero filter values. As a result, the feature maps are computed at a higher sampling rate than in conventional CNNs. Atrous convolutions enable the enlargement of the field of view of filters without increasing the number of parameters or computational burden. Atrous convolution was later called "dilated convolution". By adopting multiple parallel atrous convolutional layers with different sampling rates it is possible to capture objects at different scales, in a way similar to spatial pyramid pooling. For this reason, this technique was named Atrous Spatial Pyramid Pooling (ASPP) [Chen et al., 2016].

Table 2: Publicly available datasets in the reviewed studies

| | Name | Country | Size | Application | Adopted by reviewed studies |
|---|---|---|---|---|---|
| Roth et al. [2015] | National Institute of Health (NIH) | United States | 82 | Parenchyma | Xia et al. [2023], Tian et al. [2023], Wu et al. [2023], Tong et al. [2023], Li et al. [2023c], Liu et al. [2022a], Javed et al. [2022] Zhu et al. [2022], Huang et al. [2023b], Li et al. [2022c], Qureshi et al. [2022], You et al. [2022], Lim et al. [2022], Li et al. [2022d] Li et al. [2021a], Petit et al. [2021], Shi et al. [2022], Zhang et al. [2021b], Huang et al. [2021a], Li et al. [2021c], Ma et al. [2021a] Yan and Zhang [2021], Dogan et al. [2021], Li et al. [2021d], Wang et al. [2021a], Zhang et al. [2021a], Li et al. [2020], Zhang et al. [2021c], Boers et al. [2020] Hu et al. [2021], Li et al. [2020a], Tong et al. [2020], Xia et al. [2020], Zheng et al. [2020], Chen et al. [2020b], Xue et al. [2021], Bagheri et al. [2020] Zeng and Zheng [2019], Man et al. [2019], Gibson et al. [2018], Heinrich et al. [2018], Farag et al. [2017], Huang and Wu [2022], Chen and Wan [2022] Ju et al. [2023], Irshad et al. [2023], Huang et al. [2022c], Liu et al. [2020], Nishio et al. [2020], Chen and Wan [2022], Liu et al. [2022b] Qiu et al. [2022a], Zhang et al. [2021], Tuan et al. [2021], Chen et al. [2020], Paithane [2023], Chen et al. [2022c], Liu et al. [2022c], Zhu et al. [2020] Jain et al. [2023], Shan and Yan [2021], Wang et al. [2021a], Li et al. [2020c], Zhai et al. [2023], Long et al. [2023], Li et al. [2021], Qiu et al. [2023], Li et al. [2021e], Lu et al. [2019] Ning et al. [2020], Schlemper et al. [2019], Qiu et al. [2023b], Cui et al. [2022], Li et al. [2023d], Zhao et al. [2023a], Zhu et al. [2023], Ma et al. [2021b], Roth et al. [2018a] Zheng and Lao [2023], Yang et al. [2022b], Shen et al. [2022], Cao et al. [2023b], Cao and Li [2024], Xie et al. [2020] |
| Simpson et al. [2019] | Medical Segmentation Decathlon (MSD) | United States | 420 | Parenchyma Tumors | Tian et al. [2023], Tong et al. [2023], Fang et al. [2023], Li et al. [2023c], Zeng et al. [2022], Zhu et al. [2022], Li et al. [2021d], Shi et al. [2021] Zhang et al. [2021a], Li et al. [2020b], Xia et al. [2020], Dace et al. [2023], Wang et al. [2021a], Wang et al. [2019a] Chen et al. [2022c], Dai et al. [2023], Liu and Zheng [2023], He and Xu [2023], Li et al. [2023d], Knolle et al. [2021], Ma et al. [2023b], Isensee et al. [2020], Qu et al. [2023] Li et al. [2023c], Yang et al. [2022b], Mahmoudi et al. [2022], Cao et al. [2023b], Cao and Li [2024], Wang et al. [2021c], Turečková et al. [2020], Li et al. [2022b] |
| Ji et al. [2022] | AMOS-CT | China | 500 | Tumors (15 abdominal organs) | [Li et al., 2023f] |
| Ma et al. [2022] | AbdomenCT-1k | China | 1,112 (MSD (420), NIH (80), liver (201), kidneys (300), spleen (61), 50 from Nanjing University) | Parenchyma, Tumors (4 abdominal organs) | Tian et al. [2023], Li et al. [2023f], Francis et al. [2023], Ma et al. [2022] |
| Landman et al. [2015] | Beyond the Cranial Vault (BTCV) | United States | 50 | Parenchyma Tumors (13 abdominal organs) | Tong et al. [2020], Xia et al. [2020], Zheng et al. [2020], Gibson et al. [2018], Irshad et al. [2023], Huang et al. [2022c], Chen et al. [2022b], Yuan et al. [2023] Zhao et al. [2022b], Huang et al. [2023], Pan et al. [2022], Li et al. [2022c], Zhang et al. [2021b], Shi et al. [2021], Shen et al. [2023], Liu and Zheng [2023] |

### 3.7 Datasets

Five open datasets for pancreas segmentation, available online, were largely adopted in the reviewed studies (Table 2). The Cancer Image Archive (TCIA) from the National Institute of Health (NIH) is an online service (`https://www.cancerimagingarchive.net/`) hosting medical imaging archives. The most investigated dataset for pancreas segmentation comes from this source and consists of 82 CTs. It is known as the NIH dataset. There are also published studies using 43 CTs from TCIA-NIH. From here onward it will be referenced as TCIA dataset. The NIH dataset includes only labeled images of the pancreas parenchyma, while the Medical Segmentation Decathlon (MSD) dataset also annotations of tumors (intraductal mucinous neoplasms, pancreatic neuroendocrine tumors, and pancreatic ductal adenocarcinoma) (Fig. 9). The other three incorporate the segmentation of multiple abdominal organs, namely 15 (AMOS-CT), 13 (Beyond the Cranial Vault (BTCV)), and four (AbdomenCT-1k). Only AMOS-CT and AbdomenCT-1k are multi-vendor and multicenter, with data from two and 12 centers, respectively (Ji et al. [2022], Ma et al. [2022]). In the NIH dataset, the pancreas was manually labeled by a medical student and then verified by an experienced radiologist (Roth et al. [2015], Ma et al. [2022]). The images of the MSD dataset were provided by the Memorial Sloan Kettering Cancer Center (New York, NY, United States). The pancreatic parenchyma and pancreatic mass (cyst or tumor) were manually annotated by an expert radiologist (Simpson et al. [2019]). In the AMOS-CT dataset, 50 out of 500 CTs were initially annotated by humans. Then, one 3D UNet was trained using these 50 CTs to pre-label the remaining ones (coarse stage). Five junior radiologists refined the segmentation results. To further reduce errors, three senior radiologists with more than 10 years of experience checked and validated the results (fine stage). The process was iterated several times to reach a final consensus on the well-labeled annotations (Ji et al. [2022]). For the AbdomenCT-1k dataset, 15 junior annotators (one to five years of experience) used ITK-SNAP tool to manually segment the organs under the supervision of two board-certified radiologists. Then, one senior radiologist with more than 10 years of experience checked the annotations. After annotation, UNet models were trained to find the possible errors, which were double-checked by the senior radiologist (Ma et al. [2022]). The dataset grouped the MSD Pancreas (420 cases), the NIH (80 cases), tumors of the liver (201 cases), tumors of the kidneys (300 cases), spleen (61 cases), and 50 CT scans from Nanjing University of patients with pancreas cancer (20 cases), colon cancer (20 cases), and liver cancer (10 cases) for a total of 1,112 CTs (Ma et al. [2022]). The BTCV is a medical dataset for the MICCAI 2015 Multi-Atlas Abdomen Labelling Challenge. It consists of 50 CTs, manually labeled by two experienced undergraduate students, and verified by a radiologist. The annotations are multi-organ The Synapse dataset includes 30 CT scans of BTCV [Landman et al., 2015].

### 3.8 Metrics

This section presents a thorough mathematical formulation of the six distinct metrics identified in the systematic review for assessing model performance. These metrics are Dice Score Coefficient (DSC), Jaccard Index (JAC), Hausdorff Distance (HD), 95th percentile Hausdorff Distance (HD95), Average Surface Distance (ASD), and Normalized Surface Dice (NSD). To formally define the metrics, let us consider that the medical images are represented by a collection

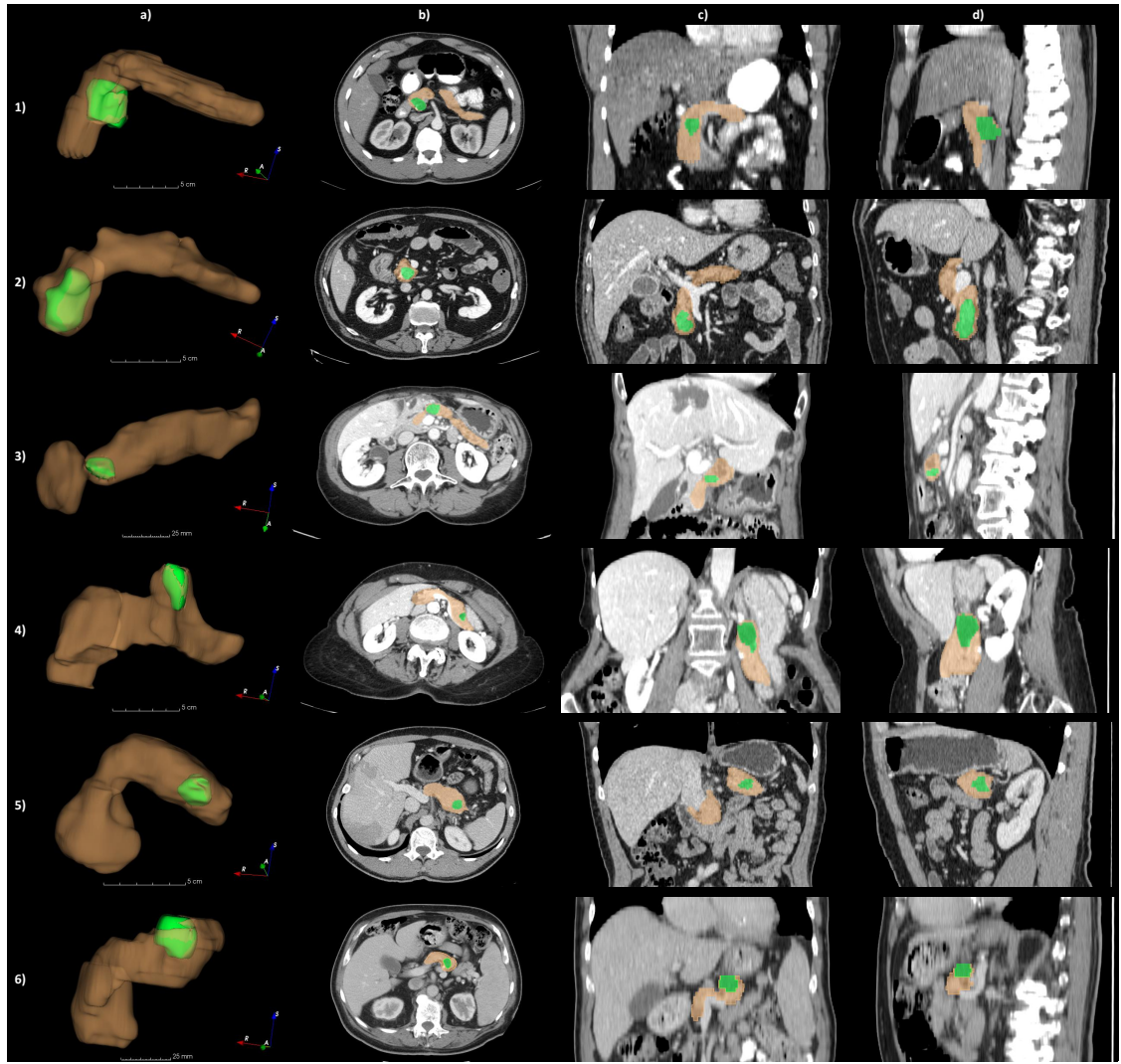

Figure 9: Six cases of pancreas anatomy, along with a tumor, from the MSD dataset (rows 1-6) to show the large morphological variability (Simpson et al. [2019]). Column a: 3D model of parenchyma (in brown) and tumor (in green). Columns b,c,d: view on the axial, coronal, and sagittal plane. Case number of MSD dataset (from row 1 to row 6): #66, #64, #334, #126, #286, and #81. Pancreas subregions grouped as follows: head (row 1 and row 2), body (row 3 and row 4), and tail (row 5 and row 6)

of points $X = \{x_1, x_2, ..., x_n\}$, where each $x_i$ corresponds to a voxel value within the image. The entire set $X$ is organized within a three-dimensional grid, such that the total number of points (voxels) is given by $|X| = N$, where $N = w \times h \times d$. Here, $w$ denotes the width, $h$ the height, and $d$ the depth of the grid, respectively. For each voxel $x \in X$, there are corresponding labels in the ground truth segmentation $S_g$ and in the automatic segmentation predicted by the model $S_p$. We define the labeling function for the ground truth segmentation as $S_g : X \rightarrow \{0, 1\}$, where $S_g(x)$ denotes the label assigned to voxel $x$ by $S_g$. Similarly, the labeling function for the predicted segmentation is defined as $S_p : X \rightarrow \{0, 1\}$, where $S_p(x)$ represents the label assigned to voxel $x$ by $S_p$.

Building on this premise, this section first defines two metrics classified as overlap-based, namely DSC and JAC. The first step is defining the four cardinalities that underlie these metrics, as delineated below:

$$TP = |\{x \in X : S_g(x) = 1 \text{ and } S_t(x) = 1\}| \tag{1}$$

$$FP = |\{x \in X : S_g(x) = 1 \text{ and } S_t(x) = 0\}| \tag{2}$$

$$FN = |\{x \in X : S_g(x) = 0 \text{ and } S_t(x) = 1\}| \tag{3}$$

$$TN = |\{x \in X : S_g(x) = 0 \text{ and } S_t(x) = 0\}| \tag{4}$$

where TP stands for true positive, FP for false positive, FN for false negative, and TN for true negative. The symbol $|\cdot|$ denotes the count of the set. The DSC, often called Dice or overlap index, is the predominant metric for validating medical volume segmentations. Beyond facilitating direct comparisons between automated and ground truth segmentations, the Dice metric is frequently employed to assess reproducibility and repeatability within these analyses (Kamnitsas et al. [2017], Ronneberger et al. [2015], Li et al. [2019]). A score of 0 indicates no overlap, while a score of 1 indicates perfect overlap, and its formulation is defined by:

$$DSC = \frac{2|S_p \cap S_g|}{|S_p| + |S_g|} = \frac{2TP}{2TP + FN + FP} \tag{5}$$

Instead, the JAC is determined by dividing the intersection of two sets by their union Jaccard [1912]. This metric quantifies the similarity between the sets, represented mathematically as:

$$JAC = \frac{|S_p \cap S_g|}{|S_p \cup S_g|} = \frac{TP}{TP + FN + FP} \tag{6}$$

DSC and JAC range between 0 and 1, where 1 means perfect overlap and 0 means null intersection between $S_p$ and $S_g$. The second part of this section defines the set of spatial distance-based metrics: HD, HD95, ASD, and NSD. These metrics represent a pivotal dissimilarity measure in evaluating image segmentation, especially when the task requires a proper edge delineation. HD was specifically designed to assess the shape similarity between two point sets within a given metric space Huttenlocher et al. [1993]. HD's evaluation is independent of point correlations, focusing only on the pairwise distances between voxels. Nevertheless, it shows a significant vulnerability to outliers in the data set. It is defined as:

$$HD(S_g, S_p) = \max(h(S_g, S_p), h(S_p, S_g)) \tag{7}$$

where $h(S_g, S_p)$ is called the directed Hausdorff distance and is given by:

$$h(S_g, S_p) = \max_{x_g \in S_g} \min_{x_p \in S_p} \|x_g - x_p\| \tag{8}$$

where $\|x_g - x_p\|$ represents a norm such as euclidean distance. Nonetheless, the HD95 introduced by Huttenlocher et al. [1993] is the quantile approach to HD providing a method to reduce the influence of outliers by considering the $q^{th}$ quantile of direct Hausdorff distances instead of the maximum distance. The choice of $q^{th}$ depends on the specific application and the characteristics of the point sets under analysis. Our systematic review focuses on the 95th percentile HD95, widely used in literature. This metric is similar to the traditional HD but is defined as follows:

$$HD95(S_g, S_p) = \max(h_{95}(S_g, S_p), h_{95}(S_p, S_g)) \tag{9}$$

where $h_{95}$ represents the 95th ranked percentile of the set of minimum distances between points from one set to the nearest points in the other. Specifically, $h_{95}(S_g, S_p)$ is defined as:

$$h_{95}(S_g, S_p) = \underset{x_g \in Sg}{rank^{95}} \min_{x_p \in S_p} \|x_g - x_p\| \tag{10}$$

where $\|x_g - x_p\|$ denotes a norm such as euclidean distance. Another metric belonging to the distance-based class is the average ASD (ASSD). It is defined as the average of all the distances from points on the boundary of the ground truth segmentation to the boundary of the predicted segmentation, and vice-versa (Yeghiazaryan and Voiculescu [2018]). The ASSD is defined by:

$$ASD(S_g, S_p) = \frac{1}{|S_g| + |S_p|} \left( \sum_{x_{sg} \in S(S_g)} d(x_{sg}, S(S_p)) + \sum_{x_{sp} \in S(S_p)} d(x_s p, S(S_g)) \right) \tag{11}$$

where $d(x_{sg}, S(S_p))$ is defined as

$$d(x_{sg}, S(S_p)) = \min_{s_{sp} \in S(S_p)} \|s_{sg} - s_{sp}\| \tag{12}$$

with $S(S_g)$ and $S(S_p)$ represent the surfaces (boundary) of $S_g$ and $S_p$ respectively. HD, HD95, and ASD are initially expressed in units of voxels and then converted into millimeters (mm) based on the voxel spacing of the medical images. Lastly, the NSD, introduced by Nikolov et al. [2021], quantifies the accuracy of segmentation boundaries by measuring the proportion that meets a specified deviation threshold, $\tau$. This threshold represents the maximum clinically acceptable error in pixels, offering a precise metric for evaluating how closely a predicted segmentation aligns with the actual boundary within a tolerable margin of error. The NSD is defined as

$$NSD = \frac{|D_g| + |D_p|}{|D'_g| + |D'_p|} \tag{13}$$

where $D_g$ and $D_p$ are the nearest neighbour distances computed respectively from the surface $S(S_p)$ to the surface $S(S_g)$ and viceversa, while $D'_g$ and $D'_p$ are respectively the subset of distances in $D_g$ and $D_p$ that are smaller or equal to acceptable deviation $\tau$ as defined by:

$$D'_g = \{d_g \in D_g \mid d_g \leq \tau\} \tag{14}$$

$$D'_p = \{d_p \in D_p \mid d_p \leq \tau\} \tag{15}$$

The NSD ranges between 0 and 1 Seidlitz et al. [2022]. A score of 0 signifies either complete inaccuracy, with all measured distances exceeding the predefined acceptable deviation threshold $\tau$, or the image's absence of the predicted class. Conversely, a score of 1 means no corrections to the segmentation boundary are needed, as all deviations from the reference boundary fall within the acceptable threshold $\tau$.

### 3.9 Loss functions

This section presents a thorough mathematical formulation of the three most commonly used loss functions identified in the systematic review. Following the conventions outlined in Section 3.8, the mathematical formulations of Binary Cross Entropy loss ($L_{BCE}$), Focal loss ($L_{Focal}$), and Dice loss ($L_{Dice}$) will be presented below. Binary Cross Entropy loss function belongs to the class of distribution-based losses, designed with the purpose of minimizing discrepancies between two probability distributions. The formulation of Binary Cross Entropy loss is given by:

$$L_{BCE} = -\frac{1}{N} \sum_{x \in X} [S_g(x) \log(S_p(x)) + (1 - S_g(x)) \log(1 - S_p(x))] \tag{16}$$

Focal loss function also belongs to the class of distribution-based losses. This loss modifies the conventional cross entropy by emphasizing misclassified pixels or voxels. It reduces the significance of the loss in well-classified samples, allowing it to effectively address imbalances between foreground and background classes. The formula below is an adaptation of the multiclass Focal loss of Lin et al. [2017] for binary classification, defined as:

$$L_{Focal} = -\frac{1}{N} \sum_{x \in X} \Big[ (1 - S_p(x))^\gamma S_g(x) \log(S_p(x)) + (1 - (1 - S_p(x)))^\gamma (1 - S_g(x)) \log(1 - S_p(x)) \Big] \tag{17}$$

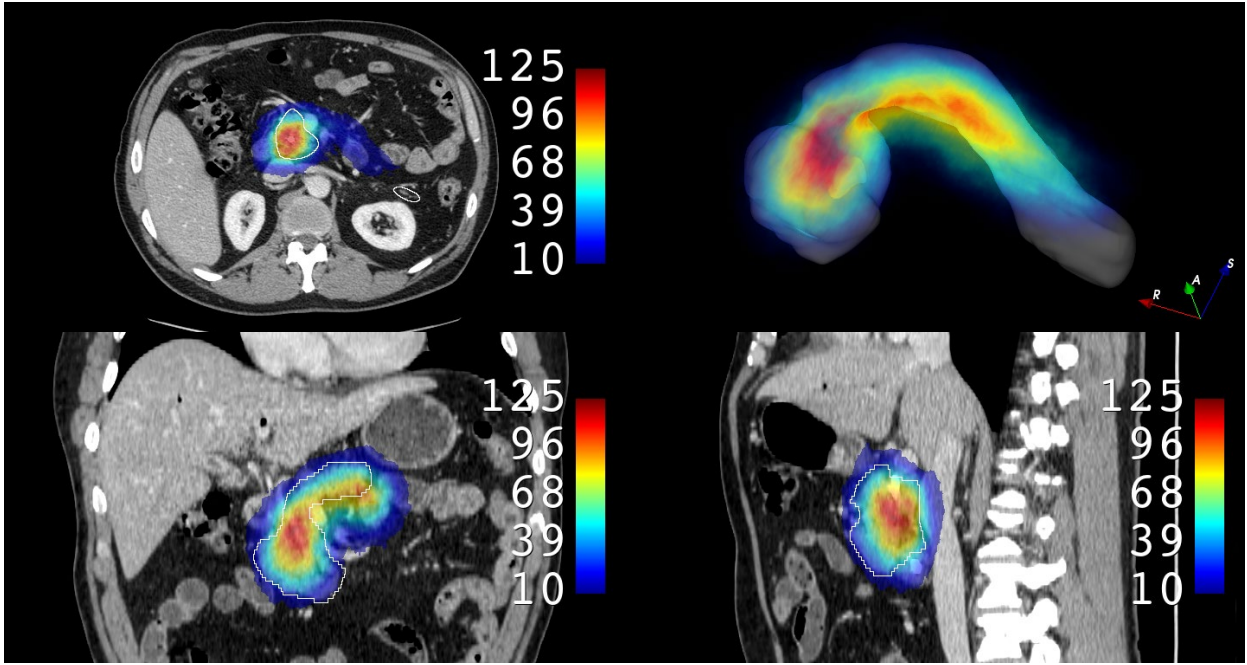

Figure 10: Spatial distribution and frequency of pancreas within the MSD dataset (Simpson et al. [2019]) (281 cases with case #29 as a reference in the image): most frequent pancreases in the dataset in red, least frequent ones in blue. Boundary of case #29 in white.

Dice loss function belongs to the class of overlap-based losses. This function aims to quantify the degree of overlap between the ground truth segmentation $S_g$ and the predicted segmentation $S_p$ Isensee et al. [2019]. It directly optimizes the DSC defined in section 3.8, and its formula is given by:

$$L_{Dice} = 1 - \frac{\sum_{x \in X} S_g(x) S_p(x)}{\sum_{x \in X} S_g(x) + \sum_{x \in X} S_p(x)} \tag{18}$$

## 4 Segmentation of the parenchyma

This section starts by showing the variability of the pancreas parenchyma in terms of size and location (Section 4.1). Then, the different approaches to the segmentation of pancreas parenchyma are analyzed. Overall, a total of 105 out of the 130 reviewed studies fall under this topic. The complete list is reported in Appendix A. Due to the high heterogeneity of the studies in terms of datasets, DL architecture, learning type, and loss functions, a comparison was not possible. Therefore we clustered the studies to obtain the largest representation. As a result, they were divided into the following groups: two-stage (coarse-fine) with single organ (pancreas) datasets (Section 4.2), multi-organ segmentation (Section 4.3), semi-supervised learning (Section 4.4), unsupervised learning (Section 4.5), generalization to a different dataset (Section 4.6), and design of new loss functions (Section 4.7). This section ends by comparing the performances of the different DL models on the publicly available datasets, described in Section 3.7, and on the private/internal ones (Section 4.8).

### 4.1 Variability of parenchyma size and location

In order to provide an example of the variability of the pancreas parenchyma in terms of size and location, a registration was performed on 281 CTs of the MSD dataset using Elastix software [Klein et al., 2010], adapting inter-subject registration parameters from the study by Qiao et al. [2016] to the CT domain. Subject #29 of MSD was considered a reference image by virtue of its high-quality image and centrality within the range of variations observed in the dataset. A Hounsfield unit (HU) from 100 to 500 was used for all the images to improve the registration process, enhancing bones and brighter abdominal structures. The results are illustrated in Fig. 10. A histogram with the frequency

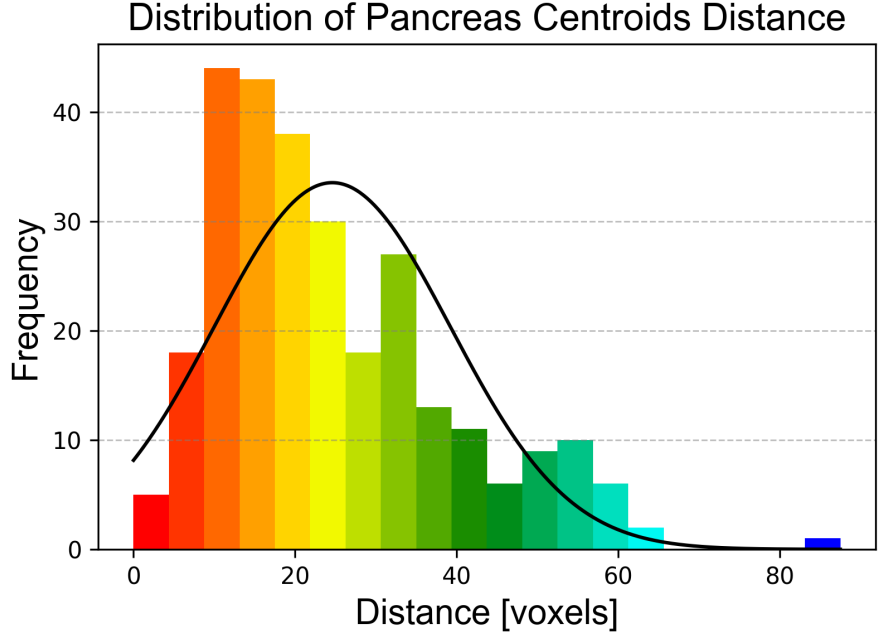

Figure 11: Frequency distribution of centroids distance of the pancreas in the MSD dataset with 281 cases. Case #29 was used as a reference to compute the distance in voxels

distribution is shown in Fig. 11. It was created by measuring the volumetric distances from the centroid of the pancreas in subject #29 of MSD to the centroids of the pancreas from all other subjects after performing the registration.

## 4.2  Two-stage methods

The DL approaches for pancreas segmentation can be divided into direct and two-stage methods [Chen et al., 2022b]. The former approach directly uses labeled images to segment the organ. In contrast, two-stage methods are cascaded. They first train a localization network to obtain the pancreas region (coarse stage), and then use the location result to train a second model for segmentation (fine stage) [Chen et al., 2022b]. Localizing the pancreas CT scans before performing segmentation has two advantages. First, the peripheral regions with very similar intensity or textural properties to the pancreas can be easily removed. Second, specifying the location of the pancreas reduces the sizes of the original CT scans, with a benefit in terms of computational costs, especially for 3D CT scans [Qureshi et al., 2022]. The reviewed studies using two-stage methods are reported in Table 3. Roth et al. [2018a] proposed a holistically-nested CNNs method for both stages on the NIH dataset. These CNNs are applied to axial, sagittal, and coronal views, and the respective probability maps are fused to generate a 3D bounding box. The segmentation works on this bounding box and incorporates organ interior and boundary mid-level cues with subsequent spatial aggregation [Roth et al., 2018a]. Man et al. [2019] introduced reinforcement learning to correct errors in localization, and deformable 3D UNet to capture the anisotropic geometry-aware information on the pancreas on the NIH dataset. Liu et al. [2020] used ResNet to generate candidate regions for pancreas localization by classifying patches based on superpixels, obtained after oversegmenting the images. Segmentation is performed via an ensemble of UNet on NIH dataset [Liu et al., 2020]. Hu et al. [2021] introduced geodesic distance-based saliency transformation to compute saliency map in the DenseASPP network during the coarse stage. Saliency information is integrated into the original DenseASPP to boost segmentation performances on the NIH dataset and 70 contrast-enhanced CT scans from Zheyi Hospital (China) [Hu et al., 2021]. Zhang et al. [2021a] employed multi-atlas registration in the coarse stage, and a patch-based 3D CNN and three slice-based 2D CNNs during the fine stage. The probability maps from the 3D CNN model are used to compute the pancreas bounding box, concatenated with the original CT image, to form the inputs to three subsequent 2D UNets. A third stage is added to refine the second one by employing a 3D level-set for better boundary delineation [Zhang et al., 2021a]. This approach was tested on 26 CTs of the 2018 International Symposium on Image Computing and Digital Medicine, NIH, and 281 CTs of MSD datasets [Zhang et al., 2021a]. Zhang et al. [2021d] proposed a prior propagation module in both stages and an encoder-decoder architecture with a scale-transferrable feature fusion module to learn rich fusion features, tested on NIH and 281 CTs of MSD dataset. In the first stage, the prior is propagated into the input slice to guide the localization, while in the second stage priors are propagated in both the input slice and feature maps to guide the segmentation. The scale-transferrable feature fusion module learns rich fusion features [Zhang et al., 2021d]. Yan and Zhang [2021] integrated spatial and channel attention modules into skip connections of 2.5 UNet in both stages. In each phase, they trained a model for axial, coronal, and sagittal views, which were then fused using majority voting.

Table 3: Studies with coarse and fine stage for pancreas parenchyma segmentation.

| Author | Application | Dataset Size | Model Architecture | Learning Strategy | Loss | Results | Main Contributions |
|---|---|---|---|---|---|---|---|
| Chen et al. [2022d] | Segmentation of pancreas | 82 (NIH) 281 (MSD) | Encoder-Decoder Attention feature fusion (Localization) Encoder-Decoder Attention feature fusion Coordinate Multi-scale Attention (Segmentation) | Supervised | Dice loss Binary cross entropy | NIH: 85.41% (DSC) 74.80% (Jaccard) 85.60% (Precision) 85.90% (Recall) MSD: 70.00-80.00% (DSC) 60.00% (Jaccard) 80.00-90.00% (Precision) 60.00-70.00% (Recall) | Attention feature fusion on low and high level features to keep context. Multi-scale attention to aggregate long-range dependencies, positional information, and exploit multi-scale spatial information |
| Dai et al. [2023] | Segmentation of pancreas | 82 (NIH) 281 (MSD) | UNet (Localization) Deformable convolution Vision Transformer (Segmentation) | Supervised | Binary cross entropy Dice loss | NIH: 89.89% (DSC) 89.59% (Precision) 91.13% (Recall) MSD: 91.22% (DSC) 93.22% (Precision) 91.35% (Recall) | Skip connections integrating: vision transformer, deformable convolutions, and scale interactive fusion (combining global and local features, and merging feature maps of different scales). Two-dimensional wavelet decomposition to solve the issue of blurred boundaries |
| Jain et al. [2023] | Segmentation of pancreas | 82 (NIH) | K-mean and Gaussian mixture model (Unsupervised) (Localization) UNet, Holistically-Nested Edge Detection, and Dense-Res-InceptionNet (Segmentation) | Unsupervised + Supervised | Dice loss | 81.75% (DSC) 83.03% (Precision) 81.70% (Recall) | Unsupervised localization of pancreas after segmenting liver and spleen using K-means and Gaussian mixture models |
| Qiu et al. [2023] | Segmentation of pancreas | 82 (NIH) | DeepUNet (Localization) Residual transformer UNet (Segmentation) | Supervised | Dice loss with Hausdorff distance term | 86.25% (DSC) | UNet like network with each convolutional block consisting of residuals blocks, residual transformers, and dual convolution down-sampling (for translational equivariance) |
| Tian et al. [2023] | Segmentation of pancreas | 200 (from AbdomenCT-1k) 82 (NIH) 281 (MSD) 50 (Jiangsu Province Hospital) (Generalization) | nnUNet (Localization) + Hybrid variational model to capture weak boundaries (Segmentation) | Supervised | Cross entropy Dice loss | AbdomenCT-1k: 89.61% (DSC) NIH: 87.67% (DSC) MSD: 87.13% (DSC) Generalization: 90.72% (DSC) | First stage: 3D CNN for coarse segmentation Second stage: a new hybrid variational model to capture the pancreas weak boundary |
| Zheng and Luo [2023] | Segmentation of pancreas | 80 (NIH) | Encoder-decoder for: (Localization) + (Segmentation) | Supervised | Weighted binary cross entropy loss | 85.58% (DSC) 74.99% (Jaccard) 86.59% (Precision) 85.11% (Recall) | Extension-contraction transformation network with a shared encoder for feature extraction and two decoders for the prediction of the segmentation masks and the inter-slice extension and contraction transformation masks |
| Ge et al. [2023] | Segmentation of pancreas | 45: (Nanjing Drum Tower Hospital) (Reconstruction) + 15 (Nanjing General PLA Hospital) for generalization + 90 (liver tumor) for generalization | Average Super Resolution GAN with: 3D CNN (Reconstruction) + 3D UNet for both Localization and Segmentation | Supervised | Mean squared error loss Dice loss Cross entropy loss | Generalization (pancreas): 84.20% (DSC) 0.54 mm (ASD) | GAN: Super resolution network to reduce anisotropy resolution. A generator reconstructs thin slices in z axis The discriminator optimizes the output of generator. The optimized generated images are sent to a dual-stage network for segmentation. Predictions on high-resolution are down-sampled to restore initial resolution |
| Li et al. [2023c] | Segmentation of pancreas | 82 (NIH) 281 (MSD) 104 (Private) | UNet with: Meta-learning (Localization) Latent-space feature flow generation (Segmentation) | Supervised | Design of adaptive loss with: Recall loss, Cross entropy and Dice loss | NIH (trained on MSD and private): 80.24% (DSC) 1.92mm (ASD) MSD (trained on NIH and private): 81.09% (DSC) 1.99 mm (ASD) Private (trained on NIH and MSD): 84.77% (DSC) 1.28 mm (ASD) | First generalization model for pancreas segmentation. Model-agnostic meta-learning to improve generalization of the coarse stage. Appearance-style feature flow generation in the fine stage to generate a sequence of intermediate representations between different latent spaces for simulating large variations of appearance-style features |

| Author | Application | Dataset Size | Model Architecture | Learning Strategy | Loss | Results | Main Contributions |
|---|---|---|---|---|---|---|---|
| Li et al. [2023d] | Segmentation of pancreas | 82 (NIH) 281 (MSD) 104 (Renji Hospital Shanghai, Private dataset) | UNet with meta-learning (Localization) 3D UNet: Global feature contrastive learning 3D UNet: Local image restoration (Segmentation) | Self-supervised | Binary cross entropy loss Dice loss Squared error loss Adversarial loss | Training on NIH Generalization on MSD: 66.73% (DSC) Generalization on Private: 73.85% (DSC) Training on MSD Generalization on NIH: 76.71% (DSC) Generalization on Private: 83.50% (DSC) Training on Private Generalization on NIH: 65.03% (DSC) Generalization on MSD: 70.08% (DSC) | Dual self-supervised generalization model to enhance characterization of high-uncertain regions. Global-feature self-supervised contrastive learning reducing the influence of extra-pancreatic tissues. Local image restoration self-supervised module to exploit anatomical context to enhance characterization of high-uncertain regions |
| Chen et al. [2022c] | Segmentation of pancreas | 82 (NIH) 281 (MSD) | VGG-16 with Attention gate (Localization) VGG-16 with Residual Multi-scale Dilated attention (Segmentation) | Supervised | Dice loss Binary cross entropy | NIH: 85.19% (DSC) 74.19% (Jaccard) 86.09% (Precision) 84.58% (Recall) MSD (generalization): 76.60% (DSC) 62.60% (Jaccard) 87.70% (Precision) 69.20% (Recall) | Attention gate used in the localization stage to suppress irrelevant background regions. Weight conversion module to transform segmentation map of the first stage into spatial weights to refine input of the second stage. Residual multi-scale dilated attention to exploit inter-channel relationships and extract multi-scale spatial information. Code available at: https://github.com/meiguiulu/TVMS |
| Khasawneh et al. [2022] | Segmentation of pancreas | 294 (from 1,917 of Mayo Clinic) | UNet-like (Localization) (Segmentation) | Supervised | – | 88.00% (DSC) 79.00% (Jaccard) | Comparison of manual segmentation by experts using 3D Slicer and automatic segmentation by CNN |
| Chen et al. [2022b] | Segmentation of pancreas | 82 (NIH) 281 (MSD) | UNet (Localization) Unet with: Fuzzy skip connection + Target attention in the decoder (Segmentation) | Supervised | Dice loss | NIH: 87.91% (DSC) 78.52% (Jaccard) 90.43% (Precision) 85.77% (Recall) MSD: 84.40% (DSC) | Fuzzy skip connections to reduce the redundant information of non-target regions. Attention to make the decoder more sensitive to target features |
| Liu et al. [2022b] | Segmentation of pancreas | 82 (NIH) 72 (ISICDM) + dataset of other organ | ResNet18 + Atrous spatial pyramid pooling for multi-scale feature extraction for both Localization and Segmentation + Saliency module for fusion | Supervised | Dice loss (Region and boundary level) Binary cross entropy (Pixel level) | NIH: 88.01% (DSC) ISICDM: 87.63% (DSC) | Segmentation network with three branches to extract pixel, boundary, and region features, fused by a saliency module. Design of a loss function integrating information from pixel-level classification, edge-level localization, and region-level segmentation |
| Qiu et al. [2022a] | Segmentation of pancreas | 82 (NIH) 281 (MSD) | UNet3+ + Multi-scale feature calibration in both Localization and Segmentation | Supervised | Dice loss | NIH: 86.30% (DSC) 76.26% (Jaccard) 85.91% (Precision) 86.85% (Recall) MSD (Generalization): 85.41% (DSC) | Dual enhancement module to multiply the coarse segmentation probability map with the input image to coarse stage. Cropping of the output by the localization model. The cropped images are sent as input to fine stage. Multi-scale feature calibration module in both stages to calibrate features vertically to preserve boundary details and avoid feature redundancy |
| Qiu et al. [2022b] | Segmentation of pancreas | 82 (NIH) | UNet-like with: Spiking neural P systems (Localization) + (Segmentation) | Supervised | Cross entropy | 81.94% (DSC) | Deep dynamic spiking neural P systems are integrated into UNet to solve memory limitation of 3D CNNs |
| Qureshi et al. [2022] | Segmentation of pancreas | 82 (NIH) | VGG-19 (Localization) + UNet (Segmentation) | Supervised | Mean Dice | 88.53% (DSC) | A morphology prior (a 3D volume template), defining the general shape and size of the pancreas, was integrated with the soft label from the second stage to improve segmentation |

This method was trained and tested on NIH dataset [Yan and Zhang, 2021]. Panda et al. [2021] used 3D UNet for both stages on an internal dataset of 1,917 CTs. They assessed model performances on subsets of 200; 500; 800; 1,000; 1,200; and 1,500 CTs. The models were tested for generalization on 41 CTs of TCIA and 80 CTs of NIH datasets

| Author | Application | Dataset Size | Model Architecture | Learning Strategy | Loss | Results | Main Contributions |
|---|---|---|---|---|---|---|---|
| Dogan et al. [2021] | Segmentation of pancreas | 82 (NIH) | Mask R-CNN (Localization) + UNet (Segmentation) | Supervised | Binary cross entropy | 86.15% (DSC) 75.93% (Jaccard) 86.23% (Precision) 86.27% (Recall) 99.95% (Accuracy) | Less powerful GPUs are required |
| Panda et al. [2021] | Segmentation of pancreas | 1,917 (Mayo Clinic) + 41 (TCIA) + 80 (NIH) | UNet for two stages: Localization + Segmentation | Supervised | Tversky loss Asymmetric dice loss | Internal dataset: 91.00% (DSC) TCIA (Generalization): 96.00% (DSC) NIH (Generalization): 89.00% (DSC) | Evaluation of dataset size on model performances: in the second stage 3D UNet was evaluated on 200; 500; 800; 1,000; 1,200; and 1,500 CTs (internal dataset). Generalization on two datasets |
| Wang et al. [2021a] | Segmentation of pancreas | 82 (NIH) | UNet (Localization) View adaptive Unet (Segmentation) | Supervised | Dice loss Weighted focal loss | 86.19% (DSC) | Data augmentation on three axes. Axial, coronal, and sagittal volumes are fed simultaneously to the network |
| Yan and Zhang [2021] | Segmentation of pancreas | 82 (NIH) | UNet + Spatial attention + Channel attention (Localization and Segmentation) | Supervised | Dice loss | 86.61% (DSC) | 2.5D UNet with spatial and channel attention integrated into skip connections. |
| Zhang et al. [2021d] | Segmentation of pancreas | 82 (NIH) 281 (MSD) | CNN (Localization) Encoder-decoder (Segmentation) Prior propagation module (both stages) Scale-transferrable feature fusion module (second stage) | Supervised | Dice loss | NIH: 84.90% (DSC) MSD: 85.56% (DSC) | Scale-transferrable feature fusion module to learn rich fusion features with lightweight architecture. Prior propagation module to explore informative and dynamic spatial priors to infer accurate and fine-level masks |
| Zhang et al. [2021a] | Segmentation of pancreas | 36 (International Symposium on Image Computing and Digital Medicine) 82 (NIH) 281 (MSD) | Multi-atlas registration (Localization) 3D patch-based and 2.5D slice-based UNet (Segmentation) 3D level set to refine the probability map (Refine stage) | Supervised | Cross entropy Dice coefficient loss | 84.40% (DSC) 73.40% (Jaccard) | Coarse stage for localization. Fine stage for segmentation: 3D patch-based and 2.5D slice-based CNN to extract local and global features. Refine stage to improve segmentation: 3D level-set for better boundary delineation. |
| Bagheri et al. [2020] | Segmentation of pancreas | 82 (NIH) | Superpixels and random forest classifier (Localization) Holistically nested neural networks (Segmentation) | Supervised | – | 78.00% (DSC) | Superpixels to get bounding boxes. Fusing holistically nested networks to generate interior and boundary |
| Hu et al. [2021] | Segmentation of pancreas | 82 (NIH) 70 (CT-Zheyi dataset) | DenseNet161 for Dense Atrous Spatial Pyramid Pooling (Localization) DenseNet161 for Distance-based saliency (Segmentation) | Supervised | Binary cross entropy | NIH: 85.49% (DSC) (CT-Zheyi): 85.48% (DSC) | Dense atrous spatial pyramid Pooling to cover larger receptive fields. Saliency map is computed through geodesic distance based saliency transformation. Both localization and saliency information are used to aid segmentation |
| Man et al. [2019] | Segmentation of pancreas | 82 (NIH) | Localization agent (Localization) + Deformable UNet (Segmentation) | Reinforcement (Localization) Supervised (Segmentation) | Dice loss | 86.93% (DSC) | First application of Deep Q Learning to medical image segmentation. Localization agent to adjust localization, by learning a localization error correction policy based on deep Q network. Deformable convolution for learnable receptive fields, instead of fix ones |
| Liu et al. [2020] | Segmentation of pancreas | 82 (NIH) | ResNet (Localization) Ensemble UNet (Segmentation) | Supervised | Dice loss Focal loss Jaccard distance loss Class balanced cross entropy Binary cross entropy | 84.10% (DSC) 72.86% (Jaccard) 84.35% (Precision) 85.33% (Recall) | Superpixes generated by oversegmentation. Classification of superpixels by ResNet. candidate regions obtained by ensemble of classification results of three different scale of superpixels. Segmentation by ensemble of multiple network with different loss functions |

[Panda et al., 2021]. Dogan et al. [2021] proposed Mask R-CNN for localization and UNet for segmentation on the NIH dataset to lower the requirements in terms of the power of the GPU. Qiu et al. [2022b] combined deep dynamic spiking neural P systems with CNNs to solve the limitation on 3D CNNs. Liu et al. [2022b] employed ResNet18, attention, and

| Author | Application | Dataset Size | Model Architecture | Learning Strategy | Loss | Results | Main Contributions |
|---|---|---|---|---|---|---|---|
| Xue et al. [2021] | Segmentation of pancreas | 82 (NIH) 59 (Fujian Medical University) | UNet for both: Localization and Segmentation | Supervised | Cross entropy Regression loss | NIH: 85.90% (DSC) 75.70% (Jaccard) 87.60% (Precision) 85.20% (Recall) Fujian: 86.90% (DSC) 77.30% (Jaccard) 91.00% (Precision) 83.50% (Recall) | Multi-task second stage. Regression (task 1 ) of object skeletons as descriptor of the shape of the pancreas to guide subsequent segmentation (task 2). Conditional random fields to remove small false segments |
| Roth et al. [2018a] | Segmentation of pancreas | 82 (NIH) | Holistically-nested networks for: Localization (fusing the three orthogonal axes) + Segmentation (boundaries and interior cues to produce superpixels aggregated by random forests) | Supervised | Cross entropy loss | 81.27% (DSC) 68.87% (Jaccard) 17.71 mm (HD) 0.42 mm (Average distance) | Segmentation incorporates deeply learned organ interior and boundary mid-level cues with subsequent spatial aggregation |

atrous spatial pyramid pooling module to facilitate multi-scale features extraction and fusion. Attention was used for the extraction of pixel features, while ASPP for edge and region features. A saliency transformation module was added after the first stage as the initialization to the fine model [Liu et al., 2022b]. Chen et al. [2022d] proposed FPF-Net, a network for feature propagation and fusion based on an attention mechanism. An attention feature fusion mechanism on low and high-level features was used in both stages to keep context. In the segmentation stage, a coordinate and multi-scale attention module was designed to aggregate long-range dependencies, and positional information, and exploited multi-scale spatial information [Chen et al., 2022d] Chen et al. [2022c] proposed attention gate into the localization stage to suppress irrelevant background regions, a weight conversion module to transform the segmentation map of the first stage into spatial weights to refine input of the second stage, and residual multi-scale dilated attention to exploit inter-channel relationships and extract multi-scale spatial information. Chen et al. [2022b] designed fuzzy operations into skip connections to reduce the redundant information of non-target regions and attention mechanism into the decoder of UNet to make the decoder more sensitive to target features in the segmentation stage in each of axial, coronal, and sagittal view. The final result is obtained as an ensemble of the three views and tested on both NIH and MSD datasets [Chen et al., 2022b]. Li et al. [2023c] addressed two issues of pancreas segmentation. First, a cluttered background may deteriorate the segmentation performance on data with large appearance-style variations. Second, since data may be collected at different centers with different scanners there may be appearance-style discrepancies among the training and testing sets. Li et al. [2023c] integrated model-agnostic meta-learning to improve the generalization ability of the coarse stage by filtering out cluttered background and an appearance-style feature flow generation in the fine stage to generate a sequence of intermediate representations between different latent spaces for simulating large variations of appearance-style features. The datasets included NIH, MSD, and 104 internal CT scans. The model was trained on two of these datasets and tested on the third one in turn. It is the first generalization model for the pancreas segmentation [Li et al., 2023c]. Tian et al. [2023] used nnUNet for localization and introduced a variational model embedding the directional and magnitude information of the boundary intensity gradient to capture weak boundary information in the segmentation stage. The model was trained on NIH, MSD, 200 out of 1,112 CTs of AbdomenCT-1k datasets, and tested for generalization on 50 CT from an internal dataset [Tian et al., 2023]. Ge et al. [2023] proposed Average Super Resolution GAN (ASRGAN) network to reduce anisotropy resolution on the z-axis w.r.t transverse plan. A GAN was designed with the generator reconstructing thin slices along the z-axis, while the discriminator optimized the output of the generator. The optimized generated images were then sent to a dual-stage network for localization and segmentation. The predicted high-resolution images were then restored to initial resolution. ASRGAN was trained on an internal dataset of 90 CTs from Nanjing Drum Tower Hospital for super-resolution reconstruction, while another dataset of 15 CTs from Nanjing General PLA Hospital was used for generalization [Ge et al., 2023]. Qiu et al. [2023] designed TRUNet, a UNet with the following changes to each of the encoder steps: residual connections between two convolutional blocks, transformers with residual connections, and dual convolution down-sampling for translational equivariance. Qiu et al. [2022a] proposed a Cascaded multi-scale feature calibration UNet (CMFCUNet) network with a dual enhancement module to jointly train coarse and fine stages. The coarse segmentation probability map was first multiplied by the input image to the coarse stage. The result was cropped by the output of the localization model. The cropped images were then sent as input to the fine stage. A multi-scale feature calibration module was inserted into skip connections of the UNet3+ network in both stages to calibrate features vertically to preserve boundary details and avoid feature redundancy. CMFCUNet was trained on NIH and tested for generalization on MSD dataset [Qiu et al., 2022a].

Jain et al. [2023] adopted unsupervised learning for localization of the pancreas after segmenting the liver and spleen. The segmentation of the liver and spleen was unsupervised and obtained after k-mean and Gaussian mixture model. UNet, Holistically-Nested Edge Detection, and Dense-Res- Inception Net were tested for pancreas segmentation on NIH dataset [Jain et al., 2023]. [Dai et al., 2023] proposed TD-Net, a trans-deformer UNet-like network. A 2D UNet was used in the first stage, followed by a modified UNet for the second stage with skip connections consisting of ViT, deformable convolutions, and scale interactive fusion. The latter combined global and local features and merged feature maps of different scales.

## 4.3 Multi-organ segmentation

The segmentation of pancreas parenchyma was investigated also in studies where datasets with multi-organ annotations were employed. These reviewed studies are reported in Table 4. The DenseVNet architecture, described in section 3.2, represents the first model applied to multi-organ segmentation from the analyzed studies. It was evaluated on a dataset of 90 CTs (43 of TCIA and 47 of BTCV) [Gibson et al., 2018]. Roth et al. [2018b] proposed a coarse-to-fine approach for multi-organ segmentation based on 3D UNet. After training on an internal dataset of 331 CTs, the model was tested on 150 CTs for segmentation of the liver, spleen, and pancreas. Tong et al. [2020] proposed self-paced DenseNet, an encoder-decoder network with a dual attention block in each encoding and decoding step. SE was used for channel attention and a convolutional layer with a 1x1x1 kernel was used for spatial attention. A self-paced learning strategy was developed for the multi-organ segmentation to adaptively adjust the weight of each class in the loss function. This architecture was evaluated on 90 CTs (43 from TCIA and 47 from BTCV datasets) [Tong et al., 2020]. Isensee et al. [2020] designed nnUNet, capable of automatically configuring itself from preprocessing, training, and post-processing. It was evaluated on 23 different datasets, both with single-labeled organs (e.g., MSD pancreas) and multi-organ (e.g., BTCV). Ma et al. [2022] implemented AbdomenCT-1k, a large dataset of 1,112 CT with annotations of four organs (liver, kidney, spleen, and pancreas) starting from datasets of single-labeled organs (Section 3.7). 3D nnUNet was used for supervised and semi-supervised learning, while 2D nnUNet and conditional random fields for weakly supervised learning, and nnUNet for continual learning. Benchmark and baseline all these types of learning were defined [Ma et al., 2022]. Partially supervised multi-organ segmentation methods were designed to train a multi-organ segmentation model from the partially labeled dataset, where few and not all organs were labeled [Shi et al., 2021]. A simple approach to address this issue was to train multiple networks by splitting the partially labeled datasets into several fully labeled subsets and training a network on each subset for a specific segmentation task. Unfortunately, these methods had large computational costs. Another solution consisted of implementing a network sharing the encoder with a specific decoder for each organ to segment. However, this solution lacked flexibility since a new decoder must be added whenever there was the need to segment a new organ [Liu and Zheng, 2023]. Shi et al. [2021] proposed marginal and exclusive loss for partially labeled datasets. nnUNet was adopted for the segmentation of four organs (liver, kidney, spleen, and pancreas) starting from four datasets of single-labeled organs, e.g. MSD pancreas, and a multi-organ dataset (BTCV). Zhang et al. [2021b] introduced conditional nnUNet by integrating an auxiliary conditional tensor into the decoder to select the specific organ to be segmented. Seven publicly available datasets were used for segmentation of the liver, kidney, spleen, and pancreas [Zhang et al., 2021b]. Liu and Zheng [2023] proposed Context-aware Voxel-wise Contrastive Learning (CVCL) to exploit the vast amount of unlabeled data in the partially labeled datasets. A context-aware voxel-wise contrastive learning was inserted into the bottleneck layer of a 3D nnUNet to increase context awareness in a patch-based strategy. This method was evaluated on the liver, spleen, kidneys, and pancreas using BTCV, MSD datasets for liver, spleen, and pancreas, and kidney tumor segmentation datasets [Liu and Zheng, 2023]. Francis et al. [2023] proposed a conditional GAN with UNet as a generator and an FCN as a discriminator. Residual dilated convolution block and spatial pyramid pooling replaced max-pooling in UNet. An attention gate module was inserted into skip connections. This model was trained, validated, and tested on the four organs of the AbdomenCT-1k dataset (liver, kidneys, spleen, and pancreas) [Francis et al., 2023]. Li et al. [2023f] designed Self-Adjustable Organ Attention UNet (SOA-Net) to adaptively adjust attention and receptive fields sized based on multiple scales of the four organs (liver, kidney, spleen, and pancreas) of both AbdomenCT-1k and AMOS-CT datasets. SOA-Net performed first a localization with a 3D UNet. Then, another UNet used a multibranch feature attention block with four branches in each encoding step and a feature attention aggregation block with two branches in each decoding step. These multi-branch modules had different kernel sizes to capture different scale features based on different scales of the four organs [Li et al., 2023f]. Pan et al. [2022] integrated a multi-layer perceptron mixer into VNet to linearize the computational complexity of transformers. This method was tested on an internal dataset of 59 patients (without pancreas segmentation) and BTCV [Pan et al., 2022]. Shen et al. [2023] designed UNet with spatial attention to highlight the location and sizes of five target organs (pancreas, duodenum, gallbladder, liver, and stomach). Deformable convolutional blocks were added to deal with variations in the shapes and sizes of the organs, while skip connections were designed with multi-scale attention to eliminate the interference of complex backgrounds. It was evaluated into 90 CTs (43 from TCIA and 47 from BTCV) [Shen et al., 2023]. Tong et al. [2023] proposed a two-stage approach using an encoder-decoder network, with a coarse stage for initial localization. Then, the fine

Table 4: Reviewed studies on multi-organ segmentation

| Author | Application | Dataset Size | Model Architecture | Learning Strategy | Loss | Results | Main contributions |
|---|---|---|---|---|---|---|---|
| Francis et al. [2023] | Segmentation of liver, kidneys, spleen, and pancreas | 1,112 (AbdomenCT-1k) | Conditional GAN: Dilated UNet + Attention gate (Generator) + Fully Convolutional Network (Discriminator) | Supervised | Adversarial loss | 86.10% (DSC) 6.65 mm (HD95) 86.80% (Precision) 86.60% (Recall) | Residual dilated convolution block and spatial pyramid pooling replacing convolutions and max pooling in UNet. Attention gate inserted into skip connections |
| Huang et al. [2023] | Segmentation of aorta, gallbladder, spleen, kidneys, liver, pancreas, spleen, stomach, ventricles, myocardium, and retina | 30 (Synapse) 100 MRI (Automated cardiac diagnosis challenge) 40 (Digital Retinal Images for Vessel Extraction) | Encoder-decoder with: transformer blocks in all encoding and decoding steps + Transformer context bridge between encoder and decoder (fusion of multi-scale information) | Supervised | – | 65.67% (DSC) | Hierarchical encoder-decoder with ReMix-FFN module in each transformer block with a convolution and a skip connection between the two fully connected layers to capture local information in addition to global dependencies. Features of different scale as output of each encoder step are concatenated, and sent to ReMixed transformer context bridge with self-attention to capture global dependencies. The output features are split into different scale feature maps and sent to ReMix-FFN modules of the decoder to mix global dependencies with local context. Code available at: https://github.com/ZhifangDeng/MISSFormer |
| Li et al. [2023f] | Segmentation of liver, kidney, spleen, and pancreas | 500 from AbdomenCT-1k 240 from AMOS-CT | 3D UNet (Localization) UNet with: Multi-branches feature attention (Encoder) + Feature attention aggregation (Decoder) (Segmentation) | Supervised | Dice loss | First dataset: 86.20% (DSC) Second dataset: 78.40% (DSC) | Network with self-adjustable attention and receptive field size to segment liver, kidney, spleen, and pancreas. Different kernel sizes to capture different scale features of different organs using: multi-branch feature attention with four branches, and feature attention aggregation with two branches |
| Liu and Zheng [2023] | Segmentation of liver, spleen, pancreas, and kidneys | 30 (BTCV) 281 (MSD) + partially labeled datasets of other abdominal organs | nnUNet | Semi-supervised | Cross entropy loss Dice loss (labeled data) Context-aware voxel-wise contrastive learning loss (unlabeled data) | 83.60% (DSC) 4.30 mm (HD95) | Exploiting unlabeled information in partially labeled datasets. Context-aware voxel-wise contrastive learning inserted into the bottleneck layer of a 3D uNet to increase context awareness in patch-based strategy |
| Pan et al. [2022] | Segmentation of spleen, kidneys, gallbladder, esophagus, liver, stomach, aorta, vena cava, and pancreas | 59 (Institutional dataset without pancreas) 30 (BTCV) | VNet with Multi-layer perceptron Mixer replacing CNN | Supervised | Cross entropy Dice loss | BTCV: 79.00% (DSC) | Multi-layer perceptron mixer was integrated into VNet to linearize the computational complexity of transformers |
| Shen et al. [2023] | Segmentation pancreas, duodenum, gallbladder, liver, and stomach | 42 (NIH) | UNet with: Spatial attention (location and size of organs) + Dilated convolution + Multi-scale attention | Supervised | Dice loss Cross entropy loss | 75.42% (DSC) 61.84% (Jaccard) 19.99 mm (HD) | Spatial attention to highlight location and sizes of target organs (pancreas, duodenum, gallbladder, liver, and stomach). Deformable convolutional blocks to deal with variations in shapes and sizes. Skip connections with multi-scale attention to eliminate interference of complex background |
| Tong et al. [2023] | Segmentation of liver, kidney, spleen, and pancreas | 511: 80 (NIH) 281 (MSD) + datasets of other organs (multi-center) | Encoder-Decoder (Localization) ResUNet and Multi-scale Attention (Segmentation) | Supervised | Dice loss (Localization) Dice loss Mean square error (Segmentation) | 59.10% (DSC) 42.20% (NSD) | Coarse stage for localization. Fine stage with multi-scale attention to segment pancreas, liver, spleen, and kidney. |
| Yuan et al. [2023] | Segmentation of aorta, gallbladder, kidneys, liver, pancreas, spleen, and pancreas | 30 (Synapse) | UNet-like with: Gated recurrent units for skip connections + Gated-dual attention (Multi-scale weighted channel attention + Transformer self attention) | Supervised | – | 62.77% (DSC) | Gate recurrent units integrated into skip connections to reduce the semantic gap between low and high-level features. Gated-dual attention to capture information on small organs and global context. Code available at: https://github.com/DAgalaxy/MGB-Net |

stage was implemented with multi-scale attention to segment the pancreas, liver, spleen, and kidney on a dataset of 511 CTs based on NIH, MSD, kidney tumor segmentation, and internal dataset of Nanjing University [Tong et al.,

| Author | Application | Dataset Size | Model Architecture | Learning Strategy | Loss | Results | Main contributions |
|---|---|---|---|---|---|---|---|
| Li et al. [2022c] | Segmentation of spleen, kidney, gallbladder, esophagus, liver, stomach, pancreas, and duodenum | 90: 43 (TCIA) + 47 (BTCV) 511: 80 (NIH) 281 (MSD) + datasets of other organs (multi-center) | 3D Encoder-Decoder (Localization) 2.5D netowrk (Segmentation) | Supervised | Design of parameter loss to remove the false positive of dice loss | First dataset: 84.00% (DSC) 5.67 mm (HD95) Second dataset: 83.00% (DSC) | Circular inference (a sort of micro-attention mechanism) and parameter Dice loss in the first stag to reduce uncertain probabilities of blurred boundaries. |
| Sundar et al. [2022] | Segmentation of pancreas and non abdominal organs | 50 (internal) | nnUNet | Supervised | – | 85.00% (DSC) | Development of Multiple-organ objective segmentation (MOOSE) framework. Code available at: https://github.com/QIMP-Team/MOOSE |
| Zhao et al. [2022b] | Segmentation of aorta, gallbladder, kidneys, liver, pancreas, spleen, and stomach | 30 (Synapse) | UNet-like with: Encoder: ResNet-50 + Progressive sampling module + Vision Transformer (Hybrid CNN-Transformer) | Supervised | Cross entropy loss Dice loss | 59.84% (DSC) | A progressive sampling module to ensure that highly relevant regions of the organ are in the same patch |
| Isensee et al. [2020] | Segmentation of heart, atrium, ventricles, myocardium, aorta, trachea, lung, hyppocampus, esophagus, liver, kidneys, pancreas, spleen, colon, gallbladder, and stomach | 281 (MSD) + datasets of other organs | nnUNet | Supervised | Cross entropy loss Dice loss Weighted binary cross entropy loss | 2D UNet: 77.38% (DSC) 3D UNet Full resolution: 82.17% (DSC) 3D UNet low resolution: 81.18% (DSC) | Original paper on the implementation of nnUNet. nnUNet has three configurations: 2D UNet, 3D UNet with full resolution, and 3D UNet with low resolution. Code available at: https://github.com/MIC-DKFZ/nnUNet ?tab=readme-ov-file |
| Ma et al. [2022] | Segmentation of liver, kidney, spleen, and pancreas | 1,112 (AbdomenCT-1k) | 3D nnUNet (Supervised and semi-supervised) 2D nnUNet + Conditional random fields (Weakly supervised) nnUNet (Continual) | Supervised Semi-supervised Weakly supervised Continual | Dice loss Cross entropy loss | Single organ (trained on MSD): 86.10% (DSC) 66.10% (NSD) Multi-organ (trained on MSD): 90.10% (DSC) 82.30% (NSD) Supervised (MSD) + liver (40) and kidney (40): 78.10% (DSC) 65.00% (NSD) Semi-supervised: 85.70% (DSC) 72.50% (NSD) Weakly supervised: 70.50% (DSC) 55.00% (NSD) Continual: 64.70% (DSC) 51.10% (NSD) | Presentation of a large dataset with the addition of multi-organ (liver, kidney, spleen and pancreas) annotations to original datasets. Definition of benchmark and baseline for supervised, semi-supervised, weakly supervised, and continual learning. Code available at: https://github.com/JunMa11/ AbdomenCT-1K |
| Shi et al. [2021] | Segmentation of liver, spleen, pancreas, and kidney | 30 (BTCV) 281 (MSD) + datasets of other organs | nnUNet | Supervised | Marginal loss Exclusive loss | 80.80% (DSC) 3.96 mm (HD) | Implementation of marginal loss (for background) label and exclusion loss (different organs are mutually exclusive) |
| Zhang et al. [2021b] | Segmentation of liver, pancreas, spleen, and kidney | 30 (BTCV) 281 (MSD) + datasets of other organs | nnUNet + Auxiliary information into decoder | Supervised | Dice loss Focal loss | 83.97% (DSC) | Four datasets with annotations of different organs (liver, pancreas, spleen, and kidney). An auxiliary conditional tensor is concatenated into the decoder to select the specific organ to segment |

2023]. Yuan et al. [2023] designed a UNet-like network with gate recurrent units integrated into skip connections to reduce the semantic gap between low and high-level features. A gated-dual attention module (multi-scale weighted channel attention and transformer self-attention) was implemented to capture information on small organs and global contexts. This approach was trained on 18 and tested on 12 CTs of the Synapse dataset [Yuan et al., 2023]. Li et al. [2022c] added circular inference (a sort of micro attention mechanism) and parameter Dice loss in the first stage of a 3D encoder-decoder network to reduce the uncertain probability of blurred edges. This model was evaluated on segmentation of eight and four organs of 90 and 511 CTs, respectively [Li et al., 2022c]. Huang et al. [2023] introduced Medical Image Segmentation tranSFormer (MISSFormer), a hierarchical encoder-decoder network with a transformer block (ReMix-FFN) in each encoding and decoding step, and a transformer context bridge (ReMixed) between encoder and decoder to fuse multi-scale information. A ReMix-FFN module was designed in each transformer block with a

| Author | Application | Dataset Size | Model Architecture | Learning Strategy | Loss | Results | Main contributions |
|--------|-------------|--------------|--------------------|--------------------|------|---------|---------------------|
| Park et al. [2020] | Segmentation of pancreas and other 16 anatomical structures | 1,150 (John Hopkins) | Two-stage Organ attention network | Supervised | – | 87.80% (DSC) | Annotation of 22 structures. Use of two-stage organ attention network: two FCN for segmentation. The first used reverse connections to get more semantic information. The results became attention-organ module to guide the second network. This architecture was applied to each view. The outputs from axial, coronal, and sagittal views were then fused |
| Tong et al. [2020] | Multi-organ Segmentation | 90: 43 (TCIA) 47 (BTCV) | Encoder-Decoder with dual attention: Squeeze and Excitation (Channel attention) Convolutional layer (Spatial attention) | Supervised | – | 79.24% (DSC) 1.82 mm (ASD) | A self-paced learning strategy for the multi-organ segmentation to adaptively adjust the weight of each class |
| Gibson et al. [2018] | Segmentation of spleen, kidney, gallbladder, esophagus, liver, stomach, pancreas, and duodenum | 90: 43 (TCIA) 47 (BTCV) | DenseVNet | Supervised | L2 regularization loss Dice loss | 78.00% (DSC) 5.9 mm (HD95) | Implementation of DenseVNet with: cascaded dense feature stacks, V-network with downsampling and upsampling, dilated convolutions, map concatenation, and a spatial prior. Application to eight abdominal organs |
| Roth et al. [2018b] | Segmentation of artery, vein, liver, spleen, stomach, gallbladder, and pancreas | 331 (internal for training) 150 (external for testing) | 3D UNet for both: Localization and Segmentation | Supervised | Weighted cross entropy loss | External dataset: 82.20% (DSC) | Application of cascaded networks for localization (coarse stage) and segmentation (fine stage) |

convolution and a skip connection between the two fully connected layers to capture local information in addition to global dependencies. Features of different scales as output of each encoder step were concatenated, and sent to the ReMixed transformer context bridge with self-attention to capture global dependencies. The output features were split into different scale feature maps and sent to ReMix-FFN modules of the decoder to mix global dependencies with local context. This model was trained on 18 and tested on 12 CTs of Synapse dataset [Huang et al., 2023]. Zhao et al. [2022b] designed an encoder-decoder architecture with a hybrid CNN-Transformer encoder evaluated on the Synapse dataset. A progressive sampling module and a ViT were inserted into the bottleneck. In contrast with ViT where the input images were linearly divided into patches without considering that this splitting may compromise the integrity of organs in CT images, a progressive sampling module was implemented to mitigate the damage to organ integrity. This would ensure that relevant regions of the organ were in the same patch as much as possible. This was achieved by updating the sampling location over four iterations using an offset vector from the previous iteration, instead of sampling at a fixed location as in the case of ViT [Zhao et al., 2022b].

## 4.4   Semi-supervised learning

Semi-supervised learning methods combine labeled with unlabeled data under the condition that labeled and unlabeled data have the same statistical distribution [Chen et al., 2022a]. Semi-supervised learning methods can be divided into consistency regularization, pseudo labeling, and generative approaches [Chen et al., 2022a]. Consistency regularization methods are based on the assumption that prediction on unlabeled data should not change significantly if perturbations like noise or data augmentation were added. An example is the mean-teacher model [Tarvainen and Valpola, 2017]. In pseudo-label methods, a semi-supervised model generates pseudo annotations for unlabeled samples. The pseudo-label examples are used jointly with labeled ones to train the model. This is an iterative process. Finally, generative models include GANs to generate high-quality samples [Chen et al., 2022a]. The reviewed studies on semi-supervised learning for pancreas parenchyma segmentation are displayed in Table 5. Liu et al. [2022a] introduced a graph-enhanced pancreas segmentation network (GEPS-Net) to overcome the limited effectiveness of previous methods on pseudo-label generation to represent pancreases with different sizes and shapes. Since the generated pseudo labels may be unreliable and noisy, they are refined using an uncertainty iterative strategy. This method was trained and tested on NIH dataset Liu et al. [2022a]. You et al. [2022] designed Simple Contrastive Voxel-Wise Representation Distillation (SimCVD), a semi-supervised framework combining contrastive distillation with geometric constraints. It imposed global consistency in object boundary contours to capture more effective geometric information. A teacher and a student model (VNet in both cases) were fed with two perturbed versions of an input image volume to generate a probability map and a boundary representation using a signed distance map. The signed distance maps were contrasted in a shared latent space distillation. Additionally, a voxel-to-voxel pair-wise distillation was performed to explore the structural relationships between voxel samples to improve spatial labeling consistency. This method was trained and tested on NIH dataset You et al. [2022]. Xia et al. [2020] introduced co-training into semi-supervised learning. Co-training was used to have different views to learn complementary information during training. The views were generated by rotations or permutation transformations. An uncertainty-weighted label fusion module, based on the Bayes network, was designed to assess the quality of each view prediction to generate reliable pseudo labels. This method was evaluated on NIH and the multi-organ dataset reported by Gibson et al. [2018] with 10%, 20% and for NIH also 100% of labeled data [Xia et al., 2020]. Since the initial pseudo labels prediction is crucial in the segmentation results, Shi et al. [2022] introduced the Conservative-Radical network (CoraNet) to reduce uncertainty. This model consisted of a module to indicate certain and uncertain region masks, a network for the segmentation of certain and another network for the segmentation of uncertain regions. For unlabeled data, the segmentation model for certain regions was used to generate pseudo labels. For the uncertain regions teacher and student models were used to impose them to have a consistent prediction on unlabeled samples. Xia et al. [2023] integrated multidimensional feature attention and improved cross-pseudo supervision into VNet for semi-supervised segmentation. This method used data disturbance consistency regularization to improve the robustness of networks. The improved cross-pseudo supervision module was designed to ensure result consistency by increasing the robustness to noise of two network branches fed with the same image as input but in one case with noise. The multidimensional feature attention module was based on CBAM to integrate low-dimensional and high-dimensional feature attention to ensure feature consistency by minimizing the difference between the maps of the two networks. This method was trained and evaluated on NIH dataset [Xia et al., 2023]. Compete-to-Win (ComWin) was introduced by Wu et al. [2023] to reduce the number of false positives and generate more accurate pseudo labels than cross-pseudo supervision. High-quality pseudo labels were generated by comparing multiple confidence maps produced by different networks to select the most confident one. A boundary-aware enhancement module was integrated to enhance boundary discriminative features. This approach was evaluated on NIH dataset [Wu et al., 2023]. Zeng et al. [2022] implemented a teacher-student model where the student learned from pseudo labels generated by the teacher network which was initially trained in supervised learning on labeled images. The teacher in turn learned from the performances of the student on the labeled images. This model was trained and tested on three datasets on different anatomical structures, one of which was the NIH dataset [Zeng et al., 2022]. Petit et al. [2021] proposed a pseudo-label method. He first used 3D spatial priors to merge the position of the pancreas with the results of segmentation by UNet on the NIH dataset. For unlabeled data, they took the output probabilities of the segmentation network on unlabeled volumes to compute a coarse position of the pancreas. Then, they randomly selected a reference CT volume in the dataset to refine the position via Kullback–Leibler divergence. They generated pseudo labels on different percentages of NIH dataset [Petit et al., 2021].

## 4.5 Unsupervised learning

The reviewed studies on unsupervised learning for pancreas parenchyma segmentation are displayed in Table 6. Zhu et al. [2022] proposed an unsupervised adversarial domain adaptation method based on multiscale progressively weighted features mapping the feature space of the target domain to source domains. A segmentation network integrating residual blocks, SE attention (with 3D convolutions), and UNet (SE-PResUNet) was designed for the segment source and target domains. Adaptation from source to target domains required three stages. In the first one, SE-PResUNet was trained in the source domain on labeled images and used to initialize the model of the target domain. In the second one, SE-PResUNet extracted the source and target domain image features from multiple scales in the upsampling layers and sent them to the discriminator separately. In the third one, the parameters of the target domain feature extraction model were updated to be closer to the feature distribution of the source domain. The second and third stages were alternating until the discriminator could not correctly distinguish the specific domain. The method was used for domain adaptation from NIH to Zheyi and from MSD to Zheyi dataset [Zhu et al., 2022]. The method proposed by Xia et al. [2020] (see Section 4.4) was tested for unsupervised domain adaptation from the the multi-organ dataset described by Gibson et al. [2018] to MSD of pancreas and MSD of liver. Zheng et al. [2020] proposed a coarse-fine method with two

Table 5: Studies on parenchyma segmentation using semi-supervised learning

| Author | Application | Dataset Size | Model Architecture | Learning Strategy | Loss | Results | Main Contributions |
|---|---|---|---|---|---|---|---|
| Xia et al. [2023] | Segmentation of pancreas | 82 (NIH) | VNet + Multi-dimensional Feature attention | Semi-supervised | Cross entropy Dice loss Mean square error | 79.55% (DSC) 66.87% (Jaccard) 7.67 mm (HD95) 1.65 mm (MSD) | Multi-dimensional feature attention and improved cross pseudo supervision to effectively use unlabeled data reducing the need of labeled data |
| Wu et al. [2023] | Segmentation of pancreas, left ventricle, myocardium, right ventricle, and colon | 80 (NIH) + datasets of other organs | V-Net + Attention | Semi-supervised | Cross entropy Dice loss | 74.03% (DSC) 59.70% (Jaccard) 2.12 voxel (ASD) 9.10 voxel (HD95) | Instead of using model predictions as pseudo labels, high-quality pseudo labels are generated by comparing multiple confidence maps produced by different networks to select the most confident one (a compete-to-win strategy). A boundary-aware enhancement module was integrated to enhance boundary discriminative features. Code available at: https://github.com/Huiimin5/comwin |
| Zeng et al. [2022] | Segmentation of pancreas, and left atrium | 82 (NIH) + datasets of other organs | V-Net | Semi-supervised | Cross entropy | 84.77% (DSC) 73.71% (Jaccard) 6.24 voxel (HD95) 1.58 voxel (ASD) | Teacher-student trained in parallel: the student learns from pseudo labels generated by the teacher learning in turn from the performances of student on the labeled images |
| Liu et al. [2022a] | Segmentation of pancreas | 82 (NIH) | Graph-enhanced nnUNet | Semi-supervised | Cross entropy Dice loss | 84.22% (DSC) 73.10% (Jaccard) 6.63 voxel(HD95) 1.86 voxel (ASD) | A graph CNN was added to nnUNet to distinguish the low contrast edges of a pancreas. Pseudo labels are refined using an uncertainty iterative strategy |
| Petit et al. [2021] | Segmentation of pancreas | 82 (NIH) | UNet | Supervised Semi-supervised | – | 77.53% (DSC) | Fusion of a FCN probability prediction volume with 3D spatial prior representing the probability of organ presence |
| You et al. [2022] | Segmentation of pancreas, and left atrium | 82 (NIH) + datasets of other organs | V-Net for knowledge distillation | Semi-supervised | Cross entropy Dice loss Mean squared error (Supervised) Design of: Boundary-aware contrastive, Pair-wise distillation, and Consistency losses | 89.03% (DSC) | Contrastive distillation model with multi-task learning (segmentation map and signed distance map from boundary). Structured distillation in the latent feature space followed by contrasting the boundary-aware features in the prediction space for better representations |
| Shi et al. [2022] | Segmentation of pancreas endocardium, right and left ventricle, and myocardium | 82 (NIH) | UNet V-Net ResNet-18 | Semi-supervised | Cross entropy loss for conservative and radical model (labeled data) Cross entropy loss for certain regions + Consistency loss for uncertain regions (Unlabeled data) | UNet: 67.01% (DSC) V-Net: 79.67% (DSC) 66.69% (Jaccard) 1.89 voxels (ASD) 7.59 voxels (HD) ResNet-18: 80.58 %(DSC) 67.91% (Jaccard) 2.27 voxels (ASD) 8.34 voxels (HD) | A conservative-radical module to automatically identify uncertain regions. A training strategy to separately segment certain and uncertain regions. Mean teacher model for uncertain region segmentation |

VNet models to initially localize the pancreas. Three 2.5D networks (each for axial, coronal, and sagittal), based on encoder-decoder architecture were used to segment the pancreas. A slice correlation module, based on attention, was

| Author | Application | Dataset Size | Model Architecture | Learning Strategy | Loss | Results | Main Contributions |
|--------|-------------|--------------|--------------------|--------------------|------|---------|--------------------|
| Xia et al. [2020] | Segmentation of pancreas | 82 (NIH) 90: 43 (TCIA) + 47 (BTCV) 281 (MSD) + MSD (liver) | Encoder-Decoder based on ResNet18 for Multi-view Co-training and Domain-adaptation | Semi-supervised Unsupervised | Combination of conventional segmentation loss (labeled) and computational function based on uncertainty-weighted label fusion (unlabeled) | NIH: 81.18% (DSC) TCIA+BTCV (External validation): 77.91% (DSC) MSD (Domain adaptation): 74.38% (DSC) | Co-training to maximize the similarity of the predictions among different views, generated by rotation or permutation transformations. Uncertainty weighted label fusion module for accurate pseudo labels generation for each view. Adaptation from multi-organ to pancreas dataset without source domain data |

designed. A pre-training unsupervised module was implemented to shuffle the slices so that the 2.5D segmentation networks had to reorder the slice. This forced the 2.5D models to learn the relationships among slices. The learned parameters were then used as initial weights of the segmentation model. This method was trained and tested on NIH dataset [Zheng et al., 2020]. Li et al. [2023d] proposed a two-stage method on which they first used the meta-learning described above for localization [Li et al., 2023c]. Then for the second stage, they implemented a dual self-supervised generalization model. Global-feature self-supervised module based on contrastive learning was developed to reduce the influence of extra-pancreatic tissues. Then, a local image restoration module based on self-supervised module learning was designed to exploit anatomical context to enhance the characterization of high-uncertain regions. This approach was trained in turn on each of NIH, MSD, and an internal dataset of 104 CTs (Renji Hospital Shanghai) and tested for generalization on the other two [Li et al., 2023d]. Zhu et al. [2023] implemented a domain transfer from a source center with labeled data to a target one with unlabeled data using a ResUNet with SE attention. This model was first trained on labeled data at the source center to obtain an initial pancreas segmentation. Then, pairs of images from labeled and unlabeled datasets were sent to the model to generate multiscale feature maps which were then trained by a discriminator of a GAN model hosted at a third center for domain identification. For labeled and unlabeled data the NIH and Zheyi (with 70 CTs) datasets were used, respectively [Zhu et al., 2023].

## 4.6 Generalization to other datasets

A typical limitation of DL models is the lack of a demonstration of how they perform on external datasets. Therefore some methods were proposed to address this limitation for pancreas segmentation. In this section we reviewed them. The two stage models developed by [Tian et al., 2023, Qiu et al., 2022a, Panda et al., 2021, Ge et al., 2023, Li et al., 2023d, Roth et al., 2018b] were tested for generalization (see Section 4.2). Knolle et al. [2021] proposed a UNet-like model with dilated convolution trained on MSD and tested for generalization on an internal dataset of 85 CTs. The method by Zhu et al. [2022] for domain adaptation (see Section 4.5) was tested for generalization from NIH to Zheyi dataset and from MSD to Zheyi dataset. Lim et al. [2022] compared four 3D architectures based on UNet on 1,006 CT from Gil Medical Center and assessed generalization on the NIH dataset. Qu et al. [2022] proposed M$^3$Net, an encoder-decoder model with a multi-scale, multi-view architecture integrating also attention for multi-phase segmentation. Two model branches were designed for arterial and venous phases. Each model consisted of a 3D encoder and a 2D decoder. Cross-phase between the models was performed via a non-local attention block. This structure was replicated for each of the axial, coronal, and sagittal axes. A multi-view ensemble strategy averaged the segmented results along with the three views. This process was repeated for a high resolution and half resolution to extract local and global features. M$^3$Net was trained on 224 CTs from Peking Union Medical College Hospital and generalized on an external dataset of 66 CTs from Hedan Cancer Hospital [Qu et al., 2022].

## 4.7 Design of loss functions

Lu et al. [2019] proposed a complex-coefficient Dice loss evaluating not only the ratio of the coincident area w.r.t the total area but also the shape similarity between the ground truth and the predicted result. Karimi and Salcudean [2020] reported the first work aiming to reduce Hausdorff distance by proposing three different losses. The first was based on distance transform, a representation where each pixel of an image has a value equal to its distance from an object of interest. Although simple, this method had a high computational cost. The second one was based on morphological erosion by considering that HD is related to the thickness between the ground truth and segmentation. The third one was based on convolutions with circular/spherical kernels [Karimi and Salcudean, 2020]. Ma et al. [2021b] presented the first work integrating geodesic active contour and CNNs to reduce boundary errors. Geodesic active contour treats image segmentation as an energy minimization problem. It is based on a level set function, defined as a signed distance function, with a value of zero at the organ contour, negative values inside, and positive values outside the organ [Ma et al., 2020]. More generally, distance transform generates a distance map with the same size as the input image, where the value on each pixel is the distance from the foreground pixel to the foreground boundary [Rosenfeld and Pfaltz, 1968]. Three loss functions were used [Ma et al., 2021b]. First, Dice loss to compute the overlap between ground truth and predicted output. Second, L1 loss to make the predicted value of the level set function close to the one of the ground truth. Third, geodesic active contour loss was capable of considering more object global information than Dice loss or cross-entropy thanks to the level set function leading to global variations in case of small segmentation errors [Ma et al., 2021b]. Xia et al. [2020] implemented a loss function for co-training of different views generated from spatial transformation of input images in a semi-supervised setting. An uncertainty-weighted label fusion module was developed for accurate pseudo-label generation of each view. Shi et al. [2021] proposed marginal loss and exclusive loss for partially supervised multi-organ segmentation to treat the unlabeled organs and the real background as an overall background, and to consider different organs are mutually exclusive. Shi et al. [2022] proposed a consistent loss for uncertain regions, based on segmentation of a teacher-student model, in addition to cross-entropy for certain regions. Both losses were used on unlabeled data of a semi-supervised learning approach [Shi et al., 2022]. Li et al. [2022c] introduced two penalty factors ($\alpha$ and $\beta$), which are learnable parameters trained together with network parameters, into the Dice loss to reduce the false positive points in blurred edges of small organs like the pancreas. Li et al. [2023c] proposed an adaptive loss to improve the generalization of a two-stage approach from the training set to the test set. The loss combined a recall loss (based on recall value) to evaluate the coarse stage, and binary cross-entropy and dice loss to evaluate the accuracy of the fine stage. The loss was adapted by applying model-agnostic meta-learning, where a set of temporary intermediate parameters ($\theta'$) computed during the meta-train stage were assessed for accuracy during the meta-test stage [Li et al., 2023c]. You et al. [2022] designed three different losses for contrastive learning. First, boundary-aware contrastive loss to enforce the consistency of the predicted signed distance map outputs on the unlabeled set during training. Second, pair-wise distillation loss to explore structural relationships between voxel samples to increase spatial labeling consistency. Third, consistency loss to improve training stability and performances on unlabeled data, by adding different perturbation operations on unlabeled input images [You et al., 2022]. Liu and Zheng [2023] designed a context-aware contrastive learning loss for CVCL for unlabeled data in addition to cross entropy and Dice loss for labeled data (see Section 4.3). The context-aware contrastive learning loss was developed by considering that the labeled organs were unlikely to be predicted in the unlabeled part of a partially labeled dataset given that all the voxels of labeled organs had already been filtered out; the higher confidence of the positive pair should be greater than a threshold to ensure the quality of the aligned features; and there may be several confidence levels for pseudo-labels for a positive voxel pair [Liu and Zheng, 2023].

Table 6: Studies on parenchyma segmentation using unsupervised learning

| Author | Application | Dataset Size | Model Architecture | Learning Strategy | Loss | Results | Main Contributions |
|---|---|---|---|---|---|---|---|
| Li et al. [2023d] | Segmentation of pancreas | 82 (NIH) 281 (MSD) 104 (Renji Hospital Shanghai Private dataset) | UNet with meta-learning (Localization): Global feature contrastive learning 3D UNet: Local image restoration (Segmentation) | Self-supervised | Binary cross entropy loss Dice loss Squared error loss Adversarial loss | Training on NIH Generalization on MSD: 66.73% (DSC) Generalization on Private: 73.85% (DSC) Training on MSD Generalization on NIH: 76.71% (DSC) Generalization on Private: 83.50% (DSC) Training on Private Generalization on NIH: 65.03% (DSC) Generalization on MSD: 70.08% (DSC) | Dual self-supervised generalization model to enhance characterization of high-uncertain regions. Global-feature self-supervised contrastive learning reducing the influence of extra-pancreatic tissues. Local image restoration self-supervised module to exploit anatomical context to enhance characterization of high-uncertain regions |
| Zhu et al. [2023] | Segmentation of pancreas | 82 (NIH) 70 (Zheyi): Zhejiang University Hospital | Adversarial network + 3D ResUNet + Attention (Squeeze-Excitation) | Supervised and Unsupervised | Dice loss Cross entropy loss | NIH (supervised): 85.45% (DSC) Zheyi (unsupervised): 75.43% (DSC) | Training with 3D ResUNet and attention module using pairs of labeled images fro one center and unlabeled ones from a different center to generate multi-scale feature maps. Labeled and unlabeled data are then trained by a discriminator for domain identification |
| Zhu et al. [2022] | Segmentation of pancreas | 82 (NIH) 281 (MSD) 70 (Zheyi) | Residual blocks + Squeeze-Excitation Attention + UNet | Domain adaptation: Supervised learning (source) Unsupervised learning (target) | Cross entropy Dice loss | NIH adapted to Zheyi 72.73% (DSC) MSD adapted to Zheyi 71.17% (DSC) | Adversarial multiscale domain adaption (from source) to generalize to external datasets (target domain) |
| Zheng et al. [2020] | Segmentation of pancreas | 82 (NIH) | 3D VNet (Localization) 2.5D Encoder-decoder (Segmentation) | Self supervised | Square root Dice loss | 78.10% (DSC) | Square Root Dice loss to deal with the trade-off between sensitivity and specificity. Slice shuffle for pre-training before input to the network which learns to reorder and understand organ shape. Capturing of non-local information through attention, pooling, and convolutional layers. Ensemble learning and recurrent refinement to improve accuracy |

## 4.8  Comparison of performances

The vast number of reviewed studies on the segmentation of pancreas parenchyma highlighted that the topic has been extensively investigated. More specifically, the application of a wide range of different DL architectures on the same datasets has made it possible to compare the studies within the groups defined in Sections 4.2-4.7. The NIH dataset was by far the most used one, recurring in 80 out of 105 studies (76.2%), as follows: in 42 studies it was the only one adopted, while in 38 it was coupled with others (in 19 cases with MSD dataset). The MSD was used in 39 studies (37.1%), BTCV in 10 (9.5%), TCIA in six (5.7%), AbdomenCT-1k in four (3.8%), and Synapse in three (2.8%). Despite the promise of transformers, either alone or in hybrid networks with CNNs, and the different architectures proposed for two-stage approaches, a UNet configuration with residual blocks in the encoder and a decoder with spatial and channel attention obtained the highest DSC score (91.37%) on the NIH dataset [Shan and Yan, 2021]. This was followed by a two-stage hybrid method, with a UNet for localization and ViT for segmentation, reporting a DSC of 89.89% on NIH and 91.22% on MSD datasets [Dai et al., 2023]. Notably, several two-stage approaches reached almost the same value of DSC (slightly above 86.0%) using UNet for localization and residual transformer with UNet for segmentation [Qiu et al., 2023], UNet3+ with multi-scale feature calibration in both stages [Qiu et al., 2022b], mask R-CNN for localization and UNet for segmentation [Dogan et al., 2021], UNet in both stages [Wang et al., 2021a], VGG with attention gate for localization) and VGG-16 with residual multi-scale dilated attention for segmentation [Chen et al., 2022c]. The majority of the studies on NIH and MSD datasets reported performances using region-based metrics like DSC and Jaccard, neglecting the importance of boundary-based metrics [Ma et al., 2022]. Of all the studies on NIH few were tested for generalization, in all cases on the MSD dataset, reaching a DSC of 85.41% Qiu et al. [2022b], 76.60% Chen et al. [2022c], and 81.09% Li et al. [2023c] using supervised learning. The latter used also an internal dataset of 104 CTs for training in addition to NIH. Likewise for model generalization from NIH to MSD, self-supervised learning with UNet for both localization and segmentation achieved a DSC of 66.73% [Li et al., 2023d]. Overall, the methods using supervised learning achieved higher DSC scores on the NIH dataset than those based on semi-supervised and unsupervised learning. VNet models obtained the highest DSC score on semi-supervised and unsupervised learning, 89.03% and 78.10% [You et al., 2022, Zheng et al., 2020]. Being a multi-organ dataset, AbdomenCT-1k (Section 3.7) was used in studies needing an annotated dataset for the pancreas and in those evaluating DL models for segmentation on four labeled organs (pancreas, spleen, kidneys, and liver). The highest DSC score (86.10%) on the full AbdomenCT-1k was reached by a conditional GAN with dilated UNet and attention gate for the generator and an FCN for the discriminator [Francis et al., 2023]. The most comprehensive analysis on AbdomenCT-1k was performed by Ma et al. [2022]. In addition to DSC, NSD was used for the assessment of segmentation results at boundary level [Ma et al., 2022]. When using the MSD subset of AbdomenCT-1k for training, nnUNet reached a DSC of 86.10% (with only annotations of the pancreas), while the metric value rose to 90.10% if nnUNet was trained on MSD (with annotations of the pancreas, liver, spleen, and kidneys) and tested on the liver tumor part of AbdomenCT-1k [Ma et al., 2022]. If trained with supervised learning with MSD plus 40 cases of liver tumors, and 40 of kidney tumors the score of DSC dropped to 78.10% when tested in 50 challenging and 50 random cases. For a semi-supervised learning setting, a DSC of 85.70% was achieved. When tested on the 50 CT scans of Nanjing University (Section 3.7) with cancers of the colon, pancreas, and liver, but keeping the same training strategy, nnUNet reached 82.50% and 82.30% for supervised and semi-supervised learning, respectively [Ma et al., 2022]. Evidence showed that DSC can vary substantially when choosing a random subset of AbdomenCT-1k. A two-stage method with nnUNet for localization and a variational model for segmentation obtained a DSC of 89.61% on 200 random cases of AbdomenCT-1k. When generalizing to an internal dataset of 50 scans it reached 90.72% [Tian et al., 2023]. A UNet for localization and another UNet for segmentation, with multi-branch feature attention in the encoder and feature attention aggregation in the decoder, obtained a DSC of 86.20% on 500 random cases AbdomenCT-1k [Li et al., 2023f]. In contrast, DSC on 240 random scans of AMOS-CT dropped to 78.40% [Li et al., 2023f]. TCIA and BTCV datasets were used alone or in combination, typically with 43 scans of the former and 47 of the latter. In all cases the proposed methods were based on UNet variants or encoder-decoder, reaching a maximum DSC of 84.00% for a two-stage model with an encoder-decoder for localization and a 2.5D network for segmentation, improving the results of the first study (78.00%) using the DenseVNet model [Li et al., 2022c, Gibson et al., 2018]. The lowest DSC scores were reported on the Synapse dataset (Table 4, and Appendix A). Even a complex model like MISSFormer (cfr. Section 4.3) was not capable of reaching a DSC of 66.00% [Huang et al., 2023]. Some large private datasets were internally curated. For instance, a two-stage model based on UNet for both localization and segmentation was applied to a dataset of 1,917

CTs from Mayo Clinic (United States). This model reached 91.00% of DSC on 41 cases of TCIA. When generalized to the NIH dataset, the DSC score was 89.00% [Panda et al., 2021]. A subset of this dataset (294 cases) reported a slightly lower DSC (88.00%) using a similar two-stage architecture on UNet [Khasawneh et al., 2022]. Another large private dataset of 1,150 CTs was curated at John Hopkins (United States) with annotation of 22 anatomical structures. A two-stage organ attention network reached a DSC of 87.80% [Wang et al., 2019b, Park et al., 2020].

## 5 Segmentation of tumors, cysts, and inflammations

Segmentation of pancreas tumors, cysts, and inflammation is quite a novel task with the first studies published in peer-reviewed journals dating back to 2020 [Turečková et al., 2020, Xie et al., 2020]. As expected, the number of reviewed studies was lower, 25 vs. 105 for parenchyma. As for pancreas parenchyma, this section starts by showing the variability of tumors in terms of size and location (Section 5.1). Then, it analyzes the studies on DL for the segmentation of tumors, cysts, and inflammation of the pancreas. Overall 25 works were reviewed. The complete list is displayed in Appendix B. The studies were subdivided into the following groups: multi-stage (Section 5.2.1), and other methods for tumors (Section 5.2.2), cysts (Section 5.3), inflammation (Section 5.4), semi-supervised learning (Section 5.5), generalization to other datasets (Section 5.6), and design of new loss functions (Section 5.7). As for parenchyma, this section ends by comparing the performances of the different DL models (Section 5.8).

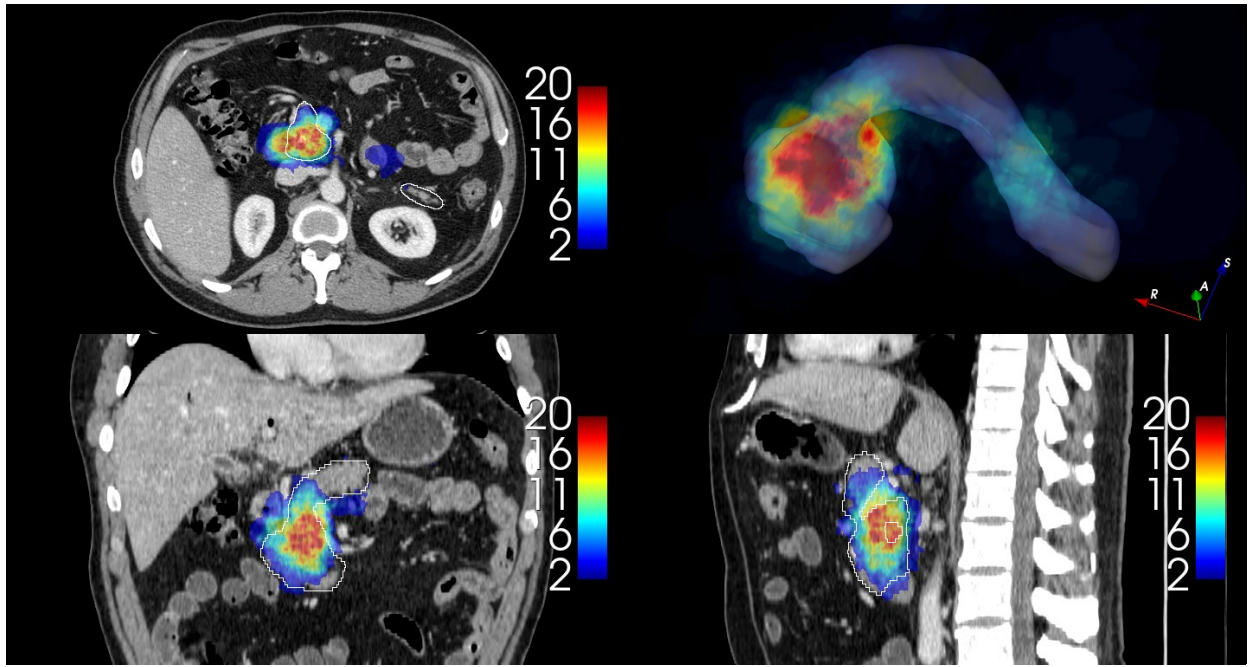

Figure 12: Spatial distribution and frequency of pancreas tumors within the MSD dataset with 281 cases and case #29 as a reference in the image (Simpson et al. [2019]): most frequent pancreases in the dataset in red, least frequent ones in blue. Boundary of case #29 in white.

### 5.1 Variability of tumors size and location

Following a similar approach described for parenchyma segmentation (cfr. Section 4), a registration was performed on 281 CTs of the MSD dataset, with case #29 as reference. The result is depicted in Fig. 12, showing the broad spatial distribution and frequency of pancreas tumors within the MSD dataset.

### 5.2 Tumors

The pancreatic ductal adenocarcinoma is a common malignant tumor of the gastrointestinal tract and generally arises in the head of the pancreas [Du et al., 2023]. The accurate segmentation of pancreas tumors is essential for the clinical integration with quantitative imaging biomarkers which have shown promising results of early detection of pancreas tumors, and for precise 3D modeling for surgical and radiotherapy planning [Mukherjee et al., 2023]. The reviewed studies on pancreas tumor segmentation are reported in Table 7.

Table 7: Studies on tumors and cysts.

| Author | Application | Dataset Size | Model Architecture | Learning Strategy | Loss | Results | Main Contributions |
|---|---|---|---|---|---|---|---|
| Cao and Li [2024] | Parenchyma and tumors | 82 (NIH) 281 (MSD) | UNet with: High resolution spatial information recovery + Multi-scale high resolution pre-segmented feature fusion + Pyramid multi-scale feature perception and fusion | Supervised | Difficulty-guided adaptive boundary-aware loss | Parenchyma (NIH): 88.96% (DSC) 89.27% (Precision) 89.98% (Recall) Parenchyma (MSD): 89.52% (DSC) 93.19% (Precision) 88.71% (Recall) Tumors (MSD): 54.38% (DSC) 69.58% (Precision) 53.17% (Recall) | High-resolution spatial information recovery module: encoder and decoder features of the same layer are sent to high resolution spatial information filtering module to extract high-resolution pre-segmented images, which are then fused. Multi-scale high-resolution pre-segmented feature fusion module: features of the encoder and decoder finely processed into a high-resolution pre-segmented feature map. Pyramid multi-scale feature perception and fusion module uses the extracted pre-segmented images to guide the network to focus on the dimensional changes of the segmented targets. Design of Difficulty-guided adaptive boundary-aware loss function to address the class imbalance and improve segmentation of uncertain boundaries |
| Cao et al. [2023b] | Parenchyma and tumors | 82 (NIH) 420 (MSD) | UNet with three attention mechanisms on skip connections: Spatial + Channel + Multi-dimensional features | Supervised | Weighted cross entropy loss | Parenchyma (NIH): 83.04% (DSC) 81.71% (Precision) 84.42% (Recall) Parenchyma (MSD): 83.39% (DSC) 85.51% (Precision) 81.37% (Recall) Tumors (MSD): 40.15% (DSC) 52.32% (Precision) 35.29% (Recall) | Design of a loss function to capture edge details of pancreas and tumors. Multi-dimensional attention gate integrated into skip connections for small target feature localization in multiple dimensions of space and channels, and for filtering redundant information in shallow feature maps, thus enhancing the feature representation of the pancreas and pancreatic tumor |
| Du et al. [2023] | Pancreas ductal adenocarcinoma | 55 (Qingdao University Hospital) 281 (MSD) | UNet with multi-scale channel attention | Supervised | Binary cross entropy | Qingdao: 68.03% (DSC) 59.31% (Jaccard) 12.04 mm (HD) MSD: 80.12% (DSC) 74.17 (Jaccard) 2.26 mm (HD) | Integration of multi-scale convolutions and channel attention into each encoder and decoder block |
| He and Xu [2023] | Parenchyma and tumors | 420 (MSD) + dataset of other organs | Hybrid CNN-Transformer Encoder: (3D Swin-Transformer + boundary extracting module) + Boundary preserving module + Decoder: CNN | Supervised | Dice loss Cross entropy loss | Parenchyma: 81.47% (DSC) 1.77 mm (ASSD) Tumor: 51.83% (DSC) 17.13 mm (ASSD) | Application of boundary awareness into 3D CNN and transformers. Swin-transformer as encoder and auxiliary boundary extracting module to obtain rich and discriminative feature representations. Boundary preserving module to fuse boundary map and features from the encoder |
| Ju et al. [2023] | Parenchyma and tumors | 82 (NIH) 281 (MSD) | UNet: Spatial visual cue fusion + Active localization offset (Localization) UNet (Segmentation) | Supervised | Dice loss Binary cross entropy loss | Parenchyma (NIH): 85.15% (DSC) Tumor (MSD): 63.36% (DSC) | Spatial visual cue fusion, based on conditional random field, learns global spatial context. It combines the correlations between all pixels in the image to optimize the rough and uncertain pixel prediction during the coarse stage. Active localization offset adjusts dynamically the localization results during the coarse stage. Code available at https://github.com/ PinkGhost0812/SANet |

| Author | Application | Dataset Size | Model Architecture | Learning Strategy | Loss | Results | Main Contributions |
|---|---|---|---|---|---|---|---|
| Li et al. [2023e] | Parenchyma and tumors | 281 (MSD) | nnUNet with attention + Balance temperature loss + Rigid temperature optimizer + Soft temperature indicator | Supervised | Balance temperature loss | Parenchyma: 85.06% (DSC) Tumors: 59.16% (DSC) | Segmentation of both pancreas and tumors. Balance temperature loss to dynamically adjust weights between tumors and the pancreas. Rigid temperature optimizer to avoid local optima. Soft temperature indicator to optimize the learning rate |
| Mukherjee et al. [2023] | Pancreas ductal adenocarcinoma | 1,151: Mayo Clinic + 152 from MSD and 41 from TCIA | 3D nnUNet | Supervised | Dice loss Cross entropy loss | Overall: 84.00% (DSC) 4.6 mm (HD) Generalization on MSD: 82.00% (DSC) 2.6 mm (HD) Generalization on TCIA: 84.00% (DSC) 4.30 mm (HD) | Bounding boxes by cropping the CT images to a 3D bounding box centered around the tumor mask. nnUNet applied to bounding boxes |
| Ni et al. [2023] | Recurrence of pancreas ductal adenocarcinoma after surgery | 205 (Internal) 64 (For recurrence prediction with radiomics) | AX-UNet with Atrous spatial pyramid pooling | Supervised | – | 85.90% (DSC) 74.20% (Jaccard) 89.70% (Precision) 87.60% (Recall) | AX-UNet combining UNet and atrous spatial pyramid pooling. Code available at: github.com/ zhangyuhong02/AX-Unet |
| Qu et al. [2023] | Parenchyma and masses (tumors, cysts) | 313 (Peking Union Medical College Hospital) 53 (Guandong General Hospital) (generalization) 50 (Jingling Hospital) (generalization) MSD (420) (generalization) | Swin Transformer and 3D CNN (Based on M3NET) Feature alignment: Transformer guided fusion + Cross-network attention (Decoder) | Supervised | Weighted cross entropy loss | Pancreas: Peking: 92.51% (DSC) Guangdong: 89.56% (DSC) Jingling: 88.07% (DSC) MSD: 85.71% (DSC) Masses: Peking: 80.51% (DSC) Guangdong: 67.17% (DSC) Jingling: 69.25% (DSC) MSD: 43.86% (DSC) | CNN and transformer branches perform separate feature extraction in the encoder. Progressive fusion between CNN and transformer in the decoder. Transformer guidance flow to address the inconsistency of the feature resolution and channel numbers between the CNN and transformer branches. Cross network attention into CNN decoder to enhance fusion capability with the transformer |
| Wang et al. [2023] | Tumors | 93 (Shanghai Changhai Hospital) | 3D UNet-like: Encoder: Multi-modal fusion downsampling block Decoder: Multi-modal mutual calibration block using attention | Supervised | Dice loss | 76.20% (DSC) 63.08% (Jaccard) 6.84 mm (HD) 75.96% (Precision) 84.26% (Recall) | Multi-modal fusion downsampling block to fuse semantic information from PET and CT, and to preserve unique features of different modal images. Multi-modal mutual calibration block to calibrate different scale semantics of one modal images guided by attention maps from the other modal images |
| Zhou et al. [2023] | Tumor | 116 abnormal pancreas 42 normal pancreas (internal) | Dual branch encoder-decoder (Pancreas segmentation) Encoder-decoder: contrast enhancement block + reverse attention block (Tumor segmentation) | Supervised | Dice loss | Abnormal: 78.72% (Jaccard) 89.07% (Precision) 87.42% (Recall) Normal: 87.74% (Jaccard) 91.47% (Precision) 95.50% (Recall) | Dual branch encoder-decoder combining semantic information extraction and detailed information extraction. Aggregation of feature maps of the two branches. Decoder to segment pancreas. Enhancement encoder-decoder network to improve segmentation accuracy of pancreatic tumors. Contrast enhancement block after each encoding step to extract the edge detail information. Reverse attention block inverting the decoder feature to guide the extraction of effective information in the encoder to generate an accurate prediction map |

### 5.2.1 Multi-stage methods

Si et al. [2021] used ResNet18 network for localization and UNet32 for segmentation. The model was trained on a dataset with different types of tumors from 319 patients from the Second Affiliated Hospital in Shanghai. It was tested on an independent dataset of 347 patients from the First and Second Affiliated Hospital in Shanghai. Mahmoudi et al. [2022] proposed a three-stage method for the segmentation of pancreatic ductal adenocarcinoma and vessels. They designed a texture attention UNet (TA-UNet) with texture attention block with scale-invariant feature transform or local binary pattern block, and attention gate block inserted into skip connections. The texture attention block provided a

| Author | Application | Dataset Size | Model Architecture | Learning Strategy | Loss | Results | Main Contributions |
|---|---|---|---|---|---|---|---|
| Zou et al. [2023] | Dilated pancreatic duct | 150 (Internal) Nanjing Drum Tower Hospital 40: Jiangsu Province Hospital of Chinese Medicine (Generalization) | 3D nnUNet for: (Localization) Terminal anatomy attention module (Segmentation) Terminal distraction attention module (Refine stage) | Supervised | Terminal Dice loss | Internal: 84.17% (DSC) 11.11 mm (HD) Generalization: 82.58% (DSC) | First work on errors on terminal regions of the dilated pancreatic duct. Terminal anatomy attention module to learn the local intensity from the terminal CT images, feature cues from the coarse predictions, and global anatomy information. Terminal distraction attention module to reduce false positive and false negative cases. Design of terminal Dice loss for segmentation of tubular structures |
| Li et al. [2023g] | Pancreatic cysts | 107 (internal) | UNet with: Atrous pyramid attention module + Spatial pyramid pooling module | Supervised | Dice loss Binary cross entropy loss | 84.53% (DSC) 75.81% (Jaccard) | Atrous pyramid attention module and spatial pyramid pooling module inserted into bottleneck layer to extract features at different scales, and contextual spatial information, respectively |
| Duh et al. [2023] | Pancreatic cysts | 335 (Internal) Spain | UNet with Attention gate in skip connections | Supervised | Dice loss | 93.10% (Recall) | Attention gate integrated into skip connections for segmentation of pancreatic cysts |
| Li et al. [2022b] | Tumors | 163 (Shanghai Jiao Tong University) 468 MRI (for style transfer) 281 (MSD) (generalization) | CycleGAN-like for: Synthetic data from MRI (Style transfer) ResNet: Extraction of knowledge from MRI (Meta-learning I) + Integration with salient knowledge from CT (Meta-learning II) | Supervised | Adversarial loss Cycle consistency loss Dice loss | Shanghai Jiao Tong University: 64.12% (DSC) MSD: 57.62% (DSC) | First study on meta-learning from one to a different modality. Random style transfer for: generation of synthetic images with continuously intermediate styles between MRI and CT to simulate domain shift. First meta-learning: the model learns the common knowledge of synthetic data, and provides pancreatic cancer-related prior knowledge for the target segmentation task. Second meta-learning: the model learns the salient knowledge of the CT data to enhance segmentation |
| Mahmoudi et al. [2022] | Tumors and surrounding vessels | 138 (MSD) | 3D local binary pattern (Localization) Ensemble of: Attention gate + Texture Attention block (Scale invariant feature transform and local binary pattern) (Segmentation) | Supervised | Generalized Dice loss Weighted Pixel-wise Cross entropy loss Boundary loss | Tumor: 60.60% (DSC) 3.73 mm (HD95) 57.80% (Precision) 78.00% (Recall) Superior mesenteric artery: 81.0% (DSC) 2.89 mm (HD95) 76.00% (Precision) 87.00% (Recall) Superior mesenteric vein: 73.00% (DSC) 3.45 mm (HD95) 68.00% (Precision) 81.00% (Recall) | Design of texture attention block with scale invariant feature transform or local binary pattern to provide a comprehensive representation of pathological tissue. Integration of attention gate and texture attention gate into skip connections of texture attention UNet. Use of a 3D CNN as an ensemble of attention UNet and texture attention UNet. Design of Generalized Dice loss, Weighted Pixel-wise Cross entropy loss, and Boundary loss to address unbalanced data, and boundary between pancreas and tumors |
| Shen et al. [2022] | Dilated pancreatic duct | 82 (NIH) for localization 30 (internal) for segmentation | 3D UNet (Localization) 3D UNet + Squeeze and excitation (Segmentation) | Supervised | Dice loss Focal loss | NIH: 75.9% (DSC) 72.4% (Recall) Internal: 49.90% (DSC) 51.90% (Recall) | First study on automated 3D segmentation of dilated pancreatic duct. Generation of an annotated dataset on dilated pancreatic duct. Attention block with squeeze and excitation inserted into the bottleneck of a 3D UNet |
| Chaitanya et al. [2021] | Tumors | 282 (MSD) | GAN + UNet | Semi-supervised | Adversarial loss | 52.90% (DSC) | Semi-supervised learning for data augmentation. Adversarial term to help two generators synthesize diverse set of shape and intensity variations present in the population, even in scenarios where the number of labeled examples are extremely low. Code available at: https://github.com/krishnabits001/task_driven_data_augmentation |

comprehensive representation of the pathological tissue. Firstly, 3D local binary patterns were employed to localize the pancreas. Secondly, attention UNet and TA-UNet were used for segmentation. Finally, a CNN aggregated attention UNet and TA-UNet [Mahmoudi et al., 2022]. Shen et al. [2022] reported the first work on segmentation of dilated

| Author | Application | Dataset Size | Model Architecture | Learning Strategy | Loss | Results | Main Contributions |
|---|---|---|---|---|---|---|---|
| Huang et al. [2021b] | Pancreatic neuroendocrine neoplasms | 98 (First Affiliated Hospital of Sun Yat-Sen University and Cancer Center of Sun Yat-Sen University) 72 (from both above centers) | UNet | Supervised | Cross entropy loss | First dataset: 81.80% (DSC) 83.60% (Precision) 81.40% (Recall) Second dataset: 74.80% (DSC) 87.20% (Precision) 68.60% (Recall) | A radiologists identified tumors by drawing bounding boxes to delineate region of interest sent as input to UNet. Radiomic analysis to predict pathohistologic grading |
| Si et al. [2021] | Pancreatic ductal adenocarcinoma and other types of tumors | 319 for training (Second Affiliated Hospital Shanghai) 347 for testing (First and Second Affiliated Hospital Shanghai) | ResNet18 (Localization) UNet32 (Segmentation) | Supervised | Cross entropy loss | 83.70% (DSC) | Three different networks used for pancreas location, segmentation, and diagnosis (presence of tumors) |
| Wang et al. [2021c] | Pancreatic ductal adenocarcinoma | 800 (John Hopkins) 281 (MSD) (generalization) | UNet with Inductive attention guidance | Semi-supervised | Cross entropy loss | John Hopkins: 60.28% (DSC) 99.75% (Recall) MSD: 32.49% (DSC) | Attention guided framework for classification and segmentation with partially labeled data (few annotated images for segmentation). Training using multiple instance learning with cancer and background regions as bags instead of per-voxel pseudo labels as in typical semi-supervised learning |
| Turečková et al. [2020] | Parenchyma and tumors | 420 (MSD) + datasets of other organs | UNet and VNet with Attention gate in skip connections | Supervised | Dice loss Cross entropy loss | Parenchyma (UNet): 81.81% (DSC) 81.21% (Precision) 84.51% (Recall) Tumors (UNet): 52.68% (DSC) 62.98% (Precision) 55.84% (Recall) Parenchyma (VNet): 81.22% (DSC) 80.61% (Precision) 84.10% (Recall) Tumors (VNet): 52.99% (DSC) 64.62% (Precision) 54.39% (Recall) | Attention gate integrated into skip connections for segmentation of pancreatic tumors |
| Xie et al. [2020] | Parenchyma and pancreatic cysts | 82 (NIH) 200 (John Hopkins: renal donors) 131 (John Hopkins: pancreatic cysts) | VGGNet with Hierarchical recurrent saliency transformation network between Localization and Segmentation | Supervised | Dice loss | NIH: 84.53% (DSC) Renal donors: 87.74% (DSC) Pancreatic cysts: 83.31% (DSC) | Saliency transformation module between first and second stage to transforms the segmentation probability map as spatial weights, iteratively, from the previous to the current iteration. Hierarchical version to segment first the pancreas and then the internal cysts. Code available at: https://github.com/198808xc/OrganSegRSTN |

pancreatic duct. A 3D UNet was first trained on the NIH dataset to localize the pancreas. A second 3D UNet with a SE attention block, inserted into the bottleneck layer, was trained on an internal dataset of 30 CTs for segmentation [Shen et al., 2022]. Ju et al. [2023] proposed a two-stage method for the segmentation of parenchyma and tumors on NIH and MSD datasets, respectively. In the localization stage, spatial-visual cue fusion and active localization offset modules were added to UNet. A spatial visual cue fusion module, based on conditional random field, learned global spatial context. It combined the correlations between all pixels in the image to optimize the rough and uncertain pixel prediction. The active localization offset module adjusted dynamically the localization results during the coarse stage [Ju et al., 2023]. Zou et al. [2023] reported the first work on errors in segmenting dilated pancreatic duct consisting of three stages. nnUNet was first used for localization. Then, for the segmentation stage, they designed a terminal anatomy attention module to learn the local intensity from the terminal part of the dilated pancreatic duct in CT images, feature cues from the coarse predictions, and global anatomy information. Finally, for the refine stage a terminal distraction attention block was developed to reduce the false positive and false negative cases. This method was initially trained and evaluated on a dataset of 150 CTs from Nanjing Drum Tower with different types of cancers, with PDAC as the most frequent. It was then assessed for generalization on an external dataset of 40 CTs of pancreas ductal adenocarcinoma from Jiangsu Province Hospital of Chinese Medicine. [Zou et al., 2023].

### 5.2.2 Other methods

Turečková et al. [2020] integrated attention gate into skip connections of UNet and VNet networks evaluated on MSD dataset for pancreas tumors in addition to segmentation of liver and spleen. Huang et al. [2021b] proposed a semi-automatic approach of segmentation of pancreas neuroendocrine tumors where a radiologist localized the tumors by drawing bounding boxes to delineate regions of interest, which were sent as input to UNet. The results of segmentation were used for radiomics analysis. Two internal datasets of 98 and 72 CTs, respectively, were used for prediction of the grading of tumors and prediction of recurrence [Huang et al., 2021b]. Yang et al. [2022b] designed AX-UNet integrating atrous spatial pyramid pooling into UNet, and a loss function to address the blurry boundary issue. NIH and MSD datasets were used for the segmentation of parenchyma and tumors, respectively [Yang et al., 2022b]. A similar approach was adopted by Ni et al. [2023] on an internal dataset of 205 CTs to predict the recurrence of pancreas ductal adenocarcinoma after surgery through radiomics on a distinct dataset of 64 patients. Li et al. [2022b] conducted the first study on meta-learning from MRI to CT domain. A CycleGAN was adopted for style transfer. It generated synthetic images with continuously intermediate styles between MRI and CT to simulate domain shift. Using a first meta-learning, the model learned the common knowledge of synthetic data and provided knowledge for the cancer segmentation task. With a second meta-learning, the model learned the salient knowledge of the CT data to enhance segmentation. This method was trained and validated on 468 MRI and 163 CT from Jiangsu Province Hospital and Ruijin Hospital, respectively. Then it was tested on 281 CTs of the MSD dataset. [Li et al., 2022b]. Zhou et al. [2023] proposed a dual branch encoder-decoder model (DB-Net) to first segment the abnormal pancreas by combining semantic information and detailed information extraction branches. The feature maps of the two branches were then aggregated. A fine-grained enhancement encoder-decoder network (FE-Net) was added to improve the segmentation accuracy of tumors. It consisted of a contrast enhancement block after each encoding step to extract the edge detail information, and a reverse attention block inverting the decoder feature to guide the extraction of effective information in the decoder to generate an accurate segmentation. An internal dataset of 116 abnormal and 42 normal pancreases was used [Zhou et al., 2023]. Wang et al. [2023] designed a 3D UNet-like model with an encoder with a multi-modal fusion downsampling block to fuse semantic information from PET to CT, and to preserve unique features of different modal images. In the decoder, a multi-modal mutual calibration block was designed to calibrate different scale semantics of one modal image guided by attention maps from the other modal images. This method was trained and tested on an internal dataset of 93 pancreatic cancer patients [Wang et al., 2023]. Qu et al. [2023] proposed transformer-guided progressive fusion network (TGPFN), an encoder-decoder architecture with transformers to complement the representation of CNN at multiple resolutions with global representation. The encoder consisted of a Swin Transformer and 3D CNN branches performing separate feature extraction. A progressive fusion between CNN and transformer was performed in the decoder. A transformer guidance flow module was designed to address the inconsistency of the feature resolution and channel numbers between the CNN and transformer branches. A cross-attention block was integrated into the CNN decoder to enhance the fusion capability with the transformer. TGPFN was evaluated on three internal datasets and MSD for the segmentation of parenchyma, tumors, and cysts. This model was trained on an internal dataset of 313 patients from Peking Union Medical College Hospital. It was then tested for generalization on 53 cases from Guangdong General Hospital, and 50 from Jingling Hospital. All these three datasets included several types of tumors. The model was also trained and evaluated on 420 cases of MSD dataset [Qu et al., 2023]. Li et al. [2023e] designed a nnUNet with attention and three modules based on temperature, i.e. balance temperature loss to dynamically adjust weights between tumors and the pancreas, a rigid temperature optimizer to avoid local optima, and a soft temperature indicator to optimize the learning rate. He and Xu [2023] designed an encoder-decoder Hybrid Transformer-CNN with Boundary-awareness (HTCB-Net) network for the segmentation of parenchyma and tumors. The encoder consisted of a Swin Transformer and an auxiliary boundary-extracting module to obtain rich and discriminative feature representations. A boundary-preserving module was inserted between the encoder and decoder to fuse boundary maps and features from the encoder. For the decoder, a CNN was used [He and Xu, 2023]. Fang et al. [2023] proposed a UMRFormer, a UNet variant with two transformer blocks embedded into the two lowest encoder-decoder steps. The transformers coupled MSA with residual convolutional block to capture both local and global features. Cao et al. [2023b] designed a multi-dimensional attention gate combining three types of attention (spatial, channel, and multi-dimensional features). The multi-dimensional attention gate was integrated into skip connections for small target feature localization in multiple dimensions of space and channels, and for filtering redundant information in shallow feature maps, thus enhancing the feature representation of the pancreas and pancreas tumor. A loss function was designed to capture edge details of the pancreas and tumors Cao et al. [2023b]. Cao and Li [2024] proposed Strongly Representative Semantic-guided Segmentation Network (SRSNet), a UNet-like network with three modules, namely the high-resolution spatial information recovery, the multi-scale high-resolution pre-segmented feature fusion, and the pyramid multi-scale feature perception and fusion. In the first module, the encoder and decoder features of the same layer were sent to a high-resolution spatial information filtering block to extract high-resolution pre-segmented images, which were then fused. In the second module, the features of the encoder and decoder were finely processed into a high-resolution pre-segmented feature map. This enabled the network not to rely on the feature information in the last layer of the decoder. The third module used the extracted pre-segmented images to guide the network to focus on the dimensional changes of the segmented targets. The pre-segmented images improved the network's ability to segment lesion areas of different sizes while reducing channel redundancy [Cao and Li, 2024]. The methods proposed by Cao et al. [2023b] and Cao and Li [2024] were trained on MSD for segmentation of parenchyma and pancreas tumors and assessed for generalization only on parenchyma using the NIH dataset.

### 5.3 Cysts

Pancreas cancer can originate from cystic lesions, which are fluid-filled sacs and are increasingly common incidental findings on abdominal imaging tests [Duh et al., 2023]. Pancreatic cysts can be nonneoplastic and neoplastic. The latter include benign lesions, such as serous cystadenomas, mucinous cystic neoplasms (MCN), and intraductal papillary mucinous neoplasm, which may degenerate into pancreas cancer [Duh et al., 2023]. The reviewed studies on cyst segmentation are listed in Table 7. Xie et al. [2020] proposed a two-stage method for the segmentation of pancreatic cysts. A saliency transformation module was inserted between the first and second stage to transform the segmentation probability map as spatial weights, iteratively, from the previous to the current iteration. A hierarchical version was designed to segment first the pancreas and then the internal cysts on NIH, and two internal datasets of 200 CTs (11 abdominal organs and five blood vessels) and 131 biopsy-proven cases with pancreatic cysts [Xie et al., 2020]. Li et al. [2023g] designed UNet with a pyramid atrous attention module and spatial pyramid pooling module inserted into the bottleneck layer to extract features at different scales, and contextual spatial information, respectively. This model was evaluated on an internal dataset of 107 CTs with pancreatic cysts. Duh et al. [2023] proposed UNet with an attention gate to segment different types of pancreatic cysts (serous cystadenomas, mucinous cystic neoplasms, and intraductal papillary mucinous neoplasm) on an internal dataset of 335 CTs.

### 5.4 Inflammations

Acute pancreatitis, an inflammation of the pancreas, is the leading cause of hospital admission for gastrointestinal disorders in the United States and several other countries [Deng et al., 2023]. The segmentation of an inflamed pancreas is more challenging than the normal pancreas since it invades the surrounding organs causing blurry boundaries, and it has higher shape, size, and location variability than the normal pancreas [Deng et al., 2023]. The reviewed studies on segmentation of pancreas inflammation are listed in Table 8. Guo et al. [2022b] adopted UNet++ to segment chronic inflammation of the common bile duct in pediatric patients. A ResUNet network was then used to classify the degree of severity of inflammation. Deng et al. [2023] performed the first study on the segmentation of acute pancreatitis on an internal dataset of 89 CTs. An FCN with a region proposal was used for the detection of pancreatitis region. The detected region was cropped and sent to the 2D U-Net for segmentation [Deng et al., 2023].

Table 8: Studies on pancreatitis

| Author | Application | Dataset Size | Model Architecture | Learning Strategy | Loss | Results | Main contributions |
|---|---|---|---|---|---|---|---|
| Deng et al. [2023] | Acute pancreatitis | 89 (Internal) | FCN + Region proposal network (Detection) UNet (Segmentation) | Supervised | Focal loss Cross entropy loss L1 regression loss | 66.82% (DSC) | FCN for detection of pancreatitis region. The detected region was cropped and sent to the 2D U-Net for segmentation. First study on segmentation on acute pancreatitis |
| Guo et al. [2022b] | Chronic inflammation of choledoch | 76 (internal) | UNet++ | Supervised | Binary cross entropy loss | 83.90% (DSC) | UNet++ to segment chronic inflammation of choledoch in pediatric patients. Then ResUNet is used to classify the degree of severity of inflammation |

### 5.5 Semi-supervised learning

Wang et al. [2021c] designed an Inductive Attention Guidance Network (IAG-Net) for classification and segmentation tasks, based on multiple-instance learning. According to multiple-instance learning, a training set consists of a group of bags, each containing several instances that are not labeled individually. In contrast, the whole bag is assigned a label [Wang et al., 2021c]. In the approach proposed by Wang et al. [2021c], pseudo labels of pancreas ductal adenocarcinoma and background regions were treated as bags instead of per-voxel pseudo labels as in conventional semi-supervised learning, addressing the problem of noise in per-voxel pseudo label. Each labeled image was sent to a UNet. The resulting feature maps were used to train an attention guidance module to learn the pancreas's location. The bag level pseudo labels were obtained by separating the pancreas location into two regions (i.e. pancreas and background) based on a threshold level of attention values. This model was trained on 800 CTs (400 with PDAC) from the John Hopkins dataset and tested for generalization on MSD [Wang et al., 2021c]. Chaitanya et al. [2021] proposed a semi-supervised method for data augmentation for tumors using two conditional GANs to model intensity and shape variations present in populations and among CT scanners from different centers. A generator of a conditional GAN received a labeled image and random vector from a Gaussian distribution as input to generate a deformation field which was applied to the original image and label. A second generator from another conditional GAN with a labeled image and random vector from a Gaussian distribution as input generated an additive intensity mask which was added to the original input image. This approach was evaluated on the MSD dataset using UNet for segmentation [Chaitanya et al., 2021].

## 5.6   Generalization to other datasets

Some studies assessed the generalization of the implemented models to external datasets. These works were reported by Si et al. [2021], Wang et al. [2021c], Li et al. [2022b], Cao et al. [2023b], Mukherjee et al. [2023], Qu et al. [2023], Zou et al. [2023], Cao and Li [2024]. The network architectures of these studies were described in Sections 5.2, 5.3, and 5.4. Mukherjee et al. [2023] initially trained and evaluated nnUNet on a dataset of 1,151 CTs of PDAC from Mayo Clinic and MD Andersen Cancer Center. It was then evaluated for generalization on 152 cases of MSD and 41 TCIA datasets. The models proposed by [Cao et al., 2023b, Cao and Li, 2024] were trained on MSD for segmentation of parenchyma and pancreas tumors and assessed for generalization only on parenchyma using the NIH dataset.

## 5.7   Design of loss functions

Mahmoudi et al. [2022] proposed a combination of cross-entropy, Dice loss, and boundary loss to address the low contrast between parenchyma and tumors. The boundary loss was a differentiable surrogate of a metric ($BF_1$) more sensitive to misalignments of boundaries than cross-entropy, Dice loss, and IoU [Bokhovkin and Burnaev, 2019]. Due to the difference of volumes of tumor and parenchyma, soft Dice loss may lead to insufficient feature extraction of tumors [Li et al., 2023e]. Li et al. [2023e] introduced balance temperature loss to dynamically adjust weights between tumors and parenchyma during training to avoid ignoring feature extraction of the tumors. This was achieved by inserting a parameter called Temperature which gradually decreased to shift the network focus from the parenchyma to tumors. Temperature started from a maximum value and was limited by a minimum value [Li et al., 2023e]. Cao and Li [2024] designed Difficulty-guided adaptive boundary-aware loss to address the class imbalance issue and increase the network sensitivity to boundary pixels. This loss used a category weight parameter to increase the misclassification penalty for pixels in small target regions, making the network focus on target regions. Adaptive boundary weights were added to improve the segmentation of uncertain boundaries [Cao and Li, 2024].

## 5.8   Comparison of performances

MSD was the most adopted dataset (13 out of 25 reviewed works or 52.0%) since it includes annotation of pancreas tumors (Section 3.7). Some studies employed only the annotated 281 CTs of the MSD dataset, while others the whole 420 scans of MSD, including 139 unlabeled CTs. For this reason, the performances of the reviewed studies on MSD varied greatly. The NIH dataset was employed in four studies, in three of which where both parenchyma and tumors were segmented, while in the fourth study, the NIH dataset was adopted by a two-stage model for the localization task (Table 7). In contrast with the reviewed studies on parenchyma, there was more use of private (internal and external) datasets as alternative sources of data for the segmentation of cancer, cysts, and inflammation (Table 7). As a consequence, a comparison among these studies was not possible. UNet-based models reported the highest DSC score on the portion of 281 annotated CTs of the MSD dataset. UNet coupled with channel attention and multi-scale convolutions achieved 80.12% of DSC. The multi-scale convolutions were embedded in each layer of the encoder to extract semantic information at different scales to localize small or very small tumors, and inserted also in the decoder layers [Du et al., 2023]. This configuration outperformed by a large margin a two-stage model, with UNet for both localization and segmentation, reaching a DSC of 63.36% [Ju et al., 2023]. This model integrated a spatial visual cue fusion module, based on the conditional random field to learn the global context, and an active localization offset module to adjust dynamically the localization results during the coarse stage [Ju et al., 2023]. When considering all 420 CTs of the MSD dataset, the highest DSC score (51.83%) was achieved by a hybrid transformer. This model consisted of a Swin-transformer as an encoder with two modules, the first as an auxiliary block for boundary extraction to obtain rich and discriminative feature representation, and the second to preserve the pancreas boundary. The decoder was a CNN [He and Xu, 2023]. When used for generalization on MSD, nnUNet reported a DSC of 82.00% on a portion of 152 cases of MSD after training on the dataset curated at Mayo Clinic of 921 CTs, described by Panda et al. [2021] [Mukherjee et al., 2023]. All the reviewed studies on tumors were based on supervised learning, except one that concerned semi-supervised learning for data augmentation using two generators of GANs. UNet achieved a DSC of 52.90% on 282 cases of MSD [Chaitanya et al., 2021]. Overall, the DSC score for tumor segmentation on the MSD dataset dropped if compared to the results for parenchyma on the same dataset, underlining the further complexity due to the particularly small size of pancreas tumors (cfr. Section 1, and Section 4.8). Segmentation of pancreatic cysts and inflammation were investigated only with internal datasets (Table 7 and Table 8). UNet with ASPP and spatial pyramid pooling (cfr. Section 5.3) reported a DSC of 84.53% on a dataset of 107 cases of cysts, while UNet++ a DSC of 83.90% on 76 CTs for pancreatitis [Li et al., 2023g].

# 6    Discussion

In this systematic review, we analyzed the published literature, consisting of 130 original studies, on DL for the segmentation of parenchyma, tumors, cysts, and inflammation of the pancreas. By looking at the geographical origin of the reviewed studies, China is leading the ranking with more than half of the published articles in peer-reviewed journals, ahead of the United States, UK, Japan, and Canada. Unexpectedly, there are countries with an established tradition in pancreatic surgery, like Italy, not present in this ranking, underlying a research gap from a technical point of view with the others [Abu Hilal et al., 2023].

## 6.1    Clinical need perspective

Pancreas diseases like tumors are aggressive and in most cases, if not promptly diagnosed, can become lethal. DL can streamline the 3D reconstruction of radiological datasets, thus helping clinicians at the diagnosis stage, provided that the quality of segmentation meets the clinical need. Unfortunately, the pancreas has been traditionally regarded as one of the toughest abdominal organs for the segmentation task due to its small volume compared with a full CT scan, blurred boundaries, and large variations among patients in terms of shape and position. DL methods are no exception, as highlighted by the present systematic review. To fill this gap many methods of DL segmentation have been proposed. The trend in the number of published studies is constantly growing, thus reflecting an increase in interest in the community. There are fewer studies on tumor segmentation as they more challenging to segment than parenchyma. DL segmentation on other tiny structures like the dilated pancreatic duct and surrounding vessels has been only recently proposed, with initial studies published in 2022 [Mahmoudi et al., 2022, Shen et al., 2022, Zou et al., 2023]. Future developments should consider the adaptability of the DL models to changes in the size and shapes of the lesions over time. Models combining different modalities, e.g., CT and MRI, should also be implemented to extract more information from the radiological data. We observe that few studies of out the reviewed ones were led by clinicians, with seven works published in journals in the medical field [Bagheri et al., 2020, Guo et al., 2022b, Li et al., 2023g, Mukherjee et al., 2023, Park et al., 2020, Si et al., 2021, Sundar et al., 2022]. In contrast, the rest of the studies were published in technical journals or cross-disciplinary ones at the boundary between medicine and computer science.

## 6.2    DL models perspective

The accuracy of pancreas segmentation by DL models improved over the years, as demonstrated by scores on DSC, and the Jaccard index on NIH dataset. However, the score on segmentation of smaller lesions, like pancreas ductal adenocarcinoma, on specific datasets for tumors like MSD is much lower. Overall, the segmentation of a small organ like the pancreas presents a class imbalance challenge with the background as the prominent class, followed by parenchyma of the pancreas, and tumor as the least present. This issue was mitigated by the adoption of the Dice loss function. As reported in Section 4.7 and Section 5.7, several studies proposed the design of new loss functions to improve metrics results. Almost all the studies used region-based metrics (e.g., DSC, and Jaccard index). Only 20 out of 105 on parenchyma, and 5 out of 26 on tumor and cysts segmentation used a boundary-based metric like HD. Two works used NSD as a region-based metric [Ma et al., 2022, Tong et al., 2023]. Our review highlights an enormous variety of DL architectures specifically designed for pancreas segmentation from standard UNet to transformers up to hybrid transformers. Likewise, many attention blocks have been designed from attention gate to SE up to reverse attention [Oktay et al., 2018, Zhou et al., 2023]. Although these models demonstrated improvements over the years on DSC, Jaccard, and HD metrics, the results may show a limited value that is not the marginal entity, as in most cases, of such improvement but the fact that the models were trained and tested on datasets of small size and mostly from a single center. By grouping the studies sharing the same application, DL approach (e.g., multi-stage), and dataset we could compare the performances in some cases. For instance, the present review highlighted that a UNet configuration with residual blocks in the encoder and a decoder with spatial and channel attention obtained the highest DSC score (91.37%) on the NIH dataset, outperforming transformers-based models and other novel architectures [Shan and Yan, 2021]. This evidence is in agreement with a recent study setting a comprehensive benchmark of current DL models for segmentation in medical imaging [Isensee et al., 2024]. However, since the data split for training, validation, and test sets were different on the same dataset we could not decree which models are the most suitable for pancreas parenchyma or tumor segmentation. Furthermore, the small size of the test set introduces result instability and questions the significance of the performance gains. Other confounding factors precluding an objective comparison are the different hardware capabilities [Isensee et al., 2024]. Additionally, the reproducibility of the results of most studies is challenging since the code is publicly available only for a few of them.

### 6.3 Datasets perspective

By looking at the tables summarizing the studies (Appendix A, and Appendix B) the first remark that stands out is that in the vast majority of studies, the DL models are trained and tested only on publicly available datasets, suggesting that there are difficulties in curating internal datasets. In contrast, some institutions were capable of collecting large datasets, e.g. with 1,917 CTs from Mayo Clinic and 1,150 CTs from John Hopkins Medical Institution for the segmentation of parenchyma and pancreas ductal adenocarcinoma, respectively [Panda et al., 2021, Wang et al., 2021c]. Other datasets, like AbdomenCT-1k, combined different datasets like NIH and MSD, and extended them by labeling other organs (liver, spleen, and kidney) in addition to the pancreas, in addition to combining data from different institutions, from multiple vendors and acquired with different stages (arterial and venous) [Ma et al., 2022]. Currently, there is a need for new datasets on pancreas tumors and vascular structures. One of the most frequent criticisms of AI models is the lack of generalization to data from different institutions or from different models of CT scanners by different vendors. Sections 4.6 and Section 5.6 reported the results of some works on private (internal or external) datasets for generalization. However, further work is required to prove the robustness of DL models to external institutions. In addition to being multicentric and multivendor, the new datasets for pancreas segmentation (including tumors and vascular structures) should also compensate for bias in the population selection, by including a more diverse range of ethnicities. Semi-supervised and unsupervised learning look promising to exploit datasets with few labeled data or with only unlabeled data, respectively. This review analyzed eight and four works on semi-supervised and unsupervised learning for parenchyma segmentation, respectively (Section 4.4, Section 4.5). For tumors only one study used semi-supervised learning (5.5). According to the literature, it seems that semi-supervised learning methods performed better than supervised ones [Chen et al., 2022b]. However, this is not the case for the pancreas segmentation. For this task, supervised learning still provided the higher scores, as documented by this systematic review.

### 6.4 Clinical translation perspective

What emerged from our analysis is a consistent push among research groups towards designing more sophisticated DL models setting new benchmarks on standard datasets. In contrast, there are also few studies using efficient and high-performing DL models like nnUNet on more variegated datasets (multi-center, multi-organ, and multi-vendor) as AbdomenCT-1k [Ma et al., 2022]. Overall, by considering the published literature, the following question arises: *"Should resources be directed towards refining existing models or harnessing established networks with extensive, meticulously curated datasets?"* The answer still remains complex. From a clinical standpoint, the tangible benefits for patients might be elusive if the DL models are not rigorously tested in real-world clinical settings. According to the most recent guidelines on minimally invasive pancreatic surgery, there exists no corpus of evidence on the impact of AI in laparoscopic or robotic pancreatic surgery. Most of the published studies assessed the technical feasibility of utilizing AI. However, there is no demonstration of clinical implementation and validation at multiple centers [Abu Hilal et al., 2023]. Moreover, transitioning from research to market poses formidable challenges, including model generalizability, explainable AI, data privacy safeguards, regulations, and certification. While federated learning seems a promising avenue for training DL models across diverse institutions without compromising data privacy, it is noteworthy that no published studies on this approach were available for pancreas segmentation at the time of our systematic review. Similarly, the application of explainable AI to pancreas segmentation remains largely unexplored.

## 7 Conclusions

This systematic review of DL applications for segmenting the pancreas and its lesions elucidates significant advancements and identifies important areas of improvement. The review highlights several critical challenges, including the limited availability of large datasets that represent the population well, and the integration of these technologies into clinical settings, requiring real-time segmentation capabilities that align with clinical workflows. This integration is further complicated by the need for standardized evaluation metrics and domain adaptation to ensure that models are generalizable across different clinical environments and imaging modalities. Efforts to address these challenges will improve the accuracy and efficiency of DL models and enhance their applicability in real-world clinical scenarios, thereby bridging the gap between technical capabilities and clinical needs.

# A    Full list of the reviewed studies on the segmentation of pancreas parenchyma

| Author | Application | Dataset Size | Model Architecture | Learning Strategy | Loss | Results | Main Contributions |
|---|---|---|---|---|---|---|---|
| Dai et al. [2023] | Segmentation of pancreas | 82 (NIH) 281 (MSD) | UNet (Localization) Deformable convolution Vision Transformer (Segmentation) | Supervised | Binary cross entropy Dice loss | NIH: 89.89% (DSC) 89.59% (Precision) 91.13% (Recall) MSD: 91.22% (DSC) 93.22% (Precision) 91.35% (Recall) | Skip connections integrating: vision transformer, deformable convolutions, and scale interactive fusion (combining global and local features, and merging feature maps of different scales). Two-dimensional wavelet decomposition to solve the issue of blurred boundaries |
| Fang et al. [2023] | Segmentation of pancreas | 281 (MSD) 91 (Clinical Proteomic Tumor Analysis Consortium Pancreatic Ductal Adenocarcinoma) | U-shaped Encoder-Decoder CNN + Transformer | Supervised | Dice loss Cross entropy loss | MSD: 77.36% (DSC) 8.34 (95HD) Tumor dataset: 85.54% (DSC) 4.05 mm (HD95) | A transformer module with multi-head self-attention and residual convolutional block was designed to capture both local and global features. Code available at: https://github.com/supersunshinefk/UMRFormer-Net |
| Francis et al. [2023] | Segmentation of liver, kidneys, spleen, and pancreas | 1,112 (AbdomenCT-1k) | Conditional GAN: dilated UNet + Attention gate (Generator) Fully Convolutional Network (Discriminator) | Supervised | Adversarial loss | 86.10% (DSC) 6.65 mm (HD95) 86.80% (Precision) 86.60% (Recall) | Residual dilated convolution block and spatial pyramid pooling replacing convolutions and max pooling in UNet. Attention gate inserted into skip connections |
| Ge et al. [2023] | Segmentation of pancreas | 45: (Nanjing Drum Tower Hospital) (Reconstruction) + 15 (Nanjing General PLA Hospital) for generalization + 90 (liver tumor) for generalization | Average Super Resolution GAN with: 3D CNN (Reconstruction) + 3D UNet for both Localization and Segmentation | Supervised | Mean squared error loss Dice loss Cross entropy loss | Generalization (pancreas): 84.20% (DSC) 0.54 mm (ASD) | GAN: Super resolution network to reduce anisotropy resolution. A generator reconstructs thin slices in z axis The discriminator optimizes the output of generator. The optimized generated images are sent to a dual-stage network for segmentation. Predictions on high-resolution are down-sampled to restore initial resolution |
| Huang et al. [2023] | Segmentation of aorta, gallbladder, spleen, kidneys, liver, pancreas, spleen, stomach, ventricles, myocardium, and retina | 30 (Synapse) 100 MRI (Automated cardiac diagnosis challenge) 40 (Digital Retinal Images for Vessel Extraction) | Encoder-decoder with: transformer blocks in all encoding and decoding steps + Transformer context bridge between encoder and decoder (fusion of multi-scale information) | Supervised | – | 65.67% (DSC) | Hierarchical encoder-decoder with ReMix-FFN module in each transformer block with a convolution and a skip connection between the two fully connected layers to capture local information in addition to global dependencies. Features of different scale as output of each encoder step are concatenated, and sent to ReMixed transformer context bridge with self-attention to capture global dependencies. The output features are split into different scale feature maps and sent to ReMix-FFN modules of the decoder to mix global dependencies with local context. Code available at: https://github.com/ZhifangDeng/MISSFormer |

| Author | Application | Dataset Size | Model Architecture | Learning Strategy | Loss | Results | Main Contributions |
|---|---|---|---|---|---|---|---|
| Irshad et al. [2023] | Segmentation of pancreas | 82 (NIH) 47 (BTCV) | UNet UNet++ Attention-UNet Two topologies: sharing encoder and decoder, but task-specific final layers; or sharing encoder with task-specific decoders | Supervised | Dice loss Binary cross entropy | NIH: 82.00% (DSC) 0.673 mm (HD) 85.90% (Precision) 82.20% (Recall) BTCV: 77.80% (DSC) 0.919 mm (HD) 81.00% (Precision) 77.00% (Recall) | Multi-task network for: region segmentation and edge prediction. Comparison of different network topologies. Code available at: https://github.com/samra-irshad/ 3d-boundary-constrained-networks |
| Jain et al. [2023] | Segmentation of pancreas | 82 (NIH) | K-mean and Gaussian mixture model (Unsupervised) (Localization) UNet, Holistically-Nested Edge Detection, and Dense-Res-InceptionNet (Segmentation) | Unsupervised + Supervised | Dice loss | 81.75% (DSC) 83.03% (Precision) 81.70% (Recall) | Unsupervised localization of pancreas after segmenting liver and spleen using K-means and Gaussian mixture models |
| Li et al. [2023c] | Segmentation of pancreas | 82 (NIH) 281 (MSD) 104 (Private) | UNet with: Meta-learning (Localization) Latent-space feature flow generation (Segmentation) | Supervised | Design of adaptive loss with: Recall loss, Cross entropy and Dice loss | NIH (trained on MSD and private): 80.24% (DSC) 1.92mm (ASD) MSD (trained on NIH and private): 81.09% (DSC) 1.99 mm (ASD) Private (trained on NIH and MSD): 84.77% (DSC) 1.28 mm (ASD) | First generalization model for pancreas segmentation. Model-agnostic meta-learning to improve generalization of the coarse stage. Appearance-style feature flow generation in the fine stage to generate a sequence of intermediate representations between different latent spaces for simulating large variations of appearance-style features |
| Li et al. [2023d] | Segmentation of pancreas | 82 (NIH) 281 (MSD) 104 (Renji Hospital Shanghai Private dataset) | UNet with meta-learning (Localization) 3D UNet: Global feature contrastive learning 3D UNet: Local image restoration (Segmentation) | Self-supervised | Binary cross entropy loss Dice loss Squared error loss Adversarial loss | Training on NIH Generalization on MSD: 66.73% (DSC) Generalization on Private: 73.85% (DSC) Training on MSD Generalization on NIH: 76.71% (DSC) Generalization on Private: 83.50% (DSC) Training on Private Generalization on NIH: 65.03% (DSC) Generalization on MSD: 70.08% (DSC) | Dual self-supervised generalization model to enhance characterization of high-uncertain regions. Global-feature self-supervised contrastive learning reducing the influence of extra-pancreatic tissues. Local image restoration self-supervised module to exploit anatomical context to enhance characterization of high-uncertain regions |
| Li et al. [2023f] | Segmentation of liver, kidney, spleen, and pancreas | 500 from AbdomenCT-1k + 240 from AMOS-CT | 3D UNet (Localization) UNet with: Multi-branches feature attention (Encoder) + Feature attention aggregation (Decoder) (Segmentation) | Supervised | Dice loss | First dataset: 86.20% (DSC) Second dataset: 78.40% (DSC) | Network with self-adjustable attention and receptive field size to segment liver, kidney, spleen and pancreas. Different kernel sizes to capture different scale features of different organs using: multi-branch feature attention with four branches, and feature attention aggregation with two branches |
| Liu and Zheng [2023] | Segmentation of liver, spleen, pancreas, and kidneys | 30 (BTCV) 281 (MSD) + partially labeled datasets of other abdominal organs | nnUNet | Semi-supervised | Cross entropy loss Dice loss (labeled data) Context-aware voxel-wise contrastive learning loss (unlabeled data) | 83.60% (DSC) 4.30 mm (HD95) | Exploiting unlabeled information in partially labeled datasets. Context-aware voxel-wise contrastive learning inserted into the bottleneck layer of a 3D nnUNet to increase context awareness in patch-based strategy |
| Paithane [2023] | Segmentation of pancreas | 80 (NIH) | ResNet18 VGG-16 VGG-19 | Supervised | Cross entropy Dice loss | 88.68% (DSC) 98.82% (Jaccard) 68.22% (Precision) 98.66% (Recall) | Data augmentation to reduce class imbalance |
| Qiu et al. [2023] | Segmentation of pancreas | 82 (NIH) | DeepUNet (Localization) Residual transformer UNet (Segmentation) | Supervised | Dice loss with Hausdorff distance term | 86.25% (DSC) | UNet like network with each convolutional block consisting of residuals blocks, residual transformers, and dual convolution down-sampling (for translational equivariance) |

| Author | Application | Dataset Size | Model Architecture | Learning Strategy | Loss | Results | Main Contributions |
|---|---|---|---|---|---|---|---|
| Shen et al. [2023] | Segmentation pancreas, duodenum, gallbladder, liver, and stomach | 42 (NIH) | UNet with: Spatial attention (location and size of organs) + Dilated convolution + Multi-scale attention | Supervised | Dice loss Cross entropy loss | 75.42% (DSC) 61.84% (Jaccard) 19.99 mm (HD) | Spatial attention to highlight location and sizes of target organs (pancreas, duodenum, gallbladder, liver, and stomach). Deformable convolutional blocks to deal with variations in shapes and sizes. Skip connections with multi-scale attention to eliminate interference of complex background |
| Tian et al. [2023] | Segmentation of pancreas | 200 (from AbdomenCT-1k) 82 (NIH) 281 (MSD) 50 (Jiangsu Province Hospital) (Generalization) | nnUNet (Localization) + Hybrid variational model to capture weak boundaries (Segmentation) | Supervised | Cross entropy Dice loss | AbdomenCT-1k: 89.61% (DSC) NIH: 87.67% (DSC) MSD: 87.13% (DSC) Generalization: 90.72% (DSC) | First stage: 3D CNN for coarse segmentation Second stage: a new hybrid variational model to capture the pancreas weak boundary |
| Tong et al. [2023] | Segmentation of liver, kidney, spleen, and pancreas | 511: 80 (NIH) 281 (MSD) + datasets of other organs (multi-center) | Encoder-Decoder (Localization) ResUNet and Multi-scale Attention (Segmentation) | Supervised | Dice loss (Localization) Dice loss Mean square error (Segmentation) | 59.10% (DSC) 42.20% (NSD) | Coarse stage for localization. Fine stage with multi-scale attention to segment pancreas, liver, spleen, and kidney. |
| Wu et al. [2023] | Segmentation of pancreas, left ventricle, myocardium, right ventricle, and colon | 80 (NIH) + datasets of other organs | V-Net + Attention | Semi-supervised | Cross entropy Dice loss | 74.03% (DSC) 59.70% (Jaccard) 2.12 voxel (ASD) 9.10 voxel (HD95) | Instead of using model predictions as pseudo labels, high-quality pseudo labels are generated by comparing multiple confidence maps produced by different networks to select the most confident one (a compete-to-win strategy). A boundary-aware enhancement module was integrated to enhance boundary discriminative features. Code available at: https://github.com/Huiimin5/comwin |
| Xia et al. [2023] | Segmentation of pancreas | 82 (NIH) | VNet + Multi-dimensional Feature attention | Semi-supervised | Cross entropy Dice loss Mean square error | 79.55% (DSC) 66.87% (Jaccard) 7.67 mm (HD95) 1.65 mm (MSD) | Multi-dimensional feature attention and improved cross pseudo supervision to effectively use unlabeled data reducing the need of labeled data |
| Yuan et al. [2023] | Segmentation of aorta, gallbladder, kidneys, liver, pancreas, spleen, and pancreas | 30 (Synapse) | UNet-like with: Gated recurrent units for skip connections + Gated-dual attention (Multi-scale weighted channel attention + Transformer self attention) | Supervised | – | 62.77% (DSC) | Gate recurrent units integrated into skip connections to reduce the semantic gap between low and high-level features. Gated-dual attention to capture information on small organs and global context. Code available at: https://github.com/DAgalaxy/MGB-Net |
| Zheng and Luo [2023] | Segmentation of pancreas | 80 (NIH) | Encoder-decoder for both Localization and Segmentation | Supervised | Weighted binary cross entropy loss | 85.58% (DSC) 74.99% (Jaccard) 86.59% (Precision) 85.11% (Recall) | Extension-contraction transformation network with a shared encoder for feature extraction and two decoders for the prediction of the segmentation masks and the inter-slice extension and contraction transformation masks |
| Zhu et al. [2023] | Segmentation of pancreas | 82 (NIH) 70 (Zheyi): Zhejiang University Hospital | Adversarial network + 3D ResUNet + Attention (Squeeze-Excitation) | Supervised and Unsupervised | Dice loss Cross entropy loss | NIH (supervised): 85.45% (DSC) Zheyi (unsupervised): 75.43% (DSC) | Training with 3D ResUNet and attention module using pairs of labeled images fro one center and unlabeled ones from a different center to generate multi-scale feature maps. Labeled and unlabeled data are then trained by a discriminator for domain identification |

| Author | Application | Dataset Size | Model Architecture | Learning Strategy | Loss | Results | Main Contributions |
|--------|-------------|--------------|--------------------|--------------------|------|---------|--------------------|
| Chen and Wan [2022] | Segmentation of pancreas | 82 (NIH) | UNet with Transformers in skip connections | Supervised | Dice loss Focal loss | 86.80% (DSC) 76.90% (Jaccard) 87.60% (Precision) 88.00% (Recall) | 3D channel transformer in the skip connections of UNet. Attention module in each encoder level with project and excite block to enhance extraction of context information. Cross attention between output of each transformer and decoder to eliminate semantic inconsistency |
| Chen et al. [2022d] | Segmentation of pancreas | 82 (NIH) 281 (MSD) | Encoder-Decoder Attention feature fusion (Localization) Encoder-Decoder Attention feature fusion Coordinate Multi-scale Attention (Segmentation) | Supervised | Dice loss Binary cross entropy | NIH: 85.41% (DSC) 74.80% (Jaccard) 85.60% (Precision) 85.90% (Recall) MSD: 70.00-80.00% (DSC) 60.00% (Jaccard) 80.00-90.00% (Precision) 60.00-70.00% (Recall) | Attention feature fusion on low and high level features to keep context. Multi-scale attention to aggregate long-range dependencies, positional information, and exploit multi-scale spatial information |
| Chen et al. [2022c] | Segmentation of pancreas | 82 (NIH) 281 (MSD) | VGG-16 with Attention gate (Localization) VGG-16 with Residual multi-scale dilated attention (Segmentation) | Supervised | Dice loss Binary cross entropy | NIH: 85.19% (DSC) 74.19% (Jaccard) 86.09% (Precision) 84.58% (Recall) MSD (generalization): 76.60% (DSC) 62.60% (Jaccard) 87.70% (Precision) 69.20% (Recall) | Attention gate used in the localization stage to suppress irrelevant background regions. Weight conversion module to transform segmentation map of the first stage into spatial weights to refine input of the second stage. Residual multi-scale dilated attention to exploit inter-channel relationships and extract multi-scale spatial information. Code available at: https://github.com/meiguiyulu/TVM |
| Chen et al. [2022b] | Segmentation of pancreas | 82 (NIH) 281 (MSD) | UNet (Localization) Unet with: Fuzzy skip connection + Target attention in the decoder (Segmentation) | Supervised | Dice loss | NIH: 87.91% (DSC) 78.52% (Jaccard) 90.43% (Precision) 85.77% (Recall) MSD: 84.40% (DSC) | Fuzzy skip connections to reduce the redundant information of non-target regions. Attention to make the decoder more sensitive to target features |
| Cui et al. [2022] | Segmentation of pancreas, and liver | 82 (NIH) + dataset of liver | UNet++ with: Channel attention (Squeeze and excitation) + Skip connection with sharpening filter | Supervised | Dice loss Cross entropy loss | 83.64% (DSC) 83.97% (Precision) 84.20% (Recall) | Laplacian sharpening filter integrated into skip connections to reduce bad artifacts. Squeeze and excitation to help the model to focus on the features beneficial for segmentation |
| Huang et al. [2022c] | Segmentation of pancreas | 82 (NIH) 47 (BTCV) | Improved Refinement Residual Block and Channel Attention Block (smooth network) + Shared bottleneck attention module + Improved Refinement Residual Block (Border network) | Supervised | Dice loss Focal loss | NIH: 82.82% (DSC) 71.13% (Jaccard) 1.69 mm (AUC) 83.16% (Precision) 83.54% (Recall) BTCV (generalization): 79.34% (DSC) 66.02% (Jaccard) 1.15 mm (ASD) | Discriminative feature attention module to address intra-class inconsistency and inter-class indistinction. Attention to avoid a node for localization. Improved refinement residual block to highlight spatial positions and aggregate contextual information |
| Huang et al. [2022b] | Segmentation of pancreas, liver, and brain | 82 (NIH) + datasets of other organs | ResNet-34 and 3D-ResUNet | Supervised | Binary cross entropy Auxiliary loss to upsample feature maps to the same spatial resolution as the inputs | 2D: 83.67% (DSC) 85.60% (Precision) 82.58% (Recall) 3D: 86.32% (DSC) 85.52% Precision) 84.51% (Recall) | Two sample balancing methods were proposed: positive-negative subset selection and hard-easy subset selection |
| Huang and Wu [2022] | Segmentation of pancreas | 82 (NIH) | MobileNet-UNet: UNet + MobileNet-V2 | Supervised | Weighted cross entropy | 82.87% (DSC) 70.97% (Jaccard) 90.54% (AUC) 89.29% (Precision) 77.37% (Recall) 99.95% (Specificity) | Implementation of MobileNet-UNet. Compared to original UNet, it adds zero padding, depthwise convolution, batch normalization, and ReLU |

| Author | Application | Dataset Size | Model Architecture | Learning Strategy | Loss | Results | Main Contributions |
|---|---|---|---|---|---|---|---|
| Javed et al. [2022] | Segmentation of pancreas | 82 (NIH) 15 (pre-cancer) 15 (cancer) | Bayes model + UNet | Supervised | Dice loss Focal loss | Pre-cancer: 95.60% (DSC) Cancer: 89.90% (DSC) | First study on segmentation of pancreas subregions (head, body, and tail). The probability map of UNet is updated with a probability map of a Bayes model indicating the three subregions |
| Khasawneh et al. [2022] | Segmentation of pancreas | 294 (from 1,917 of Mayo Clinic) | UNet-like for: Localization and Segmentation | Supervised | – | 88.00% (DSC) 79.00% (Jaccard) | Comparison of manual segmentation by experts using 3D Slicer and automatic segmentation by CNN |
| Li et al. [2022c] | Segmentation of spleen, kidney, gallbladder, esophagus, liver, stomach, pancreas, and duodenum | 90: 43 (TCIA) + 47 (BTCV) 511: 80 (NIH) 281 (MSD) + datasets of other organs (multi-center) | 3D Encoder-Decoder (Localization) 2.5D netowrk (Segmentation) | Supervised | Design of parameter loss to remove the false positive of dice loss | First dataset: 84.00% (DSC) 5.67 mm (ASD) Second dataset: 83.00% (DSC) | Circular inference (a sort of micro-attention mechanism) and parameter Dice loss in the first stag to reduce uncertain probabilities of blurred boundaries. |
| Li et al. [2022d] | Segmentation of pancreas | 82 (NIH) | Attention-guided Dual adversarial UNet (ADAU-Net) | Supervised | Basic loss of conventional segmentation Adversarial loss | 83.76% (DSC) 72.38% (Jaccard) 1.07mm (ASD) 2.17 (RMSE) | First dual adversarial network with an attention mechanism for pancreas segmentation |
| Lim et al. [2022] | Segmentation of pancreas | 1,006 (Gil Medical Center Gachon University College of Medicine) + 82 (NIH) (generalization) | UNet-like configurations | Supervised | Dice loss | Internal: 86.9% (Precision) 84.2% (Recall) 84.2% (DSC) Generalization: 77.90% (Precision) 74.90% (Recall) 73.50% (DSC) | Comparison of four 3D architectures based on UNet on a large dataset (1,006 CT) |
| Liu et al. [2022a] | Segmentation of pancreas | 82 (NIH) | Graph-enhanced nnUNet | Semi-supervised | Cross entropy Dice loss | 84.22% (DSC) 73.10% (Jaccard) 6.63 voxel(HD95) 1.86 voxel (ASD) | A graph CNN was added to nnUNet to distinguish the low contrast edges of a pancreas. Pseudo labels are refined using an uncertainty iterative strategy |
| Liu et al. [2022b] | Segmentation of pancreas | 82 (NIH) 72 (ISICDM) + dataset of other organ | ResNet18 + Atrous spatial pyramid pooling for multi-scale feature extraction for both Localization and Segmentation + Saliency module for fusion | Supervised | Dice loss (Region and boundary level) Binary cross entropy (Pixel level) | NIH: 88.01% (DSC) ISICDM: 87.63% (DSC) | Segmentation network with three branches to extract pixel, boundary, and region features, fused by a saliency module. Design of a loss function integrating information from pixel-level classification, edge-level localization, and region-level segmentation |
| Liu et al. [2022c] | Segmentation of pancreas | 82 (NIH) | Region of interest based on surrounding organs + VGG-UNet | Supervised | Dice loss | 85.40% (DSC) 73.20% (Jaccard) 18.26 mm % (HD) | Dynamic extraction of region of interest of the pancreas based on the central point of surrounding organs (liver, kidney, and spleen). Initially pre-trained on ImageNet |
| Ma et al. [2022] | Segmentation of liver, kidney, spleen, and pancreas | 1,112 (AbdomenCT-1k) | 3D nnUNet (Supervised and semi-supervised) 2D nnUNet + Conditional random fields (Weakly supervised) nnUNet (Continual) | Supervised Semi-supervised Weakly supervised Continual | Dice loss Cross entropy loss | Single organ (trained on MSD): 86.10% (DSC) 66.10% (NSD) Multi-organ (trained on MSD): 90.10% (DSC) 82.30% (NSD) Supervised (MSD) + liver (40) and kidney (40): 78.10% (DSC) 65.00% (NSD) Semi-supervised: 85.70% (DSC) 72.50% (NSD) Weakly supervised: 70.50% (DSC) 55.00% (NSD) Continual: 64.70% (DSC) 51.10% (NSD) | Presentation of a large dataset with the addition of multi-organ (liver, kidney, spleen, and pancreas) annotations to original datasets. Definition of benchmark and baseline for supervised, semi-supervised, weakly supervised, and continual learning. Code available at: https://github.com/JunMa11/ AbdomenCT-1K |

| Author | Application | Dataset Size | Model Architecture | Learning Strategy | Loss | Results | Main Contributions |
|---|---|---|---|---|---|---|---|
| Pan et al. [2022] | Segmentation of spleen, kidneys, gallbladder, esophagus, liver, stomach, aorta, vena cava, and pancreas | 59 (Institutional dataset without pancreas) 30 (BTCV) | VNet with Multi-layer perceptron Mixer replacing CNN | Supervised | Cross entropy Dice loss | BTCV: 79.00% (DSC) | Multi-layer perceptron mixer was integrated into VNet to linearize the computational complexity of transformers |
| Qiu et al. [2022a] | Segmentation of pancreas | 82 (NIH) 281 (MSD) | UNet3+ + Multi-scale feature calibration in both Localization and Segmentation | Supervised | Dice loss | NIH: 86.30% (DSC) 76.26% (Jaccard) 85.91% (Precision) 86.85% (Recall) MSD (Generalization): 85.41% (DSC) | Dual enhancement module to multiply the coarse segmentation probability map with the input image to coarse stage. Cropping of the output by the localizatio model. The cropped images are sent as input to fine stage. Multi-scale feature calibration module in both stages to calibrate features vertically to preserve boundary details and avoid feature redundancy |
| Qiu et al. [2022b] | Segmentation of pancreas | 82 (NIH) | UNet-like with: Spiking neural P systems (Localization) + (Segmentation) | Supervised | Cross entropy | 81.94% (DSC) | Deep dynamic spiking neural P systems are integrated into UNet to solve memory limitation of 3D CNNs |
| Qu et al. [2022] | Segmentation of pancreas | 224 (Internal) Peking Union Medical College Hospital 66 (External) Hehan Cancer Hospital | M3Net (3D Endoder-2D Decoder): Multi-scale Multi-view Multi-phase + Attention | Supervised | Binary cross entropy Mean square error of distance field between foreground and background | Internal: 90.29% (DSC) External: 86.34% (DSC) | Dual path segmentation models for arterial and venous phase. Each model is constituted by an encoder composed of 3D convolutions and a decoder composed of 2D convolutions. Inter-phase contextual information is explored via cross-phase non-local attention between the two models. Replication for axial, coronal, and sagittal views. Ensemble of the three views. Fusion of high and half resolution to capture local and global features |
| Qureshi et al. [2022] | Segmentation of pancreas | 82 (NIH) | VGG-19 (Localization) + UNet (Segmentation) | Supervised | Mean Dice | 88.53% (DSC) | A morphology prior (a 3D volume template), defining the general shape and size of the pancreas, was integrated with the soft label from the second stage to improve segmentation |
| Shi et al. [2022] | Segmentation of pancreas endocardium, right and left ventricle, and myocardium | 82 (NIH) | UNet V-Net ResNet-18 | Semi-supervised | Cross entropy loss for conservative and radical model (labeled data) Cross entropy based for certain regions + Consistency loss for uncertain regions (Unlabeled data) | UNet: 67.01% (DSC) V-Net: 79.67% (DSC) 66.69% (Jaccard) 1.89 voxels (ASD) 7.59 voxels (HD) ResNet-18: 80.58 %(DSC) 67.91% (Jaccard) 2.27 voxels (ASD) 8.34 voxels (HD) | A conservative-radical module to automatically identify uncertain regions. A training strategy to separately segment certain and uncertain regions. Mean teacher model for uncertain region segmentation |
| Sundar et al. [2022] | Segmentation of pancreas and non abdominal organs | 50 (internal) | nnUNet | Supervised | – | 85.00% (DSC) | Development of Multiple-organ objective segmentation (MOOSE) framework. Code available at: https://github.com/ ZhifangDeng/ MISSFormer |

| Author | Application | Dataset Size | Model Architecture | Learning Strategy | Loss | Results | Main Contributions |
|---|---|---|---|---|---|---|---|
| Yang et al. [2022b] | Segmentation of pancreas | 82 (NIH) 281 (MSD) | AX-UNet with Atrous spatial pyramid pooling | Supervised | Focal loss Dice loss | NIH: 87.70% (DSC) 78.20% (Jaccard) 92.90% (Precision) 90.90% (Recall) MSD: 85.90% (DSC) 77.90% (Jaccard) 93.10% (Precision) 86.30% (Recall) | Design of a loss function to address the blurry boundary issue. Code available at: https://github.com/zhangyuhong02/AX-Unet.git |
| You et al. [2022] | Segmentation of pancreas, and left atrium | 82 (NIH) + datasets of other organs | V-Net for knowledge distillation | Semi-supervised | Cross entropy Dice loss Mean squared error (Supervised) Design of: Boundary-aware contrastive, Pair-wise distillation, and Consistency losses | 89.03% (DSC) | Contrastive distillation model with multi-task learning (segmentation map and signed distance map from boundary). Structured distillation in the latent feature space followed by contrasting the boundary-aware features in the prediction space for better representations |
| Zeng et al. [2022] | Segmentation of pancreas, and left atrium | 82 (NIH) + datasets of other organs | V-Net | Semi-supervised | Cross entropy | 84.77% (DSC) 73.71% (Jaccard) 6.24 voxel (HD95) 1.58 voxel (ASD | Teacher-student trained in parallel: the student learns from pseudo labels generated by the teacher learning in turn from the performances of student on the labeled images |
| Zhao et al. [2022a] | Segmentation of spleen, kidneys, gallbladder, esophagus, liver, stomach, aorta, veins, pancreas, duodenum, colon, lung spinal cord, and heart | 82 (NIH) + other datasets | UNet | Supervised | Mean sqaure error loss | 88.00% (DSC) | 3D UNet used for contour interpolation |
| Zhao et al. [2022b] | Segmentation of aorta, gallbladder, kidneys, liver, pancreas, spleen, and stomach | 30 (Synapse) | UNet-like with: Encoder: ResNet-50 + Progressive sampling module + Vision Transformer (Hybrid CNN-Transformer) | Supervised | Cross entropy loss Dice loss | 59.84% (DSC) | A progressive sampling module to ensure that highly relevant regions of the organ are in the same patch |
| Zhu et al. [2022] | Segmentation of pancreas | 82 (NIH) 281 (MSD) 70 (Zheyi) | Residual blocks + Squeeze-Excitation Attention + UNet | Domain adaptation: Supervised learning (source) Unsupervised learning (target) | Cross entropy Dice loss | NIH adapted to Zheyi 72.73% (DSC) MSD adapted to Zheyi 71.17% (DSC) | Adversarial multiscale domain adaption (from source) to generalize to external datasets (target domain) |
| Dogan et al. [2021] | Segmentation of pancreas | 82 (NIH) | Mask R-CNN (Localization) + UNet (Segmentation) | Supervised | Binary cross entropy | 86.15% (DSC) 75.93% (Jaccard) 86.23% (Precision) 86.27% (Recall) 99.95% (Accuracy) | Less powerful GPUs are required |
| Hu et al. [2021] | Segmentation of pancreas | 82 (NIH) 70 (CT-Zheyi dataset) | DenseNet161 for Dense Atrous Spatial Pyramid Pooling (Localization) DenseNet161 for Distance-based saliency (Segmentation) | Supervised | Binary cross entropy | NIH: 85.49% (DSC) CT-Zheyi: 85.48% (DSC) | Dense atrous spatial pyramid Pooling to cover larger receptive fields. Saliency map is computed through geodesic distance based saliency transformation. Both localization and saliency information are used to aid segmentation |
| Huang et al. [2021a] | Segmentation of pancreas | 82 (NIH) | UNet + Deformable convolutional module | Supervised | Design of Focal generalized Dice loss | 87.25% (DSC) 88.98% (Precision) 89.97% (Recall) | Deformable convolutions for adaptive receptive fields. Focal generalized Dice loss to balance the size of foreground and background |

| Author | Application | Dataset Size | Model Architecture | Learning Strategy | Loss | Results | Main Contributions |
|---|---|---|---|---|---|---|---|
| Knolle et al. [2021] | Segmentation of pancreas, and brain | 281 (MSD) Generalization: 85 (Internal dataset) + dataset of other organ | UNet-like with Dilated convolutions | Supervised | Dice loss | 78.00% (DSC) 1.78 mm (HD) Generalization: 70.00% (DSC) | Small network with dilated convolutions designed to run on low-end hardware within federated learning |
| Li et al. [2021d] | Segmentation of pancreas | 82 (NIH) 281 (MSD) | Bi-directional recurrent UNet | Supervised | Dice similarity coefficent loss | 85.35% (DSC) 1.10 mm (ASD) 3.68 mm (HD) | Combination of a 2D slice with probabilistic map of two adjacent slice for local 3D context. The result is propagated through a 2.5D UNet. A bi-directional (forward-backward) recurrent scheme is applied to the primary segmentation to optimize the local 3D information. |
| Li et al. [2021e] | Segmentation of pancreas | 82 (NIH) | Dual GAN with UNet (Generators) and CNNs (Discriminators) + Pyramidal pooling | Supervised | Adversarial loss | 83.31% (DSC) 71.76% (Jaccard) 84.09% (Precision) 83.30% (Recall) | First GAN to preserve spatial information. Second GAN increases the preservation of spatial information and leads to more realistic segmentation results. Pyramidal pooling to replace original pooling layers in UNet |
| Li et al. [2021a] | Segmentation of pancreas | 82 (NIH) | Multi-scale selection Adversarial Multi-channel fusion UNet | Supervised | Basic loss of conventional segmentation Adversarial loss | 84.10% (Precision) 82.50% (DSC) | GAN with a generator integrating: Multi-scale field selection to grasp global spatial features; Multi-channel fusion integrating information from different locations to obtain comprehensive details |
| Li et al. [2021b] | Segmentation of pancreas | 82 (NIH) | Multi-level pyramidal pooling Adversarial UNet | Supervised | Adversarial loss | 83.03% (DSC) 84.60% (Recall) | Generator consisting of: UNet with residual blocks, and multi-level pyramidal pooling to gather contextual information |
| Li et al. [2021c] | Segmentation of pacnreas | 82 (NIH) | Dual GAN + Unet (DAUNet) + CNN for multilevel cue | Supervised | Adversarial loss | 83.08% (DSC) 71.39% (Jaccard) 82.19% (Recall) 2.22 mm (RMSE) | Two GANs with generators based on UNet. The second one fuses features from different convolutional layers to obtain additional details for segmentation. |
| Long et al. [2021] | Segmentation of pancreas | 82 (NIH) | Encoder with channel attention to enhance semantics + Feature fusion pooling attention module + Decoder | Supervised | – | 86.62% (DSC) 86.07% (Precision) 87.37% (Recall) | Parallel module in the encoder to extract semantic and spatial features. Channel attention module to enhance acquisition of semantic information. Both modules sent as input to feature fusion pooling attention to fuse semantic and spatial information |
| Ma et al. [2021a] | Segmentation of pancreas, lung, and cell contour | 82 (NIH) 281 (MSD) + datasets of other organs | UNet + Multi-scale convolutional block + Down-sampling + Context module | Supervised | Binary cross-entropy (lung) Dice loss (pancreas) | 88.48% (DSC) | Customized UNet with: convolutional modules concatenating features from three branches; a hybrid pooling consisting of max-pooling, average pooling, and convolutions; skip connections with atrous convolutions (context module) |

| Author | Application | Dataset Size | Model Architecture | Learning Strategy | Loss | Results | Main Contributions |
|---|---|---|---|---|---|---|---|
| Ma et al. [2021b] | Segmentation of pancreas, and left atrium | Combination of: 82 (NIH) and 281 (MSD) + dataset of another organ | VNet with Global active contour | Supervised | Dice loss L1 loss Geodesic active contour loss | 83.60% (DSC) 18.50 mm (HD) 1.93 mm (ASSD) | First application integrating geodesic active contour into CNN to reduce boundary errors. Geodesic active contour loss can consider more global information than dice loss or cross entropy loss because it is built on the level set function-based representation |
| Panda et al. [2021] | Segmentation of pancreas | 1,917 (Mayo Clinic) + 41 (TCIA) + 80 (NIH) | UNet for two stages: Localization + Segmentation | Supervised | Tversky loss Asymmetric dice loss | Internal dataset: 91.00% (DSC) TCIA (Generalization): 96.00% (DSC) NIH (Generalization): 89.00% (DSC) | Evaluation of dataset size on model performances: in the second stage 3D UNet was evaluated on 200; 500; 800; 1,000; 1,200; and 1,500 CTs (internal dataset). Generalization on two datasets |
| Petit et al. [2021] | Segmentation of pancreas | 82 (NIH) | UNet | Supervised Semi-supervised | – | 77.53% (DSC) | Fusion of a FCN probability prediction volume with 3D spatial prior representing the probability of organ presence |
| Shan and Yan [2021] | Segmentation of pancreas, skin lesions, and thyroid | 82 (NIH) + datasets of other organs | UNet: Encoder with residual blocks Decoder with Spatial attention + Channel attention | Supervised | Soft dice loss | 91.37% (DSC) 85.30% (Jaccard) 30.79 mm (ASSD) | Spatial attention to focus on the target spatial regions and to ignore irrelevant background. Channel attention to highlight the relevant channels and reduce the irrelevant ones |
| Shi et al. [2021] | Segmentation of liver, spleen, pancreas, and kidney | 30 (BTCV) 281 (MSD) + datasets of other organs | nnUNet | Supervised | Marginal loss Exclusive loss | 80.80% (DSC) 3.96 mm (HD) | Implementation of marginal loss (for background) label and exclusion loss (different organs are mutually exclusive) |
| Tian et al. [2021] | Segmentation of pancreas | 82 (NIH) | Markov chain Monte Carlo + UNet | Supervised | Binary cross entropy | 87.49% (DSC) 84.12% (Precision) 93.81% (Recall) | Markov Chain Monte Carlo applied to 3D UNet for patch selection during localization and segmentation. This method solved the issue of memory limit, class imbalance, and data scarcity in 3D segmentation |
| Wang et al. [2021b] | Segmentation of pancreas | 82 (NIH) | Dual input + v-mesh UNet + Attention + Spatial + Transformation and fusion | Supervised | Binary cross entropy | 87.40% (DSC) 89.50% (PPV) 87.70% (Sensitivity) | Dual input FCN: original CT and images processed by graph-based visual saliency with specific intensity features to grasp more information on the boundary. Horizontal and vertical connections with attention mechanism. Spatial transformation and fusion for deformable convolutions |
| Wang et al. [2021a] | Segmentation of pancreas | 82 (NIH) | UNet (Localization) View adaptive Unet (Segmentation) | Supervised | Dice loss Weighted focal loss | 86.19% (DSC) | Data augmentation on three axes. Axial, coronal, and sagittal volumes are fed simultaneously to the network |
| Xue et al. [2021] | Segmentation of pancreas | 82 (NIH) 59 (Fujian Medical University) | UNet for both: Localization and Segmentation | Supervised | Cross entropy Regression loss | NIH: 85.90% (DSC) 75.70% (Jaccard) 87.60% (Precision) 85.20% (Recall) Fujian: 86.90% (DSC) 77.30% (Jaccard) 91.00% (Precision) 83.50% (Recall) | Multi-task second stage. Regression (task 1) of object skeletons as descriptor of the shape of the pancreas to guide subsequent segmentation (task 2). Conditional random fields to remove small false segments |

| Author | Application | Dataset Size | Model Architecture | Learning Strategy | Loss | Results | Main Contributions |
|---|---|---|---|---|---|---|---|
| Yan and Zhang [2021] | Segmentation of pancreas | 82 (NIH) | UNet + Spatial attention + Channel attention (Localization and Segmentation) | Supervised | Dice loss | 86.61% (DSC) | 2.5D UNet with spatial and channel attention integrated into skip connections. |
| Zhang et al. [2021d] | Segmentation of pancreas | 82 (NIH) 281 (MSD) | CNN (Localization) Encoder-decoder (Segmentation) Prior propagation module (both stages) Scale-transferrable feature fusion module (second stage) | Supervised | Dice loss | NIH: 84.90% (DSC) MSD: 85.56% (DSC) | Scale-transferrable feature fusion module to learn rich fusion features with lightweight architecture. Prior propagation module to explore informative and dynamic spatial priors to infer accurate and fine-level masks |
| Zhang et al. [2021b] | Segmentation of liver, pancreas, spleen, and kidney | 30 (BTCV) 281 (MSD) + datasets of other organs | nnUNet + Auxiliary information into decoder | Supervised | Dice loss Focal loss | 83.97% (DSC) | Four datasets with annotations of different organs (liver, pancreas, spleen, and kidney). An auxiliary conditional tensor is concatenated into the decoder to select the specific organ to segment |
| Zhang et al. [2021c] | Segmentation of pancreas and brain | 82 (NIH) + dataset of other organs | Shared encoder and two decoders. Second decoder: Context residual Mapping + Context residual Attention | Supervised | Binary cross entropy Dice loss | 86.06% (DSC) | The context residual decoder takes the residual feature maps of adjacent slices produced by the decoder as its input, and provides feedback to the segmentation decoder as a kind of attention guidance |
| Zhang et al. [2021a] | Segmentation of pancreas | 36 (International Symposium on Image Computing and Digital Medicine) 82 (NIH) 281 (MSD) | Multi-atlas registration (Localization) 3D patch-based and 2.5D slice-based UNet (Segmentation) 3D level set to refine the probability map (Refine stage) | Supervised | Cross entropy Dice coefficient loss | 84.40% (DSC) 73.40% (Jaccard) | Coarse stage for localization. Fine stage for segmentation: 3D patch-based and 2.5D slice-based CNN to extract local and global features. Refine stage to improve segmentation: 3D level-set for better boundary delineation. |
| Bagheri et al. [2020] | Segmentation of pancreas | 82 (NIH) | Superpixels and random forest classifier (Localization) Holistcally nested neural networks (Segmentation) | Supervised | – | 78.00% (DSC) | Superpixels to get bounding boxes. Fusing holistically nested networks to generate interior and boundary |
| Boers et al. [2020] | Segmentation of pancreas | 82 (NIH) | UNet | Supervised | DSC-based loss weighted by voxel-specific map + Loss for volume difference | 78.10% (DSC) | Scribbles are drawn to refine initial segmentation |
| Chen et al. [2020b] | Segmentation of pancreas | 82 (NIH) | UNet, ResNet, DSN Encoder: Squeeze and excitation Decoder: Hierarchical fusion | Supervised | Weighted cross entropy | UNet: 87.04% (DSC) ResNet: 87.26% (DSC) DSN: 82.53% (DSC) | Hierarchical fusion model to retain boundary information |
| Gong et al. [2020] | Segmentation of pancreas | 40 (ISICDM) | UNet | Supervised | – | 83.00% (DSC) 85.00% (Recall) | Fractional differentiation to increase the pancreas contrast. Level set (regularization term, intensity constraint term and length term to increase accuracy at contours) |

| Author | Application | Dataset Size | Model Architecture | Learning Strategy | Loss | Results | Main Contributions |
|---|---|---|---|---|---|---|---|
| Isensee et al. [2020] | Segmentation of lung, heart, atrium, ventricles, myocardium, aorta, trachea, hyppocampus, esophagus, liver, kidneys, pancreas, spleen, colon, gallbladder, and stomach | 281 (MSD) + datasets of other organs | nnUNet | Supervised | Cross entropy loss Dice loss Weighted binary cross entropy loss | 2D UNet: 77.38% (DSC) 3D UNet Full resolution: 82.17% (DSC) 3D UNet low resolution: 81.18% (DSC) | Original paper on the implementation of nnUNet. Three configurations: 2D UNet, 3D UNet with full resolution, and 3D UNet with low resolution. Code available at: https://github.com/MIC-DKFZ/ nnUNet?tab=readme-ov-file |
| Li et al. [2020c] | Segmentation of pancreas | 82 (NIH) | UNet + Multi-scale convolution + Residual blocks | Supervised | Dice loss | 87.57% (DSC) 78.77% (Jaccard) | Three strategies to solve over-segmentation, under-segmentation, and shape inconsistency: skip network (adding residuals between encoder and decoder directly), residual network (adding residuals to the continuous convolution blocks of the encoder and decoder separately) multi-scale residual network (with multi-scale convolution block between high-resolution endocer and decoder) |
| Li et al. [2020a] | Segmentation of pancreas | 82 (NIH) | Two stages GAN based on UNet | Supervised | Conventional segmentation loss Adversarial loss | 83.06% (DSC) 71.41% (Jaccard) | Adversarial training on a model already trained with GAN |
| Li et al. [2020b] | Segmentation of pancreas | 82 (NIH) 281 (MSD) | Multi-scale Attention dense Residual UNet | Supervised | Binary cross entropy Dice loss | NIH: 86.10% (DSC) 75.55% (Jaccard) 86.43% (Sensitivity) 84.97% (Specificity) 4.40mm (HD) 1.27 mm (ASD) MSD: 88.52% (DSC) 79.42% (Jaccard) 91.86% (Sensitivity) 89.66% (Specificity) 3.78 mm (HD) 0.95 mm (ASD | Multi-scale convolution and channel attention to solve interclass indistinction. Dense residual blocks to solve intraclass inconsistency. Code available at: https://github.com/Mrqins/ pancreas-segmentation |
| Mo et al. [2020] | Segmentation of pancreas | 82 (NIH) | VGGNet with extraction of hierarchical features at different levels | Supervised | Dice loss | 82.47% (DSC) | 3D residual network to extract and aggregate hierarchical features at different levels. Concatenation of the result with features at each level to choose more discriminative features. This process is iterated. |
| Ning et al. [2020] | Segmentation of pancreas | 82 (NIH) | GAN Generator with: Autoencoder with Dilated convolution + LSTM | Supervised | Adversarial loss | 89.87% (DSC) 95.85% (Accuracy) | First application integrating dilated convolutions, GAN, and LSTM. Generator: dilated convolution autoencoder with dilated convolution layers in the encoder, and an LSTM boosting the pancreas boundary segmentation by modeling the contextual spatial correlation between neighbouring CT scan patches |
| Nishio et al. [2020] | Segmentation of pancreas | 82 (NIH) | deepUNet | Supervised | Dice loss | 78.90% (DSC) 65.80% (Jaccard) 76.20% (Recall) | Use of three data augmentation methods: conventional ones, mixup, and random image cropping and patching |
| Park et al. [2020] | Segmentation of pancreas and other 16 anatomical structures | 1,150 (John Hopkins) | Two-stage Organ attention network | Supervised | – | 87.80% (DSC) | Annotation of 22 structures. Use of two-stage organ attention network: two FCN for segmentation. The first used reverse connections to get more semantic information. The results became attention-organ module to guide the second network. This architecture was applied to each view. The outputs from axial, coronal, and sagittal views were then fused |

| Author | Application | Dataset Size | Model Architecture | Learning Strategy | Loss | Results | Main Contributions |
|---|---|---|---|---|---|---|---|
| Tong et al. [2020] | Multi-organ Segmentation | 90: 43 (TCIA) 47 (BTCV) | Encoder-Decoder with dual attention: Squeeze and Excitation (channel attention) Convolutional layer (spatial attention) | Supervised | – | 79.24% (DSC) 1.82 mm (ASD) | A self-paced learning strategy for the multi-organ segmentation to adaptively adjust the weight of each class |
| Xia et al. [2020] | Segmentation of pancreas | 82 (NIH) 90: 43 (TCIA) + 47 (BTCV) 281 (MSD) + MSD (liver) | Encoder-Decoder based on ResNet18 for Multi-view Co-training and Domain-adaptation | Semi-supervised Unsupervised | Combination of conventional segmentation loss (labeled) and computational function based on uncertainty-weighted label fusion (unlabeled) | NIH: 81.18% (DSC) TCIA+BTCV (External validation): 77.91% (DSC) MSD (Domain adaptation): 74.38% (DSC) | Co-training to maximize the similarity of the predictions among different views, generated by rotation or permutation transformations. Uncertainty weighted label fusion module for accurate pseudo labels generation for each view. Adaptation from multi-organ to pancreas dataset without source domain data |
| Zheng et al. [2020] | Segmentation of pancreas | 82 (NIH) | 3D VNet (Localization) 2.5D Encoder-decoder (Segmentation) | Self supervised | Square root Dice loss | 78.10% (DSC) | Square Root Dice loss to deal with the trade-off between sensitivity and specificity. Slice shuffle for pre-training before input to the network which learns to reorder and understand organ shape. Capturing of non-local information through attention, pooling, and convolutional layers. Ensemble learning and recurrent refinement to improve accuracy |
| Zhu et al. [2020] | Segmentation of pancreas, liver, and prostate | 82 (NIH) + datasets of other organs | UNet + Residual blocks + Attention focused modules | Supervised | – | 83.90% (DSC) | Attention modules into skip connections to focus on segmented regions and reduce influence of background. Dense connected residual blocks in down-sampling and up-sampling to reduce computational load and network parameters. Code available at: https://github.com/ahukui/SIPNet |
| Karimi and Salcudean [2020] | Segmentation of pancreas, liver, and prostate | 282 images + datasets of other organs | UNet | Supervised | Losses based on: distance transform Morphological operations Convolutions with circular/spherical kernels | 78.40% (DSC) 21.3 mm (HD) 1.84 mm (ASD) | Three different methods to reduce HD: distance-transform, morphological erosion, and convolutions with circular/spherical kernels. Three losses based on these methods for stable training |
| Liu et al. [2020] | Segmentation of pancreas | 82 (NIH) | ResNet (Localization) Ensemble UNet (Segmentation) | Supervised | Dice loss Focal loss Jaccard distance loss Class balanced cross entropy Binary cross entropy | 84.10% (DSC) 72.86% (Jaccard) 84.35% (Precision) 85.33% (Recall) | Superpixes generated by oversegmentation. Classification of superpixels by ResNet. candidate regions obtained by ensemble of classification results of three different scale of superpixels. Segmentation by ensemble of multiple network with different loss functions |
| Lu et al. [2019] | Segmentation of pancreas | 82 (NIH) | UNet + Channel attention + Spatial attention + Ring residual module | Supervised | Design of Complex-coefficient loss | 88.32% (DSC) | Ringed residual module, consisting of forward and backward residual propagation to address the boundary blur issue of pancreas. Convolutional block attention module with spatial and channel attention to improve accuracy. Complex-coefficient loss to focus not only on the ratio of the coincident area to the total area, but also on the shape similarity between the real result and the predicted result |

| Author | Application | Dataset Size | Model Architecture | Learning Strategy | Loss | Results | Main Contributions |
|---|---|---|---|---|---|---|---|
| Man et al. [2019] | Segmentation of pancreas | 82 (NIH) | Localization agent (Localization) + Deformable UNet (Segmentation) | Reinforcement (Localization) Supervised (Segmentation) | Dice loss | 86.93% (DSC) | First application of Deep Q Learning to medical image segmentation. Localization agent to adjust localization, by learning a localization error correction policy based on deep Q network. Deformable convolution for learnable receptive fields, instead of fix ones |
| Schlemper et al. [2019] | Segmentation of pancreas | 82 (NIH) | UNet + Attention gate | Supervised | Dice loss | 83.10% (DSC) 82.50% (Jaccard) 84.10% (Recall) | Attention gate in integrated into skip connections of UNet to highlight salient features and suppress irrelevant regions Code available at: https://github.com/ozan-oktay/Attention-Gated-Networks |
| Wang et al. [2019a] | Segmentation of pancreas | 281 (MSD) | UNet with: Residual blocks nested with dilations + Squeeze and excitation | Supervised | Focal loss | 84.76% (DSC) | Residual blocks nested with dilations added in the first few layers to help network adapt to targets of any size. Squeeze and excitation to boost essential features for each task |
| Zeng and Zheng [2019] | Segmentation of pancreas, hip, and lumbar intravertebral discs | 82 (NIH) + datasets of other organs | UNet + Holistic Decomposition Convolution + Dense Upsampling Convolution | Supervised | Cross entropy loss Dice loss | 83.00% (DSC) | Network agnostic segmentation approach. Holistic decomposition convolution to reduce size of data for subsequent processing: periodic down-shuffling to input to get low resolution channels, followed by convolutions on these channels. Periodic dense upsampling convolutions to recover full resolution: low resolution convolutions with periodic up-shuffling |
| Chen et al. [2018] | Segmentation of pancreas | 150 (Internal) | DRINet | Supervised | Cross entropy loss | 83.42% (DSC) 87.95% (Precision) 80.29% (Recall) | Implementation of DRINet consisting of dense connection blocks, residual inception, and unspooling blocks |
| Gibson et al. [2018] | Segmentation of spleen, kidney, gallbladder, esophagus, liver, stomach, pancreas, and duodenum | 90: 43 (TCIA) 47 (BTCV) | DenseVNet | Supervised | L2 regularization loss Dice loss | 78.00% (DSC) 5.90 mm (HD95) | Implementation of DenseVNet with: cascaded dense feature stacks, V-network with downsampling and upsampling, dilated convolutions, map concatenation, and a spatial prior. Application to eight abdominal organs |
| Heinrich et al. [2018] | Segmentation of pancreas | 82 (NIH) | UNet + Ternary weights + Ternary activations | Supervised | Weighted cross entropy | 71.00% (DSC) | Implementation of TernaryNet with ternary weigths and ternary hyberbolic tangent to reduce computational load. |
| Roth et al. [2018b] | Segmentation of artery, vein, liver, spleen, stomach, gallbladder, and pancreas | 331 (internal for training) 150 (external for testing) | 3D UNet (Localization) and (Segmentation) | Supervised | Weighted cross entropy loss | External dataset: 82.20% (DSC) | Application of cascaded networks for localization (coarse stage) and segmentation (fine stage) |
| Roth et al. [2018a] | Segmentation of pancreas | 82 (NIH) | Holistically-nested networks for: Localization (fusing the three orthogonal axes) + Segmentation (boundaries and interior cues to produce superpixels aggregated by random forests) | Supervised | Cross entropy loss | 81.27% (DSC) 68.87% (Jaccard) 17.71 mm (HD) 0.42 mm (Average distance) | Segmentation incorporates deeply learned organ interior and boundary mid-level cues with subsequent spatial aggregation |

| Author | Application | Dataset Size | Model Architecture | Learning Strategy | Loss | Results | Main Contributions |
|---|---|---|---|---|---|---|---|
| Farag et al. [2017] | Segmentation of pancreas | 80 (NIH) | Oversegmentation for superpixels (Middle-level representations) + Random forest classifier + AlexNet | Supervised | – | 70.70% (DSC) 57.90% (Jaccard) 71.60% (Precision) 74.40% (Recall) | Bottom-up approach for image segmentation, consisting of: superpixels (from oversegmentation of slices), patch labeling by random forests or deep learning, and cascaded random forests classifiers based on previous patch labeling |

## B  Full list of the reviewed studies on the segmentation of tumors, cysts, and inflammations of the pancreas

| Author | Application | Dataset Size | Model Architecture | Learning Strategy | Loss | Results | Main Contributions |
|---|---|---|---|---|---|---|---|
| Cao and Li [2024] | Parenchyma and tumors | 82 (NIH) 281 (MSD) | UNet with: High resolution spatial information recovery + Multi-scale high resolution pre-segmented feature fusion + Pyramid multi-scale feature perception and fusion | Supervised | Difficulty-guided adaptive boundary-aware loss | Parenchyma (NIH): 88.96% (DSC) 89.27% (Precision) 89.98% (Recall) Parenchyma (MSD): 89.52% (DSC) 93.19% (Precision) 88.71% (Recall) Tumors (MSD): 54.38% (DSC) 69.58% (Precision) 53.17% (Recall) | High-resolution spatial information recovery module: encoder and decoder features of the same layer are sent to high resolution spatial information filtering module to extract high-resolution pre-segmented images, which are then fused. Multi-scale high-resolution pre-segmented feature fusion module: features of the encoder and decoder finely processed into a high-resolution pre-segmented feature map. Pyramid multi-scale feature perception and fusion module uses the extracted pre-segmented images to guide the network to focus on the dimensional changes of the segmented targets. Design of Difficulty-guided adaptive boundary-aware loss function to address the class imbalance and improve segmentation of uncertain boundaries |
| Cao et al. [2023b] | Parenchyma and tumors | 82 (NIH) 420 (MSD) | UNet with three attention mechanisms on skip connections: Spatial + Channel + Multi-dimensional features | Supervised | Weighted cross entropy loss | Parenchyma (NIH): 83.04% (DSC) 81.71% (Precision) 84.42% (Recall) Parenchyma (MSD): 83.39% (DSC) 85.51% (Precision) 81.37% (Recall) Tumors (MSD): 40.15% (DSC) 52.32% (Precision) 35.29% (Recall) | Design of a loss function to capture edge details of pancreas and tumors. Multi-dimensional attention gate integrated into skip connections for small target feature localization in multiple dimensions of space and channels, and for filtering redundant information in shallow feature maps, thus enhancing the feature representation of the pancreas and pancreatic tumor |
| Deng et al. [2023] | Acute pancreatitis | 89 (Internal) | FCN + Region proposal network (Detection) UNet (Segmentation) | Supervised | Focal loss Cross entropy loss L1 regression loss | 66.82% (DSC) | FCN for detection of pancreatitis region. The detected region was cropped and sent to the 2D U-Net for segmentation. First study on segmentation on acute pancreatitis |
| Du et al. [2023] | Pancreas ductal adenocarcinoma | 55 (Qingdao University Hospital) 281 (MSD) | UNet with multi-scale channel attention | Supervised | Binary cross entropy | Qingdao: 68.03% (DSC) 59.31% (Jaccard) 12.04 mm (HD) MSD: 80.12% (DSC) 74.17 (Jaccard) 2.26 mm (HD) | Integration of multi-scale convolutions and channel attention into each encoder and decoder block |
| Duh et al. [2023] | Pancreatic cysts | 335 (Internal) Spain | UNet with Attention gate in skip connections | Supervised | Dice loss | 93.10% (Recall) | Attention gate integrated into skip connections for segmentation of pancreatic cysts |

| Author | Application | Dataset Size | Model Architecture | Learning Strategy | Loss | Results | Main Contributions |
|---|---|---|---|---|---|---|---|
| He and Xu [2023] | Parenchyma and tumors | 420 (MSD) + dataset of other organs | Hybrid CNN-Transformer Encoder: (3D Swin-Transformer + boundary extracting module) + Boundary preserving module + Decoder: CNN | Supervised | Dice loss Cross entropy loss | Parenchyma: 81.47% (DSC) 1.77 mm (ASSD) Tumor: 51.83% (DSC) 17.13 mm (ASSD) | Application of boundary awareness into 3D CNN and transformers. Swin-transformer as encoder and auxiliary boundary extracting module to obtain rich and discriminative feature representations. Boundary preserving module to fuse boundary map and features from the encoder |
| Ju et al. [2023] | Parenchyma and tumors | 82 (NIH) 281 (MSD) | UNet: Spatial visual cue fusion + Active localization offset (Localization) UNet (Segmentation) | Supervised | Dice loss Binary cross entropy loss | Parenchyma (NIH): 85.15% (DSC) Tumor (MSD): 63.36% (DSC) | Spatial visual cue fusion, based on conditional random field, learns global spatial context. It combines the correlations between all pixels in the image to optimize the rough and uncertain pixel prediction during the coarse stage. Active localization offset adjusts dynamically the localization results during the coarse stage. Code available at https://github.com/PinkGhost0812/SANet |
| Li et al. [2023g] | Pancreatic cysts | 107 (internal) | UNet with: Atrous pyramid attention module + Spatial pyramid pooling module | Supervised | Dice loss Binary cross entropy loss | 84.53% (DSC) 75.81% (Jaccard) | Atrous pyramid attention module and spatial pyramid pooling module inserted into bottleneck layer to extract features at different scales, and contextual spatial information, respectively |
| Li et al. [2023e] | Parenchyma and tumors | 281 (MSD) | nnUNet with attention + Balance temperature loss + Rigid temperature optimizer + Soft temperature indicator | Supervised | Balance temperature loss | Parenchyma: 85.06% (DSC) Tumors: 59.16% (DSC) | Segmentation of both pancreas and tumors. Balance temperature loss to dynamically adjust weights between tumors and the pancreas. Rigid temperature optimizer to avoid local optima. Soft temperature indicator to optimize the learning rate |
| Mukherjee et al. [2023] | Pancreas ductal adenocarcinoma | 1,151: Mayo Clinic + 152 from MSD and 41 from TCIA | 3D nnUNet | Supervised | Dice loss Cross entropy loss | Overall: 84.00% (DSC) 4.6 mm (HD) Generalization on MSD: 82.00% (DSC) 2.6 mm (HD) Generalization on TCIA: 84.00% (DSC) 4.30 mm (HD) | Bounding boxes by cropping the CT images to a 3D bounding box centered around the tumor mask. nnUNet applied to bounding boxes |
| Ni et al. [2023] | Recurrence of pancreas ductal adenocarcinoma after surgery | 205 (Internal) 64 (For recurrence prediction with radiomics) | AX-UNet with Atrous spatial pyramid pooling | Supervised | – | 85.90% (DSC) 74.20% (Jaccard) 89.70% (Precision) 87.60% (Recall) | AX-UNet combining UNet and atrous spatial pyramid pooling. Code available at: https://github.com/zhangyuhong02/AX-Unet |
| Qu et al. [2023] | Parenchyma and masses (tumors, cysts) | 313 (Peking Union Medical College Hospital) 53 (Guandong General Hospital) (generalization) 50 (Jingling Hospital) (generalization) MSD (420) (generalization) | Swin Transformer and 3D CNN (Based on M3NET) Feature alignment: Transformer guided fusion + Cross-network attention (Decoder) | Supervised | Weighted cross entropy loss | Pancreas: Peking: 92.51% (DSC) Guangdong: 89.56% (DSC) Jingling: 88.07% (DSC) MSD: 85.71% (DSC) Masses: Peking: 80.51% (DSC) Guangdong: 67.17% (DSC) Jingling: 69.25% (DSC) MSD: 43.86% (DSC) | CNN and transformer branches perform separate feature extraction in the encoder. Progressive fusion between CNN and transformer in the decoder. Transformer guidance flow to address the inconsistency of the feature resolution and channel numbers between the CNN and transformer branches. Cross network attention into CNN decoder to enhance fusion capability with the transformer |

| Author | Application | Dataset Size | Model Architecture | Learning Strategy | Loss | Results | Main Contributions |
|---|---|---|---|---|---|---|---|
| Wang et al. [2023] | Tumors | 93 (Shanghai Changhai Hospital) | 3D UNet-like: Encoder: Multi-modal fusion downsampling block Decoder: Multi-modal mutual calibration block using attention | Supervised | Dice loss | 76.20% (DSC) 63.08% (Jaccard) 6.84 mm (HD) 75.96% (Precision) 84.26% (Recall) | Multi-modal fusion downsampling block to fuse semantic information from PET and CT, and to preserve unique features of different modal images. Multi-modal mutual calibration block to calibrate different scale semantics of one modal images guided by attention maps from the other modal images |
| Zhou et al. [2023] | Tumor | 116 abnormal pancreas 42 normal pancreas (internal) | Dual branch encoder-decoder (Pancreas segmentation) Encoder-decoder: Contrast enhancement block + Reverse attention block (Tumor segmentation) | Supervised | Dice loss | Abnormal: 78.72% (Jaccard) 89.07% (Precision) 87.42% (Recall) Normal: 87.74% (Jaccard) 91.47% (Precision) 95.50% (Recall) | Dual branch encoder combining semantic information extraction and detailed information extraction. Aggregation of feature maps of the two branches. Decoder to segment pancreas. Enhancement encoder-decoder network to improve segmentation accuracy of pancreatic tumors. Contrast enhancement block after each encoding step to extract the edge detail information. Reverse attention block inverting the decoder feature to guide the extraction of effective information in the encoder to generate an accurate prediction map |
| Zou et al. [2023] | Dilated pancreatic duct | 150 (Internal) Nanjing Drum Tower Hospital 40: Jiangsu Province Hospital of Chinese Medicine (Generalization) | 3D nnUNet for: (Localization) Terminal anatomy attention module (Segmentation) Terminal distraction attention module (Refine stage) | Supervised | Terminal Dice loss | Internal: 84.17% (DSC) 11.11 mm (HD) Generalization: 82.58% (DSC) | First work on errors on terminal regions of the dilated pancreatic duct. Terminal anatomy attention module to learn the local intensity from the terminal CT images, feature cues from the coarse predictions, and global anatomy information. Terminal distraction attention module to reduce false positive and false negative cases. Design of terminal Dice loss for segmentation of tubular structures |
| Guo et al. [2022b] | Chronic inflammation of choledoch | 76 (internal) | UNet++ | Supervised | Binary cross entropy loss | 83.90% (DSC) | UNet++ to segment chronic inflammation of choledoch in pediatric patients. Then ResUNet is used to classify the degree of severity of inflammation |
| Li et al. [2022b] | Tumors | 163 (Shanghai Jiao Tong University) 468 MRI (for style transfer) 281 (MSD) (generalization) | CycleGAN-like for: Synthetic data from MRI (Style transfer) ResNet: Extraction of knowledge from MRI (Meta-learning I) + Integration with salient knowledge from CT (Meta-learning II) | Supervised | Adversarial loss Cycle consistency loss Dice loss | Shanghai Jiao Tong University: 64.12% (DSC) MSD: 57.62% (DSC) | First study on meta-learning from one to a different modality. Random style transfer on MRI: generation of synthetic images with continuously intermediate styles between MRI and CT to simulate domain shift. First meta-learning: the model learns the common knowledge of synthetic data, and provides pancreatic cancer-related prior knowledge for the target segmentation task. Second meta-learning: the model learns the salient knowledge of the CT data to enhance segmentation |

| Author | Application | Dataset Size | Model Architecture | Learning Strategy | Loss | Results | Main Contributions |
|---|---|---|---|---|---|---|---|
| Mahmoudi et al. [2022] | Tumors and surrounding vessels | 138 (MSD) | 3D local binary pattern (Localization) Ensemble of: Attention gate + Texture Attention block (Scale invariant feature transform and local binary pattern) (Segmentation) | Supervised | Generalized Dice loss Weighted Pixel-wise Cross entropy loss Boundary loss | Tumor: 60.60% (DSC) 3.73 mm (HD95) 57.80% (Precision) 78.00% (Recall) Superior mesenteric artery: 81.00% (DSC) 2.89 mm (HD95) 76.00% (Recall) 87.00% (Precision) Superior mesenteric vein: 73.00% (DSC) 3.45 mm (HD95) 68.00% (Precision) 81.00% (Recall) | Design of texture attention block with scale invariant feature transform or local binary pattern to provide a comprehensive representation of pathological tissue. Integration of attention gate and texture attention gate into skip connections of texture attention UNet. Use of a 3D CNN as an ensemble of attention UNet and texture attention UNet. Design of Generalized Dice loss, Weighted Pixel-wise Cross entropy loss, and Boundary loss to address unbalanced data, and boundary between pancreas and tumors |
| Shen et al. [2022] | Dilated pancreatic duct | 82 (NIH) for localization 30 (internal) for segmentation | 3D UNet (Localization) 3D UNet + Squeeze and excitation (Segmentation) | Supervised | Dice loss Focal loss | NIH: 75.9% (DSC) 72.4% (Recall) Internal: 49.90% (DSC) 51.90% (Recall) | First study on automated 3D segmentation of dilated pancreatic duct. Generation of an annotated dataset of dilated pancreatic duct. Attention block with squeeze and excitation inserted into the bottleneck of a 3D UNet |
| Chaitanya et al. [2021] | Tumors | 282 (MSD) | GAN + UNet | Semi-supervised | Adversarial loss | 52.90% (DSC) | Semi-supervised learning for data augmentation. Adversarial term to help two generators synthesize diverse set of shape and intensity variations present in the population, even in scenarios where the number of labeled examples are extremely low. Code available at: https://github.com/krishnabits001/ task_driven_data_augmentation |
| Huang et al. [2021b] | Pancreatic neuroendocrine neoplasms | 98 (First Affiliated Hospital of Sun Yat-Sen University and Cancer Center of Sun Yat-Sen University) 72 (from both above centers) | UNet | Supervised | Cross entropy loss | First dataset: 81.80% (DSC) 83.60% (Precision) 81.40% (Recall) Second dataset: 74.80% (DSC) 87.20% (Precision) 68.60% (Recall) | A radiologists identified tumors by drawing bounding boxes to delineate region of interest sent as input to UNet. Radiomic analysis to predict pathohistologic grading |
| Si et al. [2021] | Pancreatic ductal adenocarcinoma and other types of tumors | 319 for training (Second Affiliated Hospital Shanghai) 347 for testing (First and Second Affiliated Hospital Shanghai) | ResNet18 (Localization) UNet32 (Segmentation) | Supervised | Cross entropy loss | 83.70% (DSC) | Three different networks used for pancreas location, segmentation, and diagnosis (presence of tumors) |
| Wang et al. [2021c] | Pancreatic ductal adenocarcinoma | 800 (John Hopkins) 281 (MSD) (generalization) | UNet with Inductive attention guidance | Semi-supervised | Cross entropy loss | John Hopkins: 60.28% (DSC) 99.75% (Recall) MSD: 32.49% (DSC) | Attention guided framework for classification and segmentation with partially labeled data (few annotated images for segmentation). Training using multiple instance learning with cancer and background regions as bags instead of per-voxel pseudo labels as in typical semi-supervised learning |
| Turečková et al. [2020] | Parenchyma and tumors | 420 (MSD) + datasets of other organs | UNet and VNet with Attention gate in skip connections | Supervised | Dice loss Cross entropy loss | Parenchyma (UNet): 81.81% (DSC) 81.21% (Precision) 84.51% (Recall) Tumors (UNet): 52.68% (DSC) 62.98% (Precision) 55.84% (Recall) Parenchyma (VNet): 81.22% (DSC) 80.61% (Precision) 84.10% (Recall) Tumors (VNet): 52.99% (DSC) 64.62% (Precision) 54.39% (Recall) | Attention gate integrated into skip connections for segmentation of pancreatic tumors |

| Author | Application | Dataset Size | Model Architecture | Learning Strategy | Loss | Results | Main Contributions |
|---|---|---|---|---|---|---|---|
| Xie et al. [2020] | Parenchyma and pancreatic cysts | 82 (NIH) 200 (John Hopkins: renal donors) 131 (John Hopkins: pancreatic cysts) | VGGNet with Hierarchical recurrent saliency transformation network between localization and segmentation | Supervised | Dice loss | NIH: 84.53% (DSC) Renal donors: 87.74% (DSC) Pancreatic cysts: 83.31% (DSC) | Saliency transformation module between first and second stage to transforms the segmentation probability map as spatial weights, iteratively, from the previous to the current iteration. Hierarchical version to segment first the pancreas and then the internal cysts. Code available at: https://github.com/198808xc/ OrganSegRSTN |